\documentclass[12pt]{article}
\pdfoutput=1 

\usepackage[utf8]{inputenc}
\usepackage{jheppub}
\usepackage{amsmath}
\usepackage{amsfonts}
\usepackage{amssymb}
\usepackage{mathtools}
\usepackage{amsthm}
\usepackage{dsfont}
\usepackage{hyperref}
\usepackage[capitalise,sort]{cleveref}
\usepackage{pgfplots}
\usepackage{tabularx}
\usepackage{environ}
\usepackage[inline]{enumitem}
\usepackage{comment}
\usepackage{tikz}
\usetikzlibrary{decorations.pathmorphing,calc,fit,matrix,cd,shapes,arrows,positioning,chains,positioning,arrows.meta}
\usepackage[style=numeric-comp,sorting=none,maxbibnames=99,giveninits=true,backend=bibtex]{biblatex}
\usepackage{blkarray}
\usepackage{shadow}
\usepackage{fancybox}
\usepackage{bm}
\usepackage{abstract}
\usepackage[normalem]{ulem}

\DeclareFieldFormat{doilink}{
\iffieldundef{doi}{#1}
{\href{http://dx.doi.org/\thefield{doi}}{#1}}}

\DeclareBibliographyDriver{article}{%
  \usebibmacro{bibindex}%
  \usebibmacro{begentry}%
  \usebibmacro{author/translator+others}%
  \setunit{\labelnamepunct}\newblock
  \usebibmacro{title}%
  \newunit
  \printlist{language}%
  \newunit\newblock
  \usebibmacro{byauthor}%
  \newunit\newblock
  \usebibmacro{bytranslator+others}%
  \newunit\newblock
  \printfield{version}%
  \newunit\newblock
  \usebibmacro{in:}%
  \href{http://dx.doi.org/\thefield{doi}}{%
  \usebibmacro{journal+issuetitle}%
  \newunit\newblock
  \usebibmacro{byeditor+others}%
  \newunit\newblock
  \usebibmacro{note+pages}}%
  \newunit\newblock
  \iftoggle{bbx:isbn}
   {\printfield{issn}}
   {}%
  \newunit\newblock
  \usebibmacro{doi+eprint+url}%
  \newunit\newblock
  \usebibmacro{addendum+pubstate}%
  \newunit\newblock
  \usebibmacro{pageref}%
  \usebibmacro{finentry}
  }
  
\DeclareFieldFormat{url}{\mkbibacro{URL}\addcolon\space\href{#1}{Link}}

\pgfplotsset{
        compat=1.9,
        compat/bar nodes=1.8,
    }

\makeatletter
\def\@xfootnote[#1]{%
	\protected@xdef\@thefnmark{#1}%
	\@footnotemark\@footnotetext}
\makeatother

\definecolor{prhigh}{HTML}{ff0000}
\definecolor{sechigh}{HTML}{e0fbfc}
\definecolor{prcolor}{HTML}{1d3557}
\definecolor{seccolor}{HTML}{457b9d}
\definecolor{tercolor}{HTML}{98c1d9}

\newcommand\cC{\mathcal{C}}

\newcommand\bea{\begin{eqnarray}}
\newcommand\eea{\end{eqnarray}}

\newcommand\sS{{\bf S}}

\theoremstyle{plain}

\theoremstyle{definition}

\newtheorem{claim}{Claim}

\theoremstyle{remark}



\usepackage{mathtools}

\DeclareMathOperator{\U}{U}
\DeclareMathOperator{\SU}{SU}

\DeclareMathOperator{\SL}{SL}

\newcommand{\de}{\partial}

\newcommand{\PP}{\mathbb{P}}
\newcommand{\RR}{\mathbb{R}}
\newcommand{\ZZ}{\mathbb{Z}}

\newcommand{\ID}{\mathds{1}}

\newcommand{\coma}{\, , \quad}
\newcommand{\fstop}{\, .}

\newcommand{\with}{\quad\text{with}\quad}

\newcommand{\YM}{\text{\tiny YM}}
\newcommand{\cV}{\mathcal{V}}

\newcommand{\Pl}{\text{\tiny Pl}}
\newcommand{\KK}{\text{\tiny KK}}

\newcommand{\spc}{\text{\tiny sp}}
\newcommand{\Kt}{\text{\tiny K3}}
\newcommand{\WGC}{\text{\tiny WGC}}
\newcommand{\QG}{\text{\tiny QG}}

\newcommand{\het}{\text{\tiny het}}

\renewcommand{\epsilon}{\varepsilon}

\makeatletter
\newsavebox{\measure@tikzpicture}
\NewEnviron{scaletikzpicturetowidth}[1]{%
  \def\tikz@width{#1}%
  \begin{lrbox}{\measure@tikzpicture}%
  \BODY
  \end{lrbox}%
  \pgfmathparse{#1/\wd\measure@tikzpicture}%
  \BODY
}
\makeatother

\makeatletter
\newcommand{\supsetcong}{\mathrel{\mathpalette\supset@cong\relax}}
\newcommand{\supsetsim}{\mathrel{\mathpalette\supset@sim\relax}}

\newcommand{\supset@cong}[2]{%
  \vbox{\offinterlineskip\m@th
    \ialign{\hfil##\cr
      \scalebox{0.9}{$#1\sim$}\cr
      \noalign{%
        \ifx#1\displaystyle\kern-0.5pt\else
        \ifx#1\textstyle\kern-0.5pt\fi\fi
      }%
      $#1\supset$\cr
    }%
  }%
}
\newcommand{\supset@sim}[2]{%
  \vtop{\offinterlineskip\m@th
    \ialign{\hfil##\cr
      $#1\supset$\cr\noalign{\kern0.5pt}\scalebox{0.9}{$#1\sim$}\cr
    }%
  }%
}
\makeatother

\makeatletter
\newcommand{\inlineitem}[1][]{%
\ifnum\enit@type=\tw@
    {\descriptionlabel{#1}}
  \hspace{0pt}%
\else
  \ifnum\enit@type=\z@
      \hspace{-15pt} \refstepcounter{\@listctr}\fi
    \quad\@itemlabel\hspace{0pt}%
\fi}
\makeatother
\def\fnote#1#2{\begingroup\def\thefootnote{#1}\footnote{#2}
     \addtocounter{footnote}{-1}\endgroup}
\DeclareMathAlphabet{\mathdutchcal}{U}{dutchcal}{m}{n}

\catcode`\@=11   
\newdimen\@rotdimen
\newbox\@rotbox  

\def\@vspec#1{\special{ps:#1}}
\def\@rotstart#1{\@vspec{gsave currentpoint currentpoint translate
		#1 neg exch neg exch translate}}
\def\@rotfinish{\@vspec{currentpoint grestore moveto}}
\def\@rotr#1{\@rotdimen=\ht#1\advance\@rotdimen by\dp#1%
	\hbox to\@rotdimen{\hskip\ht#1\vbox to\wd#1{\@rotstart{90 rotate}%
			\box#1\vss}\hss}\@rotfinish}
\def\@rotl#1{\@rotdimen=\ht#1\advance\@rotdimen by\dp#1%
	\hbox to\@rotdimen{\vbox to\wd#1{\vskip\wd#1\@rotstart{270 rotate}%
			\box#1\vss}\hss}\@rotfinish}%
\def\@rotu#1{\@rotdimen=\ht#1\advance\@rotdimen by\dp#1%
	\hbox to\wd#1{\hskip\wd#1\vbox to\@rotdimen{\vskip\@rotdimen
			\@rotstart{-1 dup scale}\box#1\vss}\hss}\@rotfinish}%
\def\@rotf#1{\hbox to\wd#1{\hskip\wd#1\@rotstart{-1 1 scale}%
		\box#1\hss}\@rotfinish}%
\def\rotate{\@ifnextchar[{\@rotate}{\@rotate[l]}}
\def\@rotate[#1]#2{\setbox\@rotbox=\hbox{#2}\@nameuse{@rot#1}\@rotbox}

\catcode`\@=12

\tikzset{
    partial ellipse/.style args={#1:#2:#3}{
        insert path={+ (#1:#3) arc (#1:#2:#3)}
    }
}

\hypersetup{
	pdftitle={The Asymptotic Weak Gravity Conjecture in M-theory},    
	pdfauthor={\textcopyright\ Cesar Fierro Cota, Alessandro Mininno, Timo Weigand, Max Wiesner},     
	pdfsubject={HEP},   
	pdfcreator={pdfLaTex},   
	pdfproducer={LaTex}, 
	pdfkeywords={},
	colorlinks=true,
	linkcolor=blue,        
	citecolor=blue,        
	filecolor=blue,      
	urlcolor=blue, 
}

\allowdisplaybreaks[1] 

\crefname{figure}{Figure}{Figures}
\crefname{table}{Table}{Tables}
\crefname{definition}{Definition}{Definitions}
\crefname{proposition}{Proposition}{Propositions}
\crefname{claim}{Claim}{Claims}

\begin{document}
	\pagestyle{plain}

  \setlength{\sboxrule}{0.5em}
  \setlength{\sboxsep}{1.5em} 
  \setlength{\sdim}{7pt}

	\makeatletter
	\@addtoreset{equation}{section}
	\makeatother
	\renewcommand{\theequation}{\thesection.\arabic{equation}}
	\pagestyle{empty}
 \rightline{ZMP-HH/22-23}
\vspace{1.0cm}

\begin{center}
{\large \bf
The Asymptotic Weak Gravity Conjecture in M-theory 
} 

\vskip 9 mm

Cesar Fierro Cota,${}^1$ Alessandro Mininno,${}^1$ Timo Weigand,${}^{1,2}$ Max Wiesner${}^{3,4}$

\vskip 9 mm

\small ${}^{1}${\it II. Institut f\"ur Theoretische Physik, Universit\"at Hamburg, Luruper Chaussee 149,\\ 22607 Hamburg, Germany} 

\vspace{2mm}

\small ${}^{2}${\it Zentrum f\"ur Mathematische Physik, Universit\"at Hamburg, Bundesstrasse 55, \\ 20146 Hamburg, Germany  }   \\[3 mm]

\small ${}^{3}${\it Center of Mathematical Sciences and Applications, Harvard University, 20 Garden Street,\\ Cambridge, MA 02138, USA} \\[3 mm]

\small ${}^{4}${\it Jefferson Physical Laboratory, Harvard University, 17 Oxford Street, \\ Cambridge, MA 02138, USA} \\ [3mm]

\fnote{}{\hspace{-0.75cm} cesar.fierro.cota at desy.de, \\ alessandro.mininno at desy.de, \\  timo.weigand at desy.de,  \\ mwiesner at cmsa.fas.harvard.edu}

\end{center}

\vskip 4mm

\begin{abstract}

The tower Weak Gravity Conjecture predicts infinitely many super-extremal states along every ray in the charge lattice of a consistent quantum gravity theory. 
We show this far-reaching claim in five-dimensional compactifications of M-theory on Calabi--Yau 3-folds
for gauge groups with a weak coupling limit. 
We first characterize the possible weak coupling limits, building on an earlier classification of infinite distance limits in the K\"ahler moduli space of M-theory compactifications.
We find that weakly coupled gauge groups are associated to curves on the compactification space contained in generic fibers or in fibers degenerating at finite distance in their moduli space.
These always admit an interpretation as a Kaluza--Klein or winding U$(1)$ in a dual frame or as part of a dual perturbative heterotic gauge group, in agreement with the Emergent String Conjecture.
Using the connection between Donaldson--Thomas invariants and Noether--Lefschetz theory, we then show that every ray in the associated charge lattice either supports a tower of BPS states or of non-BPS states, and prove that these satisfy the super-extremality condition, at least in the weak coupling regime. 
\\
\\

\end{abstract} 

	\newpage
	\setcounter{page}{1}
	\pagestyle{plain}
	\renewcommand{\thefootnote}{\arabic{footnote}}
	\setcounter{footnote}{0}
	
	\tableofcontents

\section{Introduction and Summary}
\label{sec:intro}
One of the cornerstones of the Swampland program \cite{Vafa:2005ui} is the Weak Gravity Conjecture \cite{Arkani-Hamed:2006emk}.\footnote{See \cite{Palti:2020mwc,Harlow:2022gzl} for reviews on the Weak Gravity Conjecture and \cite{Brennan:2017rbf,Palti:2019pca,vanBeest:2021lhn,Grana:2021zvf,Agmon:2022thq} for reviews on the Swampland program more generally.} This conjecture asserts that in order for an effective field theory (EFT) to allow for a consistent completion to a theory of quantum gravity (QG), there needs to exist a state in the spectrum for which the charge-to-mass ration is larger than that of the corresponding black hole. Whereas the original Weak Gravity Conjecture (WGC) requires only a finite number of such super-extremal states, refinements of the WGC, such as the tower WGC (tWGC)~\cite{Heidenreich:2015nta,Heidenreich:2016aqi,Montero:2016tif,Andriolo:2018lvp}, predict that any consistent effective theory of gravity must exhibit an infinite tower of super-extremal states in every direction of the charge lattice. The motivation for this refinement stems from the requirement that the original form of the WGC is preserved under dimensional reduction. More precisely, if in a $(d+1)$-dimensional EFT there is a tower of super-extremal states for the $(d+1)$-dimensional gauge sector (in the sense that each site in the charge lattice or some multiple of it is populated by a super-extremal state \cite{Heidenreich:2019zkl}), then after circle compactification the convex hull condition \cite{Cheung:2014vva} for the gauge sector in the $d$-dimensional theory including the Kaluza--Klein (KK) $\U(1)$ is automatically satisfied. 

Much support for the tWGC comes from string constructions where towers of super-extremal states can be identified explicitly. In this context, one typically restricts to supersymmetric theories that can be described as compactifications of string theory on some geometric backgrounds. To verify the tWGC one must reliably compute the charge-to-mass ratio of candidates of super-extremal states. In practice, this can be achieved by considering BPS or non-BPS states whose charge-to-mass ratio is known in certain asymptotic corners of the field space. For BPS states, it is sufficient to show their existence, since the BPS bound implies that a BPS state is (super-) extremal. Furthermore the charge-to-mass ratio of a BPS state is protected by supersymmetry. Modulo potential wall-crossing phenomena (see, e.g., \cite{Palti:2021ubp}), for a super-extremal BPS tower the tWGC can therefore in principle be shown to hold at any point in field space. 
Successful checks of the Weak Gravity Conjecture using BPS states have been performed in particular in \cite{Grimm:2018ohb,Gendler:2020dfp,Bastian:2020egp} for Calabi--Yau compactifications to four dimensions with ${\cal N}=2$ supersymmetry, embarking from the asymptotic region in complex structure moduli space,
and in \cite{Alim:2021vhs,Gendler:2022ztv}
for M-theory compactifications to five dimensions.

The situation is different if the tower predicted by the tWGC consists of non-BPS states. Their charge-to-mass ratio is not protected by supersymmetry, and hence can change as we move in field space. The charge-to-mass ratio of non-BPS states can typically be calculated reliably only near asymptotic, weakly coupled points in field space where the gauge coupling $g_\text{\tiny YM}$ is small. The WGC in weakly coupled regimes of field space is referred to as the      
 \textit{asymptotic} WGC. Since points where $g_\text{\tiny YM}\rightarrow 0$ lie at infinite distance in moduli space, there is a natural relation between the tower WGC and the Distance Conjecture \cite{Ooguri:2006in}. The Distance Conjecture predicts that as we approach an infinite distance point in field space, a tower of states becomes asymptotically massless in Planck units. And indeed, if it exists, a tower of super-extremal states charged under the gauge group that becomes weakly coupled has to become massless at these infinite distance points. It is therefore natural to expect that at least a subsector of the tower of light states predicted by the Distance Conjecture consists of super-extremal states.

In this article, we address the asymptotic tWGC in M-theory compactifications to five dimensions. A detailed study checking the WGC for BPS states in this context also away from asymptotic regions has been performed in \cite{Alim:2021vhs,Gendler:2022ztv}. Our main focus will be placed on the role of super-extremal non-BPS towers. These will be shown to make up for the potential lack of super-extremal BPS states \cite{Alim:2021vhs} in all directions in the charge lattice dual to gauge groups with a weak coupling limit, hence guaranteeing the asymptotic tWGC.

The asymptotic WGC in theories with minimal supersymmetry has previously been investigated in detail in compactifications of F-theory to six and four dimensions \cite{Lee:2018urn,Lee:2018spm,Lee:2019tst,Klaewer:2020lfg,Cota:2022yjw}. In these theories, there are no BPS particle states. Instead, to prove the asymptotic WGC, one resorts to the non-BPS excitations of a weakly coupled BPS string. As shown in \cite{Lee:2018urn,Lee:2018spm}, in six-dimensional compactifications of F-theory all BPS strings with weak coupling limits  in which also some gauge groups in the effective theory become asymptotically perturbative are dual to heterotic strings. Based on the modular properties of the elliptic genus of these six-dimensional strings, the existence of super-extremal states charged under the perturbative heterotic gauge groups was established in \cite{Lee:2018urn,Lee:2018spm}.\footnote{These results hold even in the presence of non-perturbative effects such as NS5-branes due to which the explicit worldsheet description may not be known. Perturbative worldsheet arguments for the existence of super-extremal states were also given in \cite{Arkani-Hamed:2006emk,Heidenreich:2016aqi,Heidenreich:2017sim}, and \cite{Montero:2016tif} argues based on the modularity of a holographically dual CFT.} Similarly, in four-dimensional ${\cal N}=1$ theories super-extremal states can be shown to exist in emergent string limits where a BPS string dual to a heterotic string becomes light. However, unlike in six dimensions, in four dimensions there also exist weak coupling limits for gauge theories that cannot be identified with the perturbative heterotic gauge group in a dual frame. Such more general weak coupling limits have been investigated in \cite{Cota:2022yjw} in the context of the tWGC. The strings which become light in Planck units in such limits are not critical strings, but axionic  \cite{Heidenreich:2021yda,Kaya:2022edp} or EFT strings \cite{Lanza:2020qmt,Lanza:2021udy,Marchesano:2022avb,Grimm:2022sbl,Wiesner:2022qys}. It is found in \cite{Cota:2022yjw} that in certain cases, there are no particle-like excitations of BPS strings that can furnish a tower of super-extremal tower of states. This phenomenon occurs in weak coupling limits $g_\text{\tiny YM}\rightarrow 0$ in which the ratio 
\begin{equation}\label{WGCtosp}
    \frac{\Lambda_{\text{\tiny WGC}}^2}{\Lambda_{\text{\tiny sp}}^2} \equiv \frac{g_{\YM}^2 M_{\Pl}^{d-2}}{\Lambda_{\text{\tiny sp}}^2}\,
\end{equation}
does not vanish asymptotically, with $d=4$. Here, $\Lambda_\text{\tiny WGC}$ is the scale associated to the magnetic form of the WGC \cite{Arkani-Hamed:2006emk}. It serves as the cut-off for the gauge theory, where additional charged states are expected to appear. On the other hand, $\Lambda_{\text{\tiny sp}}$ is the species scale that acts as an effective quantum gravity cut-off \cite{Dvali:2007hz,Dvali:2009ks,Dvali:2010vm}. Accordingly from a quantum gravity point of view, these limits do not correspond to weak coupling since the mass scale set by the coupling (and hence the scale at which we expect a tower of super-extremal states) is at or above the quantum gravity cut-off. 

In this paper, we extend the discussion of the asymptotic WGC to theories with minimal supersymmetry in five dimensions. More precisely, we consider M-theory compactified on a Calabi--Yau 3-fold, leading to an effective five-dimensional theory with ${\cal N}=1$ supersymmetry. When the 3-fold is elliptically fibered, this five-dimensional theory can be viewed as the circle reduction of the six-dimensional F-theory setting analyzed in \cite{Lee:2018spm,Lee:2018urn} in the context of the WGC. In the five-dimensional M-theory context, the Abelian gauge factors arise from reducing the M-theory 3-form over the 2-cycle classes in the 3-fold. Unlike its six-dimensional relative, the five-dimensional ${\cal N}=1$ theory has BPS particles that can in principle furnish a tower of (super-)extremal states. The role of BPS particles for the WGC in five-dimensional M-theory compactifications has been analyzed in detail in \cite{Alim:2021vhs}. The BPS states here correspond to M2-branes wrapping certain 2-cycles in the internal Calabi--Yau and are hence charged under the $\U(1)$s obtained by reducing the M-theory 3-form $C_3$ over the 2-cycles. Interestingly, \cite{Alim:2021vhs} has shown that the BPS bound does not necessarily coincide with the extremality bound on the entire charge lattice. On the other hand, the BPS and extremality bound do agree in the sub-cone of the charge lattice corresponding to M2-branes on movable curves \cite{Alim:2021vhs}. Whereas \cite{Alim:2021vhs} conjectured that for any direction in the charge lattice within the movable cone there needs to exist a tower of BPS states, this is not necessarily the case for any direction of the charge lattice outside this cone. Indeed, outside the movable cone the BPS bound and the extremality bound do not necessarily agree, such that also non-BPS states can be (super-)extremal \cite{Alim:2021vhs}. In fact, for many directions in the charge lattice, there are only finitely many BPS states.  This is for instance the case for $\U(1)$s associated to shrinkable curves in the 3-fold.\footnote{By shrinkable we refer, as is customary, to curves that can shrink without enforcing a divisor to shrink as well.} 
As a result, whether the tWGC is satisfied along such directions is still an open question.

Our main focus lies on those directions in the charge lattice that are not populated by towers of BPS states. 
Our goal is to show explicitly that at least the {\it asymptotic} tower WGC holds even along such non-BPS directions. 
In other words, we aim to show that 
all directions in the charge lattice are populated by super-extremal towers -- BPS or non-BPS -- provided the dual directions in the space of gauge groups admit a weak coupling limit.

Following the lessons from the classification of weak coupling limits in four-dimensional ${\cal N}=1$ compactifications of F-theory \cite{Cota:2022yjw}, the correct criterion for a weak coupling limit is to require that the ratio \eqref{WGCtosp} between the WGC scale, $\Lambda_\text{\tiny WGC}$, and the species scale, $\Lambda_\text{\tiny sp}$, vanishes asymptotically.
 Given that the gauge couplings of the $\U(1)$ factors are controlled by vector multiplets, we can entirely focus on the vector multiplet sector of the effective five-dimensional ${\cal N}=1$ theory, i.e., on the K\"ahler deformations of the internal Calabi--Yau  3-fold. Infinite distance limits in the vector multiplet sector of Calabi--Yau 3-fold compactifications of M-theory have previously been classified in \cite{Lee:2019wij}: Such limits only exist if the Calabi--Yau 3-fold allows for either a $T^2$-fibration or a surface fibration with generic fiber a K3 or $T^4$.\footnote{In \cite{Heidenreich:2020ptx}, it has been shown that every infinite distance limit in the vector multiplet space is a weak coupling limit and vice versa.} In both cases, the infinite distance limit corresponds to the limit where the respective fiber shrinks relative to the base. Whereas the $T^2$-type limit is a decompactification limit from five to six dimensions, limits with a shrinking surface fiber can be interpreted as emergent string limits, since a critical string obtained by an M5-brane wrapped on the generic fiber becomes tensionless and weakly coupled. This classification serves as a starting point for our analysis of the asymptotic WGC in five-dimensional M-theory Calabi--Yau compactifications. 

\paragraph{Summary of Results}\mbox{}
\vspace{2mm}

In this work, we refine the classification of \cite{Lee:2019wij} for tests of the asymptotic WGC: Given any of the aforementioned infinite distance limits, we classify the linear combinations of $\U(1)$ factors that become weakly coupled in the respective limit. Naively one might have expected that for a given infinite distance limit, any $\U(1)$ obtained by reducing $C_3$ over a curve contained in a shrinking fiber leads to an asymptotically weakly coupled $\U(1)$ as the dual divisor becomes large in the asymptotic limit. However, this logic is not quite correct. In fact, there are many such `fibral' $\U(1)$s that do not become weakly coupled. As we find in this work, in general $\U(1)$s can only become weakly coupled in a given infinite distance limit if they are obtained from reducing $C_3$ over either of the following kinds of curves: \\

\begin{enumerate}[label={\it\roman*)}]
 \item a curve that is contained in the generic fiber or 
 \item 
 a curve that is localized in a degenerate fiber associated to a \textit{finite} distance degeneration in the deformation space of the generic fiber. \end{enumerate}
In particular, any curve that arises in a degenerate fiber corresponding to an \textit{infinite} distance degeneration does not yield a weakly coupled $\U(1)$.

For $T^2$-type limits, it follows from our results that the only $\U(1)$ that becomes weakly coupled is the $\U(1)$ associated to the generic $T^2$-fiber; this gauge group is the KK $\U(1)$ of the lift to F-theory on the base of the Calabi--Yau 3-fold times $S^1$. In this case, a tower of super-extremal states is furnished by M2-branes multi-wrapped on the $T^2$-fiber. These are BPS states, reflecting the fact that the generic $T^2$ fiber is a movable curve, and are hence already contained in the analysis of \cite{Alim:2021vhs}. The situation is more interesting for limits with vanishing surface fiber. Here we indeed identify weakly coupled $\U(1)$s whose associated direction in the charge lattice is not populated by BPS states. To argue for the existence of super-extremal non-BPS states charged under these weakly coupled $\U(1)$s we first show that these gauge groups can be identified with perturbative gauge groups of the critical string that becomes weakly coupled in the asymptotic limits. This is to be contrasted with those `fibral' $\U(1)$s which do not become weakly coupled in the emergent string limit. The latter arise from the circle reduction of the 2-forms coupling to non-critical E- or M-like strings. It is therefore consistent that such $\U(1)$s do not admit a perturbative limit.

Of particular interest for us are setups where the generic fiber is a K3 surface; the critical weakly coupled, emergent string describes a heterotic string compactified in five dimensions. The super-extremal states with respect to the perturbative heterotic gauge group are then expected to be excitations of the heterotic string, which are not necessarily BPS. Our approach to show that there is indeed a tower of super-extremal states arising from the heterotic string parallels the procedure in six and four dimensions \cite{Lee:2018spm,Lee:2018urn,Lee:2019tst,Klaewer:2020lfg}, but involves a number of interesting differences: We first show that there exist excitations of the heterotic string for which
\bea \label{nQrelation-1}
n_{\text{\tiny L}}= - \frac{1}{2} \bm{Q}^2 \,,
\eea
where $n_{\text{\tiny L}}$ is the left-moving excitation number and $\bm{Q}$
the quantized charge vector. The second step then amounts to showing that these states are super-extremal, where we use that in the weak coupling limit and for particle-like excitations, the WGC is equivalent to the Repulsive Force Conjecture \cite{Palti:2017elp,Lee:2018urn,Heidenreich:2019zkl}. To accomplish the first step, we make use, in certain cases, of the modular properties of the elliptic genus for the five-dimensional string  obtained by wrapping an M5-brane on the K3-fiber. This allows us to relate the degeneracy of these non-BPS states in five dimensions to certain Donaldson--Thomas invariants counting indices of BPS bound states of D4-D2-D0-branes obtained after circle compactification to Type IIA~\cite{Gaiotto:2006wm,Bouchard:2016lfg}. The existence of states with the property \eqref{nQrelation-1} then follows from the connection to Noether--Lefschetz theory~\cite{Katz:1999xq,Pandharipande:2014qoa}.

In other words, we show the existence of non-BPS states in five dimensions by relating them to BPS states in one lower dimension. This is similar in spirit to, e.g., the six-dimensional analysis \cite{Lee:2018spm,Lee:2018urn} where the non-BPS excitations of the six-dimensional heterotic string can be related to BPS M2-branes in five dimensions of the kind considered also in \cite{Alim:2021vhs}. 

To prove that these non-BPS states are indeed super-extremal (at least in the asymptotic region) we next show that they are self-repulsive. Compared to the analogous problem in six-dimensional ${\cal N}=(1,0)$ theories, here we must take into account that five-dimensional ${\cal N}=1$ theories have access to the Coulomb branch. As a consequence, in addition to a mass term proportional to the string tension, the mass of the charged excitations of the heterotic string also receives a contribution proportional to their charge and the Coulomb branch parameters. Taking this into account, we find that, indeed, the WGC is satisfied for a tower of charged heterotic states. Since our five-dimensional heterotic string setup can be viewed as a circle compactification of the six-dimensional heterotic string, our results constitute a non-trivial test for the validity of the WGC under circle compactifications. 

A slight variation of the result for the heterotic string further shows that, also in the case of a $T^4$-type limit, there exist super-extremal states for all weakly coupled $\U(1)$s. 

In short, the results of this paper can be summarized as the following statement, which we formalize in Claims~\ref{mainclaim} and \ref{claim2} in Section~\ref{sec:existencetower}:

\vspace{4mm}

\noindent\shadowbox{
\begin{minipage}{0.962\textwidth}
Any direction in the charge lattice of five-dimensional M-theory that is not populated by a tower of BPS states either carries charge under a $\U(1)$ without a weak coupling limit, or there are super-extremal non-BPS states arising as excitations of a critical string.
\end{minipage}
}

\vspace{2mm}

\paragraph{Structure of the Paper}\mbox{}
\vspace{2mm}

The remainder of this article is organized as follows. In Section \ref{sec:MtheoryCY3}, we set our conventions by reviewing central aspects of M-theory compactifications on Calabi--Yau 3-folds. In Section \ref{sec:existencetower}, we state the main results of the paper, which are summarized in Claims \ref{mainclaim} and \ref{claim2}. In Section \ref{sec:WeakCoupllims}, we study the appearance of weakly coupled gauge symmetries in different limits in the vector moduli space. In Section \ref{sec_TowerWGC}, we argue that there always exists a tower of (non-)BPS particles whenever it is possible to realize a weak coupling limit and complete the proof of the asymptotic tower WGC in five-dimensional M-theory. We illustrate the main features of our analysis in an example presented in Section \ref{sec:examples}. In Section \ref{sec:discussion}, we discuss our findings and end with some speculative remarks.

\newpage

\section{M-theory on Calabi--Yau 3-folds}
\label{sec:MtheoryCY3}

To set the stage for our analysis, we summarize some of the most relevant aspects of compactifications of eleven-dimensional M-theory  on a Calabi--Yau 3-fold $X_3$ to five dimensions.

 Our main interest is in the gauge theory sector of the five-dimensional ${\cal N}=1$ effective action, starting from the bosonic part of the $11$d M-theory effective action\footnote{Our conventions can be obtained from those in \cite{Polchinski:1998rr}, where $2\kappa_{11}^2=(2\pi)^8M_{\text{\tiny 11d}}^{-9}$, by rescaling $M_{\text{\tiny 11d}}$ by $2\pi$.}
\begin{equation}
    S_\text{\tiny 11d} = 2\pi M_\text{\tiny 11d}^9 \int_{\RR^{1,10}}\left(R\star \ID -\frac{1}{2}d C_3\wedge \star d C_3\right)+\ldots \,.
    \label{eq:S11}
\end{equation}
Upon compactifying on a 3-fold $X_3$,  we can decompose the 3-form gauge potential as 
\bea
C_3=(2\pi)^{-1}M_{\text{\tiny 11d}}^{-1}A^\alpha \wedge J_\alpha+\ldots \,,
\eea
where 
\begin{equation}
\{J_\alpha\} \coma \alpha = 1, \ldots, h^{1,1}(X_3) \,,
\end{equation}
denotes a basis of $H^{1,1}(X_3,\ZZ)$.
This defines a basis $\{\U(1)^\alpha\}$ of abelian gauge group factors, with gauge potentials $A^\alpha$. 
Note that to each curve class $C \in H_2(X_3)$ one can  associate a linear combination $\U(1)_C$ of such gauge group factors; it is the gauge group whose gauge potential arises by reducing $C_3$ along the curve $C$. In terms of the dual basis $\{\cC^\alpha\}$
of $H_4(X_3, \mathbb Z)$, which obeys
\bea
J_\alpha \cdot \cC^\beta = \delta_\alpha^\beta \,,
\eea
we define this linear combination
as
\bea\label{U1C}
\U(1)_C = c_\alpha \U(1)^\alpha  \coma \text{for} \quad C= c_\alpha \, \cC^\alpha   \in H_2(X_3) \,. 
\eea

The fluctuations of the K\"ahler form are
encoded in real expansion parameters $v^\alpha$
with respect to the basis $J_\alpha$,
\begin{equation}\label{Jexpand}
    J = \sum_{\alpha} v^\alpha J_\alpha\, .
\end{equation}
Since the overall volume of $X_3$ is part of a universal hypermultiplet, this yields $h^{1,1}(X_3)-1$ independent scalar fields $\Phi^A$ which can be viewed as functions of the $v^\alpha$ and which define the scalar fields inside the five-dimensional vector multiplets.

The kinetic terms of these scalars $\Phi^A$ and of the abelian gauge fields $A^\alpha$ in the five-dimensional action are given by
\begin{equation}\label{5daction}
   S_{5d} = \frac{M_\Pl^3}{2}\int_{\RR^{1,4}} \left(   R\star \ID  - \mathfrak{g}_{AB}  d\Phi^A \wedge \star   d \Phi^B \right) -\frac{1}{2g_5^2}\int_{\RR^{1,4}}  f_{\alpha\beta}F^\alpha \wedge \star F^\beta+\ldots\coma
    \end{equation}
with\footnote{Our convention matches the one in \cite{Alim:2021vhs} identifying their $\kappa_5^{2}$ with our $M_\Pl^{-3}$.} 
\begin{equation}
    M_\Pl^3 = 4\pi M_\text{\tiny 11d}^3\mathcal{V}\coma \frac{1}{g_5^{2}} = \frac{ M_\Pl}{(2\pi)(4\pi)^{1/3}}\fstop
    \label{eq:MplM11}
\end{equation}
The gauge kinetic matrix $f_{\alpha\beta}$ is defined as a dimensionless quantity and obtained by dimensionally reducing the kinetic term of $C_3$ as
\begin{equation}
\begin{split} \label{fabMth}
     f_{\alpha\beta} &= \frac{1}{\mathcal{V}^{1/3}}\int_{X_3}J_\alpha \wedge \star J_\beta \\
             &= \frac{1}{\mathcal{V}^{1/3}}\left(\frac{3}{2}\dfrac{(\int_{X_3}J_\alpha\wedge J^2)(\int_{X_3}J_\beta\wedge J^2)}{\int_{X_3}J^3}-\int_{X_3}J_\alpha\wedge J_\beta \wedge J \right)\\
             &= \frac{1}{\mathcal{V}^{1/3}}\left(\frac{\mathcal{V}_\alpha \mathcal{V}_\beta}{\mathcal{V}}-\mathcal{V}_{\alpha\beta}\right) 
          = \left(\widehat{\mathcal{V}}_\alpha \widehat{\mathcal{V}}_\beta-\widehat{\mathcal{V}}_{\alpha\beta}\right)\fstop
\end{split}
\end{equation}
Here we have introduced the abbreviations
\begin{equation}
\begin{array}{*1{>{\displaystyle}c}p{5cm}}
 \mathcal{V} = \frac{1}{6}\int_{X_3} J^3 = \frac{1}{6}\kappa_{\alpha\beta\gamma}v^\alpha v^\beta v^\gamma \coma\\
 \mathcal{V}_\alpha =\frac{1}{2}\int_{X_3} J_\alpha \wedge J^2  = \frac{1}{2}\kappa_{\alpha\beta\gamma}v^\beta v^\gamma\coma    \mathcal{V}_{\alpha\beta} = \int_{X_3}J_\alpha \wedge J_\beta \wedge J = \kappa_{\alpha\beta\gamma}v^\gamma
\end{array}
\label{eq:volumesJ}
\end{equation}
for the volume of $X_3$, of the divisor dual to $J_\alpha$ and of the intersection curve of two such divisors, respectively, in units of $M_{\text{\tiny 11d}}$. 
We have also introduced the intersection form
\begin{equation}\label{eq:intnumb}
    \kappa_{\alpha\beta\gamma} = \int_{X_3} J_\alpha \wedge J_\beta \wedge J_\gamma  
\end{equation}
and the
re-scaled coordinates
\begin{equation}\label{eq:rescaledcoord}
     \hat{v}^\alpha = \frac{v^\alpha}{\mathcal{V}^{1/3}}\,,
\end{equation}
so that
\begin{equation} \label{rescaledvolumes}
    \widehat{\mathcal{V}}_\alpha=\frac{1}{2}\kappa_{\alpha\beta\gamma}\hat{v}^\beta\hat{v}^\gamma\coma \widehat{\mathcal{V}}_{\alpha\beta}=\kappa_{\alpha\beta\gamma}\hat{v}^\gamma\fstop 
\end{equation}
The coordinates $\hat{v}^\alpha$ 
satisfy the relation $\frac{1}{6}\kappa_{\alpha \beta\gamma} \hat v^\alpha \hat v^\beta  \hat v^\gamma =1$ and so we should think of them as functions of the independent vector multiplet scalars $\Phi^A$. 

Note that the eigenvalues of the gauge kinetic matrix correspond to the inverse-squares of gauge couplings. To obtain the gauge couplings (in units of the five-dimensional Planck mass),
one introduces the 
inverse matrix $f^{\alpha\beta}=(f_{\alpha\beta})^{-1}$  
given by 
\begin{equation} \label{finv}
\begin{split}
    f^{\alpha\beta} = \mathcal{V}^{1/3}\left(\frac{1}{2}\frac{v^\alpha v^\beta}{\mathcal{V}}-\mathcal{V}^{\alpha\beta}\right)
      = \frac{1}{2}\hat{v}^\alpha\hat{v}^\beta-\widehat{\mathcal{V}}^{\alpha\beta}\coma
\end{split}
\end{equation}
where $\mathcal{V}^{\alpha\beta}$ is the inverse of $\mathcal{V}_{\alpha\beta}$. 
The interaction strength of a linear combination $\U(1)_C$, defined in \eqref{U1C}, of abelian gauge group factors is then set by the (dimensionful) quantity\footnote{Despite discussing abelian gauge groups, we refer to the dimensionful gauge coupling as $g_\YM$ to point out that the results carry over {\it mutatis mutandis} to non-abelian settings as well. We trust that this does not confuse the reader.}
\bea\label{gYMCafabCb}
g_{\text{\tiny YM,C}}^2 = g_5^2 \,  c_\alpha f^{\alpha\beta} c_\beta \,.
\eea
Finally, the scalar metric  $\mathfrak{g}^{AB}$ is related to $f^{\alpha\beta}$ as
\begin{equation} \label{eq:Ifabwithgalbe}
  f^{\alpha\beta} = \frac{1}{2}\mathfrak{g}^{AB} \frac{\partial}{\de \Phi^A} \hat v^\alpha \frac{\partial}{\de \Phi^B} \hat v^\beta + \frac{1}{3} \hat v^\alpha \hat v^\beta \fstop
\end{equation}

The above expressions hold for an arbitrary choice of basis $\{J_\alpha\}$ and $\{\cC^\alpha\}$. In this paper, it will be convenient to take 
$\{\cC^\alpha\}$ to consist of 
generators of the Mori cone of effective curves on $X_3$.
If the Mori cone is simplicial, then the dual basis $\{J_\alpha\}$ consists of the dual K\"ahler cone generators. In this case, the intersection numbers \eqref{eq:intnumb} are non-negative.
Irrespective of whether the $J_\alpha$ are K\"ahler cone generators, we can expand
the K\"ahler form $J$ of $X_3$
as a positive linear combination \eqref{Jexpand},
where the expansion parameters $v^\alpha \geq 0$ give the volumes of the Mori cone generators $\cC^\alpha$.
If the Mori cone is not simplicial, then the choice of basis of Mori cone generators is not unique, and correspondingly it takes several patches to cover all possible K\"ahler forms in this way.\footnote{If the Mori cone is generated by an infinite union of cones, our analysis applies for any fixed choice of basis $\{J_\alpha\}$ and $\{\cC^\alpha\}$.} 

Furthermore, the $v^\alpha$ are then subject to additional constraints which ensure positivity of all effective divisor volumes and of the volume of $X_3$.
For ease of notation, we will oftentimes refer to the expansion parameters $v^\alpha$ as K\"ahler moduli even though we stress that this interpretation is strictly speaking correct only if the Mori cone is simplicial.

\section{Criteria for (Non-)BPS Towers and Weak Coupling Limits}
\label{sec:existencetower}

Given a basis of $\U(1)$ gauge factors in five-dimensional M-theory, the tower WGC \cite{Heidenreich:2015nta,Heidenreich:2016aqi,Montero:2016tif,Andriolo:2018lvp,Heidenreich:2019zkl} predicts an infinite tower of super-extremal particle states along all rays in the charge lattice $\Lambda_{\bm{Q}}$.
By this we mean that given
a charge vector $\bm{Q} \,  \in \Lambda_{\bm{Q}}$, 
there must exist a super-extremal state of charge
\begin{equation}
 n \, \bm{Q} \coma   \forall n \in {\cal I} \coma 
\end{equation}
where the index set ${\cal I}$ is a subset of $\mathbb N$. 
A particle is super-extremal
if its charge-to-mass ratio equals or exceeds that of an extremal black hole. 
A stronger version of this conjecture, the Sublattice WGC \cite{Heidenreich:2016aqi}, requires these super-extremal states to even populate a sublattice of the full charge lattice.

A related conjecture, the Repulsive Force Conjecture (RFC), replaces the requirement of super-extremality by that of self-repulsiveness \cite{Arkani-Hamed:2006emk,Palti:2017elp,Heidenreich:2019zkl}, i.e., for two particles of a given species, the repulsive {\it long-range} forces must not be weaker than the sum of attractive long-range forces,
\begin{equation} \label{RFC1}
F_{\rm Coulomb} \geq F_{\rm Grav.} + F_{\rm Yukawa} \,.
\end{equation}

Both conditions, super-extremality and self-repulsiveness, are fulfilled by BPS states since a BPS black hole is always extremal 
and BPS states exert an exactly vanishing net force between one another. Hence, if there exists a tower of BPS states, this tower satisfies both the WGC and the RFC.

For non-BPS states, the relation between the WGC and the RFC is more intricate \cite{Heidenreich:2019zkl}.
However,
under certain assumptions, the two conditions agree in the infinite distance regime corresponding to a weak coupling limit \cite{Lee:2018spm}.
If  we consider a particle of mass $M_k$, then the RFC requires
\begin{equation}\label{eq:5dRFC}
   \frac{M_\Pl g_{\text{\tiny YM}}^2}{M_k^2/M_\Pl^2} \equiv  \frac{(M_\Pl g^2_5) (Q_\alpha  f^{\alpha\beta} Q_\beta)}{M_k^2/M_\Pl^2} \geq  \left.\frac{d-3}{d-2}\right|_{d=5}+ \frac{1}{4}\frac{M_\Pl^4}{M_k^4}\mathfrak{g}^{AB}\frac{\de}{\de \Phi^A} \left(\frac{M_k^2}{M_\Pl^2}\right)\frac{\de}{\de \Phi^B}\left(\frac{M_k^2}{M_\Pl^2}\right)\coma
\end{equation}
where we recall that $\Phi^A$ are scalar fields with $A = 1,\ldots, h^{1,1}-1$, and $\mathfrak{g}^{AB}$ is the inverse of the scalar metric appearing in \eqref{5daction}. 
The two terms on the right-hand side account for the attractive gravitational \cite{Arkani-Hamed:2006emk} and Yukawa  \cite{Palti:2017elp} forces, respectively.
In the following, we will express $M_k$ as a function of the coordinates $\hat{v}^\alpha$, so it is more practical to write \eqref{eq:5dRFC} as
\begin{equation} \label{eq:5dRFC-b}
\begin{split}
      \frac{(M_\Pl g^2_5) (Q_\alpha  f^{\alpha\beta} Q_\beta)}{M_k^2/M_\Pl^2} &\geq \left.\frac{d-3}{d-2}\right|_{d=5}+ \frac{1}{4}\frac{M_\Pl^4}{M_k^4}\mathfrak{g}^{AB}\frac{\de \hat{v}^\alpha}{\de \Phi^A}\frac{\de \hat{v}^\beta}{\de \Phi^B} \de_\alpha\left(\frac{M_k^2}{M_\Pl^2}\right)\de_\beta\left(\frac{M_k^2}{M_\Pl^2}\right)\\
      & \geq \left.\frac{d-3}{d-2}\right|_{d=5}+ \frac{1}{2}\frac{M_\Pl^4}{M_k^4}\left( f^{\alpha\beta}-\frac{1}{3} \hat{v}^\alpha\hat{v}^\beta\right) \de_\alpha\left(\frac{M_k^2}{M_\Pl^2}\right)\de_\beta\left(\frac{M_k^2}{M_\Pl^2}\right)\coma
\end{split}
\end{equation}
where we have expressed the scalar metric in terms of the inverse of the gauge kinetic metric, as in \eqref{eq:Ifabwithgalbe}.

Not every ray in the charge lattice of a Calabi--Yau 3-fold admits a tower of BPS states.
In fact, \cite{Alim:2021vhs} has shown in examples that when no such BPS tower exists, the black hole extremality condition and the BPS condition do not agree. This leaves room for the existence of a tower of  non-BPS states along the ray in question which can nonetheless be super-extremal, though no such non-BPS tower has been identified explicitly in \cite{Alim:2021vhs}.
At the same time, one faces the possibility that the tower WGC or RFC might be violated along certain rays in the charge lattice.

In this paper we will show 
that at least the asymptotic tower WGC or RFC is satisfied along every ray in the charge lattice.\footnote{Since we are focusing on the asymptotic weak coupling limit, the two versions -- WGC and RFC -- are equivalent, and we will use both terms interchangeably.} This means that whenever we can take a weak coupling limit, there exists either a BPS tower of (super-)extremal (or self-repulsive) states or a non-BPS tower which satisfies the tower RFC in the asymptotic weak coupling limit. In order to test the asymptotic WGC, we first need to define what we mean by weak coupling. To this end, notice that the magnetic WGC associates a scale $\Lambda_{\WGC}$ to a gauge theory in $d$ dimensions with (dimensionful) gauge coupling $g_{\YM}$, which is given by 
\begin{equation}
 \Lambda_{\WGC}^2 = g_{\YM}^2 M_{\Pl}^{d-2}\,.
\end{equation} 
In the present case, $d=5$ and the gauge theories of interest are linear combinations of the M-theory $\U(1)$s. Given a basis $\{\U(1)^\alpha\}$ we can define the WGC scale for any $\U(1)_C$ as in \eqref{U1C} as 
\begin{equation}\label{defLambdaWGC}
    \Lambda_{\WGC}^2\left(\U(1)_C\right) = g_{\text{\tiny YM,C}}^2 M_{\Pl}^3 = g_5^2\left(c_\alpha f^{\alpha\beta} c_\beta \right) M_{\Pl}^3 \,,
\end{equation}
where we used \eqref{gYMCafabCb}. The weak coupling limit for the gauge group $\U(1)_{C}$ now corresponds to the limit 
\begin{equation} \label{weakcouplingcond}
    \frac{\Lambda^2_{\WGC}\left(\U(1)_C\right)}{\Lambda^2_{\QG}} \rightarrow 0\,. 
\end{equation}
 Here $\Lambda_{\QG}$ is the quantum gravity cut-off, i.e., the scale at which gravity becomes strongly coupled.

 With this preparation, we can state the main result of this paper:

\vspace{6mm }

\noindent\shadowbox{
\begin{minipage}{0.962\textwidth}

\begin{claim}\label{mainclaim}
Consider M-theory compactified to five dimensions on a Calabi--Yau $X_3$.
Suppose there exists a primitive charge vector ${\bm{Q}}^0\in \Lambda_{\bm{Q}}$ such that $\{ \lambda {\bm{Q}}^0\}_{\lambda\in \mathbb{R}}\cap \Lambda_{\bm{Q}}$ is not populated by a BPS tower of super-extremal states. Defining $\U(1)_{{\bm{Q}}^0}= Q_a^0 \U(1)^a$ one of the following two holds:
\begin{enumerate}[label={C\arabic*.},ref={C\arabic*},leftmargin=*,labelindent=0pt]
    \item\label{list:claim1}  There exists no limit in moduli space in which 
\begin{equation}\label{qfabq}
 \frac{\Lambda^2_{\text{\tiny WGC}}\left(\U(1)_{{\bm{Q}}^0}\right)}{\Lambda^2_{\QG}} \rightarrow 0\, 
\end{equation}
with $\Lambda^2_{\WGC}\left(\U(1)_{{\bm{Q}}^0}\right)$ defined as in \eqref{defLambdaWGC}.

\item\label{list:claim2} Alternatively, there does exist a non-BPS tower of  states along the ray in the charge lattice which is part of the tower of excitations of a critical string becoming weakly coupled in the limit  \eqref{qfabq}, and this tower of states is self-repulsive in the asymptotic limit in the sense of satisfying condition \eqref{RFC1}.
\end{enumerate}
\end{claim}

\end{minipage}
}

\vspace{4mm }

To show this, we will prove an equivalent converse statement:
Whenever there does exist a limit of the type
\eqref{qfabq}, either there exists a BPS tower of super-extremal states along the given ray in the charge lattice, or we can identify a non-BPS tower of self-repulsive states originating in the excitation spectrum of an asymptotically weakly coupled critical string.
Notice that since $\Lambda_{\QG}\leq M_{\Pl}$ a necessary condition for the existence of a weak coupling limit of the form \eqref{qfabq} is the existence of a limit 
\begin{equation}\label{necessarycondition}
 Q_\alpha^0 f^{\alpha\beta} Q_\beta^0 \rightarrow 0\,. 
\end{equation} 
In the following, we therefore first classify all possible limits in the five-dimensional moduli space for which \eqref{necessarycondition} can be achieved and subsequently compare with the scaling of $\Lambda_{\QG}$. In order to engineer a limit of the form \eqref{necessarycondition}, note that \eqref{finv} is invariant under a homogeneous rescaling of all the K\"ahler moduli 
\begin{equation} \label{homo-resc}
v^\alpha \to \lambda \tilde{v}^\alpha \coma \forall \alpha = 1, \ldots, h^{1,1}(X_3)\,.
\end{equation}
This means that to classify the possible weak coupling limits, without loss of generality, we can assume that the overall volume ${\cal V}$ stays constant along the trajectory in moduli space because otherwise we can simply perform a suitable rescaling \eqref{homo-resc} without affecting the gauge coupling matrix $f^{\alpha\beta}$. 
Hence it suffices to analyze limits in the vector moduli space. Such limits have been classified in \cite{Lee:2019wij} and leave us with two qualitatively different scenarios: 
\begin{enumerate}[label={{(\arabic*)}},ref={{\arabic*}}]
     \item\label{pos1} {\it Limits of Type $T^2$}: $X_3$ admits a torus fibration 
     \begin{equation}
     \pi: X_3 \to B_2
     \end{equation}
     and the weak-coupling limit corresponds to a limit in which the volume of the generic fiber $T^2$ shrinks as
     \begin{equation}  \label{T2limit-def}
      {\cal V}_{\text{\tiny $T^2$}} \sim \frac{1}{\lambda} \coma {\cal V}_{\text{\tiny $B_2$}} \sim \lambda \coma \lambda \to \infty \,.
     \end{equation}
     \item \label{pos2} {\it Limits of Type K3/$T^4$}: $X_3$ allows for a surface fibration \begin{equation}
     \rho: X_3 \to \mathbb P^1
     \end{equation}
     with generic fiber $\sS$ being either
     \begin{enumerate*}[label={(\ref*{pos2}.\alph*)},ref={\ref*{pos2}.\alph*}, labelindent=0pt,topsep=0pt,parsep=0pt,labelsep=-0pt,itemjoin=\\ ]
         \item\label{pos2a} \hspace{-8pt} a K3 surface, or \inlineitem\label{pos2b} an abelian surface, $T^4$.
     \end{enumerate*}
     In the weak coupling limit, the volume of the surface fiber shrinks as
     \begin{equation}
      {\cal V}_{\text{\tiny \text{K}$3/T^4$}} \sim \frac{1}{\lambda^2} \coma {\cal V}_{\text{\tiny $\PP^1$}} \sim \lambda^2 \coma \lambda \to \infty \,.
     \end{equation}
\end{enumerate}

A number of comments concerning this classification are in order \cite{Lee:2019wij}.
First, consider a limit of Type \ref{pos1} and suppose that $X_3$ admits in addition a K3 or $T^4$-fibration. If in the limit also the volume of 
the generic surface fiber shrinks,
then, by definition, the shrinking rate of the $T^2$-fiber volume is faster than the square-root of the shrinking rate of the K3 or $T^4$-fiber volume, otherwise the limit is said to be of Type \ref{pos2}. 
In other words, the classification into Type \ref{pos1} or \ref{pos2} is according to the type of fiber which shrinks at the fastest rate, where in case a $T^2$-fiber volume shrinks at the same rate as the square root of a K3 or $T^4$-fiber volume, the limit is classified as Type \ref{pos2}.
Second, in the limits of Type \ref{pos2} it can be shown that there always exists a {\it unique} K3 or $T^4$-fiber shrinking at a rate strictly faster than that of any other K3 or $T^4$-fiber. Similarly in limits of Type \ref{pos1}, the 3-fold $X_3$ may well exhibit several $T^2$-fibrations whose fiber volumes shrink, but there is always a unique such fiber which shrinks at the fastest rate. See \cite{Lee:2019wij} for more details and proofs.

Limits of Type $T^2$ are effective decompactification limits from five to six dimensions, with the BPS tower of KK states furnished by M2-branes wrapping the shrinking $T^2$-fiber an arbitrary number of times.
Limits of Type K3/$T^4$ are effectively five-dimensional limits, in which the heterotic/Type II string obtained by wrapping an M5-brane once around the K3/$T^4$-fiber sets the new duality frame and becomes asymptotically tensionless with respect to the five-dimensional Planck scale.

As we will show in Section \ref{sec:WeakCoupllims}, in the limits of Type $T^2$, the only $\U(1)$ which becomes asymptotically weakly coupled is the $\U(1)_{\cal E}$ associated with the \textit{generic} torus fiber ${\cal E}$, in the sense of \eqref{U1C}.
Consistently, there exists a tower of (BPS) states charged exclusively under $\U(1)_{\cal E}$.

On the other hand, in limits of Type \ref{pos2a} only $\U(1)$s associated to curves in the generic K3-fiber or curves associated to degenerations of the K3-fiber at finite distance in the K3 moduli space become weakly coupled. In particular, curves associated to degenerations of the K3-fiber that are at infinite distance in the K3 moduli space do not give rise to $\U(1)$ gauge factors with a weak coupling limit. As a consequence, the charged states satisfying \eqref{qfabq} in the limits of Type \ref{pos2a} can all be interpreted as excitations of the heterotic string obtained by wrapping an M5-brane on the generic K3-fiber. A similar interpretation holds for limits of Type \ref{pos2b}.

We can summarize these findings as

\vspace{6mm}

\noindent\shadowbox{
\begin{minipage}{0.962\textwidth}

\begin{claim}\label{claim2}

In M-theory compactified on a Calabi--Yau $X_3$, the only $\U(1)$s which admit a weak coupling limit in the sense of \eqref{qfabq} are obtained by reducing $C_3$ over a curve in a generic torus or K3/$T^4$ fiber of $X_3$ or curves contained in a degenerate fiber arising at finite distance in the fiber moduli space.

\end{claim}

\end{minipage}
}

\vspace{4mm}

For example, all degenerations of an elliptic fiber  correspond to the point $\tau=-i\infty$ (or $\SL(2,\mathbb{Z})$ images thereof) in the complex structure moduli space of the torus and are hence at infinite distance (the elliptic points at $\tau=i, (-1)^{1/3}$ do not correspond to degenerate tori). This reflects our above claim that for limits of Type \ref{pos1} the only $\U(1)$ that can become asymptotically weakly coupled is the KK $\U(1)_{\cal E}$, as one would expect from M-/F-theory duality.

\section{Weak Coupling Limits in five-dimensional M-theory}
\label{sec:WeakCoupllims}

In this section, we analyze the different types of weak coupling limits in the vector moduli space of five-dimensional M-theory and prove Claim \ref{claim2}. 
We proceed by examining the possible infinite distance limits of Types \ref{pos1}, \ref{pos2a} and \ref{pos2b} listed in Section \ref{sec:existencetower} for the appearance of a gauge sector satisfying the weak coupling criterion \eqref{qfabq}.

\subsection{Weak Coupling in Type \texorpdfstring{$T^2$}{} Limits} 
\label{sec:WeakCouplT2lim}

    Let $X_3$ be a $T^2$-fibered Calabi--Yau 3-fold over a base $B_2$ with projection $\pi: T^2\rightarrow B_2$. According to the Shioda--Tate--Wazir theorem, 
\begin{equation}
    h^{1,1}(X_3) = h^{1,1}(B_2) + 1 + (n-1) \,,
\end{equation}
where $(n-1)$ is the sum of the Mordell--Weil rank and of the total rank of the non-abelian symmetry algebras associated with the codimension-one degenerations. 
A general basis $\{\cC^\alpha \}$ of $H_2(X_3,\mathbb Z)$ composed of Mori cone generators then splits as 
\bea
\{\cC^\alpha \} = \left\{\cC^a, \cC^i_f \right\} \coma  a=1,\dots,h^{1,1}(B_2)\coma  i= 1, \ldots, n \,,
\eea
where $\left\{\cC^i_f\right\}$ is composed of generators of the 
relative Mori cone ${\bf M}(X_3/B_2)$, i.e., of the Mori subcone of $X_3$ spanning all purely fibral curves, while the remaining curve classes
$\{\cC^a\}$ lie on a section or multi-section of the fibration.

The dual basis of $H^{1,1}(X_3)$ correspondingly splits as 
\begin{equation} \label{JaJiT2split}
\{J_\alpha \} = \{J_a, J_i\} \coma a=1,\dots,h^{1,1}(B_2)\coma  i= 1, \ldots, n \,,
\end{equation}
where 
\begin{equation}
J_a=\pi^*(j_a) \with j_a \in H^{1,1}(B_2)
\end{equation}
denote $\pi$-vertical divisor classes, i.e.,
the set $\{j_a\}$ forms 
a basis of $H^{1,1}(B_2)$. 
Note that if the Mori cone is simplicial, the 
$\{j_a\}$ are the K\"ahler cone generators of $B_2$.

Given this basis, we can expand the M-theory 3-form $C_3$ as
\bea
(2\pi) M_{\text{\tiny 11d}}   C_3 =  A^\alpha \wedge J_\alpha = A^a \wedge  J_a + A^i \wedge J_i \,.
\eea
The gauge potentials are associated with the abelian gauge group factors $\U(1)^a$ and $\U(1)^i$, respectively.

The generic elliptic fiber $C_{\mathcal{E}}$ degenerates over the discriminant locus $\Delta\subset B_2$ into a union of curves, which can be written as positive linear combinations of the fibral curves $\cC_{f}^i$.
This implies that there exists a homological relation
\begin{equation} \label{CEsplit}C_{\mathcal{E}} =  \sum_{i=1}^n c_i \, \cC^i_f\,,   
\end{equation} 
for some coefficients $c_i$ and, by construction, the curves $\cC^i_f$ are localized in special fibers over the discriminant $\Delta$. 
   Moreover, notice that $J_i$ and $J_j$ only differ over the locus $\Delta \in B_2$ over which $C_{\mathcal{E}}$ splits and
    \begin{equation}
    \pi^*(j_a)\cdot \pi^*(j_b) = n_{ab} [C_{\mathcal{E}}]\,,
    \label{eq:jajbCE}
\end{equation}
for $n_{ab} = j_a \cdot_{\text{\tiny $B_2$}} j_b$.
From \cref{eq:jajbCE,CEsplit}, it follows that 
\begin{equation}\label{identitycase1}
   c_j J_i\cdot \pi^*(j_a)\cdot \pi^*(j_b) =  c_i  J_j\cdot \pi^*(j_a)\cdot \pi^*(j_b)\coma 
\end{equation}
for all $(a,b)$ and $(i,j)$.
We will show momentarily that as a consequence of \eqref{identitycase1}, the leading part of the gauge kinetic matrix \eqref{fabMth}  has lower rank in the $T^2$ limit, and the only linear combination of  $\U(1)^i$ that becomes asymptotically weakly coupled is
\begin{equation} \label{U1+def}
    \U(1)_{\cal E} = \sum_{i=1}^n c_i \U(1)^i\,.
\end{equation}

This will be seen to be a consequence of the fact that in the weak coupling limit of Type $T^2$, the elliptic fiber shrinks as in \eqref{T2limit-def}, so that the volume of the divisors $J_i$ is dominated by terms of the form $v^a v^b$, where $v^{a}$ and $v^{b}$ are the volumes of curves in the base $B_2$. Together with \eqref{identitycase1} this results in a rank reduction of the gauge kinetic matrix to leading order.

The weakly coupled combination \eqref{U1+def} is precisely $\U(1)_{\KK}$, i.e., the actual KK $\U(1)$ obtained by compactifying six-dimensional F-theory on $B_2$ on an additional circle. All other eigenvectors of the gauge kinetic function correspond to linear combinations of $\U(1)^i$ that do not become weakly coupled, at least not in the asymptotic effectively six-dimensional frame. This, of course, is nothing but the well-known statement that in the F-theory lift of M-theory only the KK $\U(1)$ becomes weakly coupled, whereas the gauge coupling of all gauge theories that lift to gauge theories in six dimensions remains finite. 
This means that the (only) gauge symmetry that becomes weakly coupled in the limit is associated with a charge vector whose lattice is populated by a BPS tower of super-extremal states, i.e., the KK states. 

In order to prove these claims, we 
  explicitly parametrize the infinite distance limit of Type $T^2$ following the classification in \cite{Lee:2019wij}. In this work, two possible types of limits for the K\"ahler form were identified to give rise to a behavior as in Type $T^2$ limits, called $J$-class A and $J$-class B limits, respectively. We begin with the asymptotic parametrization of 
$J$-class A, which is of the form
\begin{align} \label{JclassA}
J = v^\alpha J_\alpha = \sqrt{\lambda} \tilde v^a J_a + \frac{1}{\lambda} \tilde v^i J_i \, \quad \text{for} \quad \lambda \to \infty \,.
\end{align}
Here the split of the divisor basis is as in \eqref{JaJiT2split}, and the expansion parameters $\tilde v^a$ and $\tilde v^i$ are of order one or smaller and normalized such that $\cV = \frac{1}{6} \int_{X_3} J^3$ remains finite. 
In particular, $J_a\cdot J_b\cdot J_c = 0$ for all labels of type $a,b,c$.
Note that in \cite{Lee:2019wij}, \eqref{JclassA} denotes the asymptotic expansion for the K\"ahler form in a basis of {\it K\"ahler cone} generators, while in general, our basis \eqref{JaJiT2split} is not composed of K\"ahler cone generators.
However, the fibration structure guarantees that a basis of K\"ahler cone generators can be obtained as $\hat J_a = m^b_a J^a$ and $\hat J_i = k^i_j J_j$ for suitable matrices $m^b_a$ and $k^i_j$.\footnote{Such matrices can be chosen as the identity when the K\"ahler and Mori cones are simplicial.} This means that the scaling of the expansion parameters with $\lambda$ remains as in \eqref{JclassA} even if we have no K\"ahler cone basis.

The volumes appearing in the gauge kinetic function \eqref{fabMth},
\begin{equation}
f_{\alpha \beta} = \widehat{\mathcal{V}}_\alpha \widehat{\mathcal{V}}_\beta - \widehat{\mathcal{V}}_{\alpha \beta} \,, 
\end{equation}
evaluate to 
\begin{equation}
    \begin{split} \label{volumescalings-T2}
        \widehat{\mathcal{V}}_a & = \lambda^{-\frac{1}{2}}\kappa_{abi}\hat{\tilde{v}}^b\hat{\tilde{v}}^i+\lambda^{-2}\frac{1}{2}\kappa_{aij}\hat{\tilde{v}}^i\hat{\tilde{v}}^j    =: \lambda^{-\frac{1}{2}} \,  \widehat{\widetilde{\mathcal{V}}}_a \coma \\
        \widehat{\mathcal{V}}_i & = \lambda\frac{1}{2}\kappa_{iab}\hat{\tilde{v}}^a\hat{\tilde{v}}^b+\lambda^{-\frac{1}{2}}\kappa_{iaj}\hat{\tilde{v}}^a\hat{\tilde{v}}^j+\lambda^{-2}\frac{1}{2}\kappa_{ijk}\hat{\tilde{v}}^j\hat{\tilde{v}}^k  =: \lambda \,  \widehat{\widetilde{\mathcal{V}}}_i\coma \\
        \widehat{\mathcal{V}}_{ab} & = \lambda^{-1}\kappa_{abi}\hat{\tilde{v}}^i\coma\\
        \widehat{\mathcal{V}}_{ai} & = \lambda^{1/2}\kappa_{aib}\hat{\tilde{v}}^b+\lambda^{-1}\kappa_{aij}\hat{\tilde{v}}^j\coma\\
        \widehat{\mathcal{V}}_{ij} & = \lambda^{1/2}\kappa_{ija}\hat{\tilde{v}}^a+\lambda^{-1}\kappa_{ijk}\hat{\tilde{v}}^k\fstop
    \end{split}
\end{equation}
Here we are making use of the rescaled volumes defined in \eqref{eq:rescaledcoord}, starting from the K\"ahler parameters $\tilde v^a$ and $\tilde v^i$ defined in \eqref{JclassA}.

The components of the gauge kinetic matrix therefore become
\begin{equation} \label{fcomponentsT2limit}
    \begin{split}
    f_{ij} &= \lambda^2 \widehat{\widetilde{\mathcal{V}}}_i \widehat{\widetilde{\mathcal{V}}}_j - \sqrt{\lambda} \kappa_{ija} \hat{\tilde v}^a + {\cal O}(1/\lambda) \,,\\
f_{a i} &= \sqrt{\lambda}\left( \widehat{\widetilde{\mathcal{V}}}_a \widehat{\widetilde{\mathcal{V}}}_i -\kappa_{iab} \hat{\tilde v}^b\right) + {\cal O}(1/\lambda) \,, \\
f_{a b} &= {\cal O}(1/\lambda) \coma
    \end{split}
\end{equation}
where $\widehat{\widetilde{\mathcal{V}}}_i$ and $\widehat{\widetilde{\mathcal{V}}}_a$ remain finite in the limit $\lambda \to \infty$.
By \eqref{identitycase1}, we note that
\begin{align}
c_j \widehat{\widetilde{\mathcal{V}}}_i = c_i  \widehat{\widetilde{\mathcal{V}}}_j + {\cal O}\left(1/\lambda^{3/2}\right)  \coma c_j  \kappa_{iab} \hat{\tilde v}^a = c_i \kappa_{jab} \hat{\tilde v}^a  \coma  \forall i, j, b \fstop
\end{align}
Hence if we introduce the vectors
$m^{(i)}$ with components 
\bea
m^{(i)}_j = c_i \delta_{i,j} -  c_{i+1} \delta_{i+1,j} \coma m^{(i)}_a  =0   \,,
\eea
then
these vectors satisfy the relations
\begin{equation}
        f_{aj} m^{(i)}_k \delta^{k,j} = {\cal O}(1/\lambda) \coma 
f_{ij} m^{(i)}_k \delta^{k,j} = {\cal O}\left(\sqrt{\lambda}\right) \, 
\end{equation}
because the leading order terms cancel.
This means that the $(n-1)$ linear combinations 
\begin{equation}\label{eq:SClinearcomb}
\U(1)^{(i)}_-  = m^{(i)}_\alpha \U(1)^\alpha 
=  c_i \U(1)^i - c_{i+1} \U(1)^{i+1}   \quad \text{(no sum over i)}\coma \quad  i = 1, \ldots, n-1 \,,
\end{equation}
are not weakly coupled in the infinite distance limit, in the following sense: These $\U(1)^{(i)}_-$ gauge groups satisfy  
\begin{equation}\label{eq:qfijqstronglycoupled}
m_\alpha^{(i)} f^{\alpha \beta} m_\beta^{(i)} = {\cal O}\left(\sqrt{1/\lambda}\right) \,
\end{equation}
and hence fulfill  the {\it necessary} condition \eqref{necessarycondition} for the existence of a weak coupling limit. However, in order to check the stronger requirement \eqref{qfabq} for these gauge groups in the infinite distance limit, we need to compare \eqref{eq:qfijqstronglycoupled} to the scaling of the asymptotic quantum gravity cut-off. 

An infinite distance limit of Type $T^2$ corresponds to an effective decompactification limit from five- to six dimensions.
The quantum gravity cutoff 
in \eqref{qfabq} is hence the
species scale associated to the KK tower. 
For a general number $n$ of decompactifying dimensions, this species scale is given by\footnote{One can find this relation, as  in \cite{Montero:2022prj,Cota:2022yjw}, from the general definition of the species scale in five dimensions, i.e., $\Lambda_\text{\tiny sp}^3 =\frac{M_\text{\tiny Pl}^3}{N_\text{\tiny sp}}$, with $N_{\text{\tiny sp}}$ being the number of KK states with mass $m\leq k_{\text{\tiny max}} M_\text{\tiny KK}$, for a certain $k_{\text{\tiny max}}\in \mathbb{N}$.}
\begin{equation}
    \frac{\Lambda_\text{\tiny sp,KK}^3}{M_\Pl^3} = \left(\frac{M_\text{\tiny KK}^3}{M_\Pl^3}\right)^{\frac{n}{3+n}}\fstop
\end{equation}
In the present case, $n=1$ and $M_{\text{\tiny KK}}$ is set by the volume of the shrinking $T^2$-fiber wrapped by M2-branes multiple times, i.e., 
\begin{equation}
  \frac{M_\KK}{M_\text{\tiny 11d}}= 2\pi \mathcal{V}_{\text{\tiny $T^2$}} \Rightarrow \frac{M_{\KK}^3}{M_\Pl^3} = 2\pi^2\frac{\mathcal{V}_{\text{\tiny $T^2$}}^3}{\mathcal{V}}  = 2\pi^2\widehat{\mathcal{V}}^3_{\text{\tiny $T^2$}} \sim \frac{1}{\lambda^3}\fstop
\end{equation}
This leads to 
\begin{equation}\label{eq:5dKKspecscale}
    \frac{\Lambda_\text{\tiny sp,KK}^3}{M_\Pl^3} \sim \frac{1}{\lambda^{3/4}}\fstop
\end{equation}
Inserting \cref{eq:5dKKspecscale,eq:qfijqstronglycoupled} into \eqref{qfabq} we find 
\begin{equation}
    \frac{\Lambda_{\WGC}^2\left(\U(1)^{(i)}_-\right)}{\Lambda_\text{\tiny sp,KK}^2} \sim \frac{\lambda^{1/2}}{\lambda^{1/2}} \sim \text{const.}\,,
\end{equation}
such that, indeed, the $T^2$ limit is not a weak coupling limit for the $\U(1)^{(i)}_-$ gauge theories.

On the other hand, 
within the space $\U(1)^i$, 
the only linear combination of $\U(1)$s, orthogonal to 
the combinations $\U(1)^{(i)}_-$,
is the diagonal combination $\U(1)_{\cal E}$ defined in \eqref{U1+def}, which obeys the relation
\begin{equation}
\frac{\Lambda_{\WGC}^2\left(\U(1)_{\cal E}\right)}{\Lambda_\text{\tiny sp,KK}^2} \sim \frac{1}{\lambda^{3/2}} \,.
\end{equation} 
This $\U(1)_{\cal E}$  is the KK $\U(1)$ associated with the decompactification limit and, consistently, becomes weakly coupled in the limit.

Finally, it is clear from the scaling of the components $f_{ab}$ in \eqref{fcomponentsT2limit} that all $\U(1)^a$ for $a=1, \ldots, h^{1,1}(B_2)$, associated with the  base divisors, do not become weakly coupled in the limit of Type $T^2$. 
Unless $X_3$ admits in addition a limit of Type K3 or $T^4$ or a different type of $T^2$ limit,
these $\U(1)$s can never become asymptotically weakly coupled.

One may repeat a similar computation also for the $J$-class B limits introduced in \cite{Lee:2019wij}.  In fact, it follows from the uniqueness of the scaling behavior in the Type $T^2$ limit that the scalings of the divisor volumes $\mathcal{V}_i$ and $\mathcal{V}_a$ must be the same as in \eqref{volumescalings-T2}. As a result, our conclusions concerning the weak coupling nature of the abelian gauge fields carry over to limits of Type $T^2$ which are technically described as limits of $J$-class B.
Note that such a limit cannot be described as a  $J$-class A limit in particular when the base does not have any K\"ahler cone generators that do not square to zero. This is the case for a $\PP^1\times \PP^1$ base, where both K\"ahler generators of the base square to zero.

The analysis of the (non-)BPS towers associated with the gauge symmetries that become weakly coupled in a Type $T^2$ limit will be carried out in Section \ref{sec:tWGClimT2}.

\subsection{Weak Coupling in Type K3 Limits}
\label{sec:WeakCouplK3lim}

The next type of infinite distance limit is a limit of Type K3.
A Calabi--Yau 3-fold $X_3$ admitting such a limit is a K3-fibration
over a rational curve $\PP^1$ with projection
\bea \label{rhodef}
\rho:X_3\rightarrow \mathbb{P}^1 \,.
\eea
Let us choose a basis of $H_2(X_3)$ composed of Mori cone generators
\bea \label{MorisK3-gen}
\left\{\cC^\alpha \right\} = \left\{\cC^0, \cC^i, \quad i=1,\ldots, h^{1,1}(X_3)-1\right\} \,,
\eea
where $\cC^0 = \mathbb P^1$ is the base of the K3-fibration and $\cC^i$ are generators of the $\rho$-relative Mori cone of curves in the fiber. 
We furthermore introduce the dual basis of $H^{1,1}(X_3)$, 
\bea \label{K3basis}
\left\{J_\alpha \right\} = \left\{J_0, J_i, \quad i=1,\ldots, h^{1,1}(X_3)-1\right\} \,,
\eea
via the property that $J_\alpha \cdot \cC^\beta = \delta_\alpha^\beta$. 
In particular, $J_0$ is the class of the generic K3-fiber.
Recall that if the Mori cone is simplicial, the $J_\alpha$ are the generators of the dual K\"ahler cone. 
More generally, thanks to the K3-fibration structure
 one can always find a basis of K\"ahler cone generators $\{\hat J_0, \hat J_i\}$ where $\hat J_i \cdot \cC^0 =0$ and  $\hat J_0 \cdot \cC^i =0$   and furthermore $\hat J_0 = J_0$, $\hat J_i = m_i^j J_j$ for some non-degenerate matrix $m_i^j$.

As in Section \ref{sec:WeakCouplT2lim}, even though the basis $\{J_\alpha \}$ is composed of K\"ahler cone generators only if the cones are simplicial, it will be convenient for our purposes to expand the K\"ahler form $J$ in this basis. In the infinite distance limit of Type K3, the K\"ahler form can be parametrized as
\bea \label{JscalingK3a}
J = v^\alpha J_\alpha   = \lambda \tilde v^0 J_0 + \frac{1}{\sqrt{\lambda}} \tilde v^i J_i \coma \lambda \to \infty \,,
\eea
where $v^\alpha \geq 0$ can be interpreted as volumes of the basis of Mori cone generators $\cC^\alpha$. Note that $\tilde v^0$ does not scale with $\lambda$ and $\tilde v^i$ does either not scale with $\lambda$  or vanishes in the limit $\lambda \to \infty$, but in such a way that the total volume $\cV = \frac{1}{6}\int_{X_3} J^3$ remains finite. 

Indeed in \cite{Lee:2019wij} it was shown that every limit of Type K3 is characterized by a K\"ahler form
$J = \lambda \tilde v^0 \hat J_0 + w^i \hat J_i$, where $\hat J_\alpha$ is a basis of K\"ahler cone generators such that in particular $\hat J_0^2=0$ and with $w^i \prec \lambda$ for all remaining K\"ahler parameters $w^i$. 
The scaling $v^0 = \lambda \tilde v^0$
reflects the expansion rate of the  volume of the base $\cC^0 =\mathbb P^1$.  
Furthermore, the K3-fiber volume scales like $1/\lambda$ homogeneously, meaning that the volumes of all curves in the fiber scale to zero at a rate $1/\sqrt{\lambda}$ or faster.\footnote{Otherwise, the K3-fiber would have to admit an elliptic fibration and the limit $\lambda \to \infty$ would describe a limit of Type $T^2$ \cite{Lee:2019wij}.} We can then transform to the basis $J_\alpha$ as explained above and obtain the scaling \eqref{JscalingK3a} with the infinite distance parameter $\lambda$.

To determine the gauge kinetic function \eqref{fabMth},
we compute the divisor and curve volumes 
taking into account the relation 
\bea
\kappa_{00i}=0\,
\eea
that follows from $J_0^2=0$. 
In terms of the expansion parameters appearing in \eqref{JscalingK3a}
 and the intersection form of the generic K3-fiber,
\begin{equation} \label{etadef}
    \kappa_{0ij} = \eta_{ij} \,,
\end{equation}
one finds 
\begin{equation}
\begin{split} \label{eq:volumes-K3fibr}
    \mathcal{V}   &= \frac{1}{6}\kappa_{\alpha\beta\gamma}v^\alpha v^\beta v^\gamma = \frac{1}{2}v^0 \eta_{ij}v^iv^j+\frac{1}{6}\kappa_{ijk}v^iv^jv^k\coma\\
    \mathcal{V}_0 &= \frac{1}{2}\kappa_{0\beta\gamma}v^\beta v^\gamma = \frac{1}{2}\eta_{ij}v^iv^j\coma\\
    \mathcal{V}_i &= \frac{1}{2}\kappa_{i\beta\gamma}v^\beta v^\gamma = v^0 \eta_{ij}v^j+\frac{1}{2}\kappa_{ikl}v^kv^l\coma\\
    \mathcal{V}_{\alpha\beta} & = \begin{pmatrix} \mathcal{V}_{00} & \mathcal{V}_{0j} \\ \mathcal{V}_{i0} & \mathcal{V}_{ij} \end{pmatrix} = \begin{pmatrix} 0 & \eta_{jk}v^k \\ \eta_{ik}v^k & v^0\eta_{ij} + \kappa_{ijk}v^k  \end{pmatrix} \,.
\end{split}
\end{equation}
These formulae hold
even before taking any asymptotic limit.
Expressed in the volumes rescaled as in \eqref{eq:rescaledcoord}, the components of the gauge kinetic matrix become
\begin{equation}
    \begin{split} \label{gaugekinK3limita}
        f_{00} &=  \frac{1}{4}(\eta_{kl}\hat{v}^k\hat{v}^l)^2 \coma \\
        f_{0j} & =  \frac{1}{2}(\eta_{kl}\hat{v}^k\hat{v}^l)\left(\hat{v}^0 \eta_{jk}\hat{v}^k+\frac{1}{2}\kappa_{jkl}\hat{v}^k\hat{v}^l\right)-\eta_{jk}\hat{v}^k \coma\\
        f_{i0} &=  \frac{1}{2}(\eta_{kl}\hat{v}^k\hat{v}^l)\left(\hat{v}^0 \eta_{ik}\hat{v}^k+\frac{1}{2}\kappa_{ikl}\hat{v}^k\hat{v}^l\right)-\eta_{ik}\hat{v}^k\coma \\
        f_{ij} & = \left(\hat{v}^0 \eta_{ik}\hat{v}^k+\frac{1}{2}\kappa_{ikl}\hat{v}^k\hat{v}^l\right)\left(\hat{v}^0 \eta_{jk}\hat{v}^k+\frac{1}{2}\kappa_{jkl}\hat{v}^k\hat{v}^l\right)-(\hat{v}^0\eta_{ij} + \kappa_{ijk}\hat{v}^k) \fstop
    \end{split}
\end{equation}

In the infinite distance limit \eqref{JscalingK3a}
one finds the scaling
\begin{equation} \label{K3-fcomponents2}
    f_{00} \sim \lambda^{-2} \coma f_{0j} \sim f_{i0} \sim \lambda^{-1/2} +\mathcal{O}(\lambda^{-1}) \coma f_{ij} = \lambda ((\hat{\tilde v}^0)^2 \eta_{ik}\hat{\tilde v}^k \eta_{j l}\hat{\tilde v}^l -  \hat{\tilde v}^0 \eta_{ij}) +\mathcal{O}(\lambda^{-1/2})\fstop
\end{equation}

We now determine the linear combinations of abelian gauge fields which become weakly coupled in this limit.
Let us suppose for the moment that the components of $f_{ij}$ scaling with $\lambda$ in \eqref{K3-fcomponents2} define a submatrix  of maximal rank. 
Under this assumption, any charge vector $\bm{Q}=(0,Q_i)$ satisfies the relation 
\bea\label{eq:qfabqK3fib}
Q_\alpha f^{\alpha\beta} Q_\beta = Q_i f^{ij} Q_j \sim \frac{1}{\lambda} \, \to 0 \,.
\eea
This means that each of the gauge groups $\U(1)^i$ associated with the  basis elements  $J_i$
satisfies the necessary condition \eqref{necessarycondition} to become weakly coupled.
To test for the sufficient condition \eqref{qfabq},
recall that 
in the infinite distance limit, the duality frame changes to that of the perturbative heterotic string obtained from an M5-brane wrapped on the generic K3 surface.
In the limit, 
 the new heterotic string scale decreases at the rate
 \bea
 M_\het \sim  \frac{1}{\sqrt{\lambda}}M_{\text{\tiny 11d}} \,.
 \eea
 The excitations of the heterotic string provide a tower of non-BPS particles whose mass scales with $\lambda$ in the same way as the KK tower \cite{Lee:2019wij}, namely
\begin{equation}
    \frac{M_{n_{\text{\tiny L}}}^2}{M_\Pl^2} \sim n_{\text{\tiny L}} \frac{T_\het}{M_\Pl^2} = 2\pi n_{\text{\tiny L}} \mathcal{V}_{\text{\tiny K3}}\frac{M_{\text{\tiny 11d}}^2}{M_\Pl^2} = 2\pi \frac{n_{\text{\tiny L}}}{\lambda}\mathcal{V}^{1/3}\coma
\end{equation}
where $n_{\text{\tiny L}}$ is the left-moving excitation level. The masses of the particles of this tower, then, set the species scale $\Lambda_{\text{\tiny sp}}$ \cite{Dvali:2009ks,Dvali:2010vm} of the effective heterotic string theory, which is the quantum gravity cutoff scale $\Lambda_{\text{\tiny{QG}}}$ appearing in \eqref{weakcouplingcond}.

More precisely, $\Lambda_{\text{\tiny sp}}$ lies above the heterotic string scale due to a logarithmic scaling factor between species scale and string scale \cite{Marchesano:2022axe,Castellano:2022bvr},
\bea
\Lambda_\text{\tiny sp}^2 \sim M_\het^2 \log\left(\frac{M_{\Pl}}{M_\het}\right)\,.
\eea

We can then compute \eqref{qfabq} for a $\U(1)$ of the form
$\U(1) = Q_i \U(1)^i$
with the help of \eqref{eq:qfabqK3fib}, which results in
\begin{equation}
    \label{weakcoupling-hetlimit}
    \frac{\Lambda^2_{\WGC}\left(\U(1)\right)}{\Lambda^2_{\spc}} \sim \frac{1/\lambda}{1/\lambda(1+{\rm log}(\lambda))}   \to 0 \fstop
\end{equation}
Therefore every linear combination of $\U(1)^i$, $i= 1, \ldots, h^{1,1}(X_3)-1$, is indeed weakly coupled in the infinite distance limit, provided the leading part of $f_{ij}$ in \eqref{K3-fcomponents2} defines a maximal rank matrix.

By contrast, any combination of gauge factors containing $\U(1)^0$, the abelian gauge group associated with the K3-fiber class $J_0$, remains strongly coupled in the limit. In fact, in the K3 limit, $\U(1)^0$ can be identified with the graviphoton $\U(1)_{\rm grav.}$. To see this, recall that the linear combination of the $\U(1)$s that is identified with the graviphoton is moduli-dependent in such a way that the central charge of wrapped M2-branes gives the charge under the $\U(1)_{\rm grav.}$. In the limit \eqref{JscalingK3a} the central charges of M2-branes vanish asymptotically unless the wrapped curve contains the base $\mathbb{P}^1$ and hence carry charge under $\U(1)^0$. Accordingly, $\U(1)^0$ is asymptotically identified with the graviphoton $\U(1)_{\rm grav.}$. This explains why  $\U(1)^0$ does not admit a weak coupling limit in the vector multiplet moduli space. In fact, it does not admit a weak coupling limit at all because its coupling strength is always set by the Planck scale.

The discussion so far is valid under the assumption that the leading components of the matrix  $f_{ij}$ define a maximal rank matrix.
It therefore remains to understand the invertibility properties of the leading components of $f_{ij}$ in more detail.
To this end, we must distinguish 
fibrations in which the K3-fiber does not degenerate anywhere over the base $\mathbb P^1$ and those where degenerations occur over points.

\subsubsection{Fibrations without Degenerations}

If the K3 fiber does not degenerate anywhere, the divisors $J_i$ other than the generic K3-fiber of class $J_0$ are obtained by fibering holomorphic curves of the generic fiber over the base $\mathbb P^1$.
More precisely, if $\iota: \text{K3} \xhookrightarrow{} X_3$ denotes the embedding of the K3-fiber into $X_3$, then the image of 
the Picard group ${\rm Pic}(\text{K3}):= H^2(\text{K3},  \mathbb Z) \cap H^{1,1}(\text{K3})$ of the generic fiber under $\iota_\ast: H_2(\text{K3}) \to H_2(X_3)$ defines a set of holomorphic curve classes on $X_3$. Under our assumption that the fiber does not degenerate anywhere, their span is the $\rho$-relative Mori cone of $X_3$.
We can obtain the basis of Mori generators \eqref{MorisK3-gen} of $X_3$ by taking a basis of generators of 
this relative Mori cone and completing it with the base $\mathbb P^1$. The dual basis of divisors $J_\alpha$ of $X_3$ is then the basis \eqref{K3basis}, where each $J_i$, $i=1, \ldots, h^{1,1}(X_3)$, indeed has the structure of a fibration over $\mathbb P^1$ by a curve in $\iota_\ast({\rm Pic}(\text{K3}))$.

The pullback $\iota^\ast(H^{1,1}(X_3))$ defines the so-called polarization lattice
\bea \label{Lambda0-def}
\Lambda_0 \subset \Gamma^{3,19} = U^{\oplus 3} \oplus (E_8 \oplus E_8) \,
\eea
of signature $(1,k_0)$ for some $0 \leq k_0 \leq 19$.
The curves in $\iota_\ast({\rm Pic}(\text{K3}))$ then span the lattice $\Lambda_0^\ast$ dual to $\Lambda_0$.
In \eqref{Lambda0-def}, $U$ denotes a copy of the hyperbolic lattice and $E_8$ represents the eponymous Lie algebra lattice.
Indeed, recall that $\Gamma^{3,19}$ is the lattice $H^2(\text{K3}, \mathbb Z)$ of a K3 surface, and for every algebraic K3 surface,
${\rm Pic}(\text{K3})$ is a sublattice thereof of signature $(1,r)$ with $r \leq 19$.

The intersection matrix \eqref{etadef} coincides with the non-degenerate intersection matrix on the lattice $\Lambda_0$ and therefore has full rank. This shows that also the leading components of $f_{ij}$ in \eqref{K3-fcomponents2} define a matrix of maximal rank, as claimed.

\subsubsection{Fibrations with Degenerations}

Fibrations with degenerate
fibers over points on the base
can be divided in two main blocks, depending on whether the degenerations  occur at \textit{finite} or \textit{infinite} distance in the K3 fiber moduli space. 
Up to base change and birational transformations,
degenerations at finite distance in K3 moduli space correspond to K3 degenerations of Kulikov Type I, while those at infinite distance are of Kulikov Type II or III \cite{Kulikov1,Kulikov2,PerssonPink}. Following the reviews of semi-stable degenerations and Kulikov models in \cite{Braun:2016sks,Lee:2021qkx}, let us discuss how the degenerate K3-fibers fit in Kulikov's classification. 
   Therefore consider the K3-fibration $X_3$ and its projection $\rho$ 
   defined in \eqref{rhodef}.
We can zoom-in to a small disk $D$ centered around the point $p$ and focus on  
    \begin{equation}
        \rho_D: X_D \rightarrow D\,,
    \end{equation}
    the restriction of the K3-fibration to $D$. The fiber $\sS_u$ over a generic point $0\neq u\in D$ is a smooth K3. As we vary $u$ we generate a one-parameter family of K3-surfaces $\sS_u$. 
    Under the assumption of semi-stability,
    the central fiber $\sS_0 = \sS_{u}|_{u=0}$ degenerates into a union
    \begin{equation}\label{S0split}
        \sS_0 = \bigcup_{M=1}^N \sS_M \,
    \end{equation}
     of reduced surfaces $\sS_M$ with at worst normal crossing singularities. Notice that for the local 3-fold $X_D$ this semi-stability can always be achieved by performing a suitable base change, which amounts to a reparametrization of the local fibration $\rho_D$, possibly together with a birational transformation.
    For compact fibrations $X_3$, such a base change is a more drastic operation which would change the geometry.
    For now, we assume that $X_D$ is embedded in the compact 3-fold in a way consistent with the semi-stable degeneration without having to perform a base change on $X_3$, and we will comment on more general situations below.
    
    The main difference between Type I and Type II, III Kulikov models is that for the former, the number of components is $N=1$ and the special fiber is smooth, whereas the latter cases have $N>1$. Hence for a Type I Kulikov model the fiber $\sS_0$ is irreducible, whereas for Type II and III it splits into multiple components. For our following discussion, the distinction between Type II and III Kulikov fibers is inessential.\footnote{Fibers of Kulikov Type II and III are distinguished, among other things, by the way how the different components $\sS_M$ intersect. For Type II Kulikov models the $\sS_M$ form a chain with $\sS_M\cap \sS_{M+1}$ being elliptic curves with fixed complex structure. For Type III Kulikov models, instead, the $\sS_M$ form a triangulation of $S^2$ and, if non-empty, $\sS_M\cap \sS_K$ are rational curves.}  Instead the main question is whether the degenerate fiber is reducible or irreducible and hence 
    whether the degeneration is at 
    finite distance (Type I) or at infinite distance (Type II and III) in the K3 moduli space. 

In the presence of degenerations, the set of Mori cone generators $\cC^i$ (not including the base $\cC^0 = \mathbb P^1$) 
splits into different types of curves,
\bea
\{ \cC^i  \}= \{  \cC^\iota, \cC^m, \cC^\mu   \} \fstop
\eea
Here we denote by
\begin{itemize}
\item
$\cC^\iota$, $\iota \in {\cal I}_0$,
Mori cone generators located in the generic K3-fiber,
\item
$\cC^m$, $m \in {\cal I}_{\rm I}$, Mori cone generators localized in special fibers of Kulikov Type I or which can be ensured to lie in Kulikov Type I fibers upon deformations of $X_3$,\footnote{\label{footnote_Kulikov}It can happen that a Type I and a Type II/III fiber coincide on $X_3$. The distinction between curves $\cC^m$ and $\cC^\mu$ must then be made after separating the fiber types.}
\item
$\cC^\mu$, $\mu \in  {\cal I}_{\rm II/III}$ Mori cone generators localized in special fibers of Kulikov Type II or III and which cannot be deformed out of such fibers.
\end{itemize}

This split induces a corresponding distinction between the dual divisors $J_i$, $J_m$ and $J_\mu$.

The Mori cone generators $\cC^\iota$, $\iota \in {\cal I}_0$, and their dual divisors $J_\iota$, behave in the same way as the generators in the absence of any degenerations.
Generators of the form $\cC^m$ arise when a curve in a generic fiber splits into one or several components over a special point $p_{\rm I} \in \mathbb P^1$ where the fiber undergoes a Type I Kulikov degeneration. 
Over the degeneration point $p_{\rm I}$, the rank of the Picard lattice of the fiber may enhance due to the appearance of such extra holomorphic curves. What is important for us is that the intersection form restricted to the index set $\{\iota, m\} \in {\cal I}_0 \cup {\cal I}_{\rm I}$ continues to be non-degenerate because it can be identified with the intersection form of a -- possibly higher-rank  -- sublattice $\Lambda_{\rm 0+I}$ of $\Gamma^{3,19}$. 
Hence again all linear combinations of abelian gauge potentials $\U(1)^\iota$ and $\U(1)^m$ are asymptotically weakly coupled because they satisfy the relation \eqref{weakcoupling-hetlimit}.

 Let us now assume that the K3-fiber degenerates at \textit{infinite} distance (in the K3 moduli space) over a point $p_{\rm II/III}\in \mathbb{P}^1$ in the base. As reviewed above, at an infinite distance degeneration a K3 surface in general splits into a union of surfaces intersecting over curves. These surfaces give rise to additional effective divisors of $X_3$ which are localized over one of the points $p_{\rm II/III}$.
We denote by $D_\rho$ the set of all such divisor classes such that we can write the class of the generic fiber $J_0$ as
\begin{equation}\label{eq:J0inDrho}
   J_0 = \sum_{\rho} a_\rho D_\rho\,. 
\end{equation}
Importantly, since the divisors $D_\rho$ are localized in the fibers over one of the points $p_{\rm II/III}$, their intersection with the generic fiber $J_0$ vanishes.

There must also exist one or several curve classes $C$ in the generic K3-fiber whose class splits 
into various components over one or several of the points $p_{\rm II/III}$ (if necessary after performing a deformation to separate the fibers, as in Footnote \ref{footnote_Kulikov}).
Let us define by ${\bf M}(X_3, \infty)$ the set of curve classes which {\it only} exist in the fiber over one of the points $p_{\rm II/III}$.
Each element in ${\bf M}(X_3, \infty)$ can be written as a positive linear combination of the subset of Mori cone generators $\cC^\mu$.
A general curve class in the fiber can be expressed as 
\bea
C = \sum_\mu c_\mu \cC^\mu + C_{\rm rest}  \coma
\eea
for some coefficient $c_\mu$ and if $c _\mu =0$ for all $\mu$, then $C$ does not (partially) split over any $p_{\rm II/III}$.
The divisors $J_\mu$ dual to $\cC^\mu$
have the property   
\begin{equation}
  c_\nu  J_\mu \cdot C = c_\mu J_\nu \cdot C \,, 
\end{equation}
and furthermore, 
\bea
J_\mu \cdot C =0   \quad \text{if} \quad c_\mu = 0 \,.
\eea
In particular, if for a given curve all $c_\mu = 0$, this curve can be moved away from any of the points $p_{\rm II/III}$.
Hence, away from the points $p_{\rm II/III}$ all generators $J_\mu$ agree with each other.

In other words, we can relate any $J_\mu$ and $J_\nu$ up to divisors $D_\rho$ localized in the fiber over $p_{\rm II/III}$, i.e.,
\begin{equation}
  c_\nu  J_\mu = c_\mu J_\nu + \sum_\rho \alpha_\rho D_\rho\coma
\end{equation}
where  $\alpha_\rho\in \mathbb{\mathbb Q}$. Since the $D_\rho$ are localized in the fiber over $p_{\rm II/III}$, they integrate to zero over the generic fiber, i.e., 
\begin{equation}\label{eq:intgenfiber}
    0=\int_{\rm generic\;fiber} D_\rho \cdot D_\alpha\,,
\end{equation}
where $D_\alpha$ is any other effective divisor.\footnote{In particular \eqref{eq:intgenfiber} applies to all K\"ahler cone generators.} Since $J_0$ is the class of the generic fiber we have
\begin{align}\label{identity}
  c_\nu J_\mu \cdot J_0 \cdot J_\alpha =c_\mu J_\nu \cdot J_0 \cdot J_\alpha  \coma \forall \alpha \in \{0\} \cup {\cal I}_{0} \cup {\cal I}_{\rm I} \cup {\cal I}_{\rm II/III} \,.
\end{align}
Therefore, in the limit    \eqref{JscalingK3a},
the components of the gauge kinetic matrix \eqref{gaugekinK3limita} satisfy the relations
\begin{align}
  c_\nu f_{\mu \alpha} \sim  c_\mu f_{\nu \alpha} \coma   \forall \mu, \nu \in {\cal I}_{\rm II/III} \coma \forall \alpha \coma
\end{align}
to leading order in $\lambda$. As a result, the rank of
the submatrix $f_{ij}$ is in fact reduced to leading order.
This implies that for each curve $C$ that splits in a Type II/III degeneration, there is a single linear combination 
\begin{equation}\label{eq:weaklycoupledU1K3IIt}
 \U(1)_C= \sum_\mu c_{\mu} \U(1)^\mu + \U(1)_{C_{\rm rest}}
\end{equation}
which becomes weakly coupled in this limit. All orthogonal linear combinations of $\U(1)^\mu$ do not become weakly coupled. 
For $\U(1)_C$, on the other hand, the discussion around 
\eqref{eq:qfabqK3fib} applies and 
\begin{equation}
    \Lambda^2_{\WGC}(\U(1)_C) = g_{\text{\tiny YM,C}}^2 M_{\Pl}^3 \sim \frac{1}{\lambda}\fstop
\end{equation}
One therefore finds a relation as in \eqref{weakcoupling-hetlimit} for $\U(1)_C$, which confirms that it becomes weakly coupled.

In Section \ref{sec:examples} we illustrate the different types of $\U(1)$s and their weak coupling limits in a K3 fibration with a Type II degenerate fiber.

The discussion so far has made use of the Kulikov classification of semi-stable K3 degenerations, and we have already pointed out that in general the degenerations can be put in semi-stable form only up to base change. The appearance of non-Kulikov type fibers on compact Calabi--Yau 3-folds has been stressed in particular in \cite{Braun:2016sks}, to which we refer for some examples of what the authors call would-be Type I, II, or III fibers prior to performing such base change and birational transformations to make all fibers semi-stable.
However,  the details of the degenerate fibers do not matter for our arguments, except for the distinction between degenerations in which the K3-fiber remains irreducible versus those where it splits into several components over points. 
The latter types of degenerations do not give rise to any additional weakly coupled gauge groups, even if they are not yet in semi-stable form.

We conclude that the $\U(1)$s
with a weak coupling limit fall into the classification of Claim \ref{claim2}. They arise from the perturbative gauge sector of the emergent heterotic string \cite{Lee:2019wij} associated with an M5-brane along the K3-fiber, including possible Kaluza--Klein or winding $\U(1)$s as well as (a subgroup of) the ten-dimensional heterotic gauge group. This interpretation and 
the origin of the associated (non-)BPS towers in the five-dimensional effective theory will be elaborated on in Section \ref{sec:tWGClimK3}.

\subsection{Weak Coupling in Type \texorpdfstring{$T^4$}{} Limits}   
\label{sec:WeakCouplT4lim}

The last type of infinite distance limit  is a limit of Type $T^4$, in which the shrinking fiber is an Abelian surface. The infinite distance limits of Type $T^4$ in M-theory are equi-dimensional in the sense of \cite{Lee:2019wij}. The pattern of asymptotically weakly  gauge groups generally parallels the Type K3 limits studied in Section \ref{sec:WeakCouplK3lim}. We can therefore be brief and focus on where the discussions differ.

Let us consider a Calabi--Yau $X_3$ that admits an Abelian surface fibration over a base $\mathbb P^1$, with general fiber $\sS$. The pullback of $H^{1,1}(X_3)$ to the generic fiber induces a polarization lattice (cf. \cite{HulekKlaus2011MSoA}).
In order to study the possible degenerations of the fibration,
note that 
there exist no $(-2)$-curves on the generic fiber $\sS$ (which has $c_2({\sS}) =0$) that could shrink to zero size at finite distance in the fiber moduli space and would thereby lead to a local degeneration of $\sS$.

If the fiber degenerates over a special point $p\in \mathbb{P}^1$ at infinite distance in the fiber moduli space, it becomes reducible as ${\sS}_0=\bigcup_{i=1}^n {\sS}_i$. The surface components ${\sS}_i$ lead to divisors of $X_3$ localized in the degenerate fiber, such that we can apply the logic of Section \ref{sec:WeakCouplK3lim} and infer that the additional $\U(1)$s associated to degenerations of the Abelian surface fiber do not become asymptotically weakly coupled in a limit of Type $T^4$. As a result, for an Abelian surface fibration the potential asymptotically weakly coupled 
$\U(1)$s  are in one-to-one correspondence with the polarization lattice induced by the pullback of $H^{1,1}(X_3)$ to the generic fiber.

This, of course, matches expectations based on the interpretation of the Limit of Type $T^4$ given in \cite{Lee:2019wij}:
The M5-brane wrapping the generic fiber gives rise to a Type IIB string defining a dual frame. Unlike the heterotic string, this string does not support a perturbative gauge group in ten dimensions. The potentially weakly coupled gauge group factors therefore always play the role of Kaluza--Klein or winding $\U(1)$s, which, in Type II string theory, cannot enhance to a non-abelian gauge group. This explains why there cannot occur any $(-2)$-curves on the generic fiber, whose shrinking at finite distance in the surface moduli space would signal such a non-abelian enhancement.
On the other hand, $(-2)$-curves can of course localize in the multiple component fibers, as is common for instance in suitable Schoen manifolds \cite{Schoen} with Kodaira type fibers. The corresponding directions in the charge lattice, however, are not associated with asymptotically weakly coupled $\U(1)$s.

\section{(Non-)BPS Tower and Weak Gravity Conjecture} 
\label{sec_TowerWGC}

In Section \ref{sec:WeakCoupllims}, we have analyzed which abelian gauge groups become weakly coupled in each of the possible weak coupling limits of Type \ref{pos1}, \ref{pos2a} and \ref{pos2b} and shown that these are classified as in Claim \ref{claim2}. We are therefore ready to
return to our original goal of proving Claim \ref{mainclaim}. With this in mind, we now
analyze the (non-)BPS towers that exist with charges under the gauge groups admitting a weak coupling limit, considering each type of limit separately.

 \subsection{(Non-)BPS Towers in Type \texorpdfstring{$T^2$}{} Limits}
\label{sec:tWGClimT2}

For each limit of Type $T^2$,
there exists only a single asymptotically weakly coupled $\U(1)_{\cal E}$, defined in \eqref{U1+def}. This $\U(1)_{\cal E}$ clearly satisfies the asymptotic tower WGC conjecture because the ray in the charge lattice along $n \bm{Q}_{\cal E}$ is populated by a tower of BPS particles.
These are the BPS states obtained by wrapping $n$ M2-branes along the generic fiber, which furnish the KK tower for the effective five-dimensional to six-dimensional limit. The BPS invariants, more precisely the genus-zero Gopakumar--Vafa invariants, for the curve $n{\cal E}$ are well known to coincide with the Euler characteristic\footnote{When $\chi(X_3)=0$, this indicates a cancellation of BPS states in the index computed by the BPS invariants due to an underlying supersymmetry enhancement.} of $X_3$, i.e., \cite{Oehlmann:2019ohh,Kashani-Poor:2019jyo}
\bea
N^0_{n \cal E} = -\chi(X_3) \,.
\eea

Unless $X_3$ also admits a K3/$T^4$ fibration, 
all other curve classes are necessarily charged under linear combinations of $\U(1)$s without a weak coupling limit. 
 For example, this is true for any curve class charged only under the combinations $\U(1)^{(i)}_-$ defined in \eqref{eq:SClinearcomb}. These curve classes are contractible and as such they do not lie inside the movable cone of $X_3$ (because they are localized in special fibers over the discriminant of the fibration). According to the criteria of \cite{Alim:2021vhs}, these charge rays do not necessarily give rise  to BPS towers, in agreement with explicit computations of their BPS indices.
Since one cannot take a weak coupling for the $\U(1)_i^{(i)}$ the absence of an obvious candidate tower -- BPS or non-BPS -- is not in contradiction with the {\it asymptotic} WGC. We will come back to this point at the end of Section \ref{sec:discussion}.
 
This does not mean that BPS towers are excluded along rays whose generating curves are charged only under $\U(1)$s without a weak coupling limit. 
For instance, for an elliptic fibration over $B_2 = \mathbb P^2$ the ray along the generator of the curve lattice of the zero-section does admit a tower of BPS states even though it is charged under a $\U(1)$ without a weak coupling limit (in particular, $X_3$ cannot undergo a limit of Type K3 or $T^4$ in this case). 

\subsection{(Non-)BPS Towers in Type K3 Limits}
\label{sec:tWGClimK3}

We now investigate the existence of asymptotically light towers for Type K3 limits. We have shown in Section \ref{sec:WeakCouplK3lim} that the only abelian gauge groups which become asymptotically weakly coupled in limits of Type K3 are those associated with curves in the generic K3-fiber or curves in irreducible degenerate fibers. Such curves form a sublattice $\Lambda^\ast_{\rm 0+I}$ of rank $(1,k_{\rm 0+I})$ of $\Gamma^{3,19}$, which we will henceforth abbreviate as
\begin{equation}
\Lambda^\ast:= \Lambda^\ast_{\rm 0+I} \,.
\end{equation}
This lattice is the dual of the polarization lattice $\Lambda$ of the K3-fibration 
$\rho: X_3 \rightarrow \mathbb{P}^1$ and will also be referred to as the charge lattice.

In the sequel, we will denote by $\Lambda^\ast_\pm$ the (anti-)self-dual part of $\Lambda^\ast$ inside the orthogonal decomposition 
\begin{equation} \label{eq:Lambdapm}
\Lambda^\ast_{\mathbb{R}}= \Lambda^\ast_+ \oplus \Lambda^\ast_- \,,
\end{equation}
where $\Lambda^\ast_+$ maps into $\mathbb{R}^{1,0}$ and $\Lambda^\ast_-$ into $\mathbb{R}^{0,\text{rk}\Lambda^\ast-1}$
 under the isometry from $\Lambda^\ast_{\mathbb{R}}$ to $\mathbb{R}^{1,\text{rk}\Lambda^\ast-1}$. 
 Here our convention is such that an element $\boldsymbol{Q}_+ \in \Lambda^\ast_+$ satisfies $\boldsymbol{Q}_+^2 > 0$, while $\boldsymbol{Q}_-^2  < 0$ for $\boldsymbol{Q}_- \in \Lambda^\ast_-$.

Let us consider first the sublattice associated with
curves $\Sigma$ with $\Sigma \cdot_{\text{\tiny K3}} \Sigma \geq 0$ in $\Lambda^\ast$. These define rays in the charge lattice admitting towers of BPS states: M2-branes wrapped $n$ times along such curves give rise to BPS particles with associated genus-zero Gopakumar--Vafa (or BPS) invariants 
\bea \label{BPSinvariants-K3fiber}
N^0_{n \Sigma} \neq 0 \quad  \text{for} \quad  n \in \mathbb N  \quad \text{if} \quad \Sigma \cdot_{\text{\tiny K3}} \Sigma \geq 0 \,.
\eea

The non-vanishing of $N^0_{n\Sigma}$ is a consequence of Noether--Lefschetz theory \cite{MR3114953} and also fits perfectly with the results of \cite{Alim:2021vhs}: Such curves lie in the movable cone of $X_3$ because they are movable within the K3-fiber. 
In light of the duality between M-theory on the K3-fibration $X_3$ and the heterotic string in five dimensions, the BPS states counted by the invariants \eqref{BPSinvariants-K3fiber} represent certain
modes of the dual heterotic string. The non-vanishing of their multiplicities reflects the fact that the BPS invariants $N^0_{n \Sigma}$  are the expansion coefficients of  modular forms \cite{Harvey:1995fq,MR3114953}. 
This property was already used in
\cite{Lee:2019wij} to argue for the appearance of the asymptotically light associated BPS towers 
in infinite distance limits of Type K3.

By contrast, all curves with $\Sigma \cdot_{\text{\tiny K3}} \Sigma < 0$ define rays in the charge lattice which do not support a BPS tower. 
Again, this is in agreement with the results of \cite{Alim:2021vhs} because such curves are rigid within the K3-fiber and consequently lie outside the movable cone of $X_3$. At the same time, the associated directions in the charge lattice are charged only under abelian gauge groups with a weak coupling limit, and hence the asymptotic tower WGC calls for a super-extremal tower of states in these directions.

As we will argue in this section, the missing BPS tower is replaced by a tower of non-BPS states which satisfy the asymptotic tower RFC, \eqref{eq:5dRFC}. The non-BPS states are excitations of the emergent heterotic string, which becomes light in the weak coupling limit, as stressed already in \cite{Lee:2019wij}.
The existence of this asymptotically self-repulsive non-BPS tower proves our Claim \ref{mainclaim}.

The arguments leading to the super-extremal strings differ depending on whether $\bm{Q}$ receives a contribution from $\Lambda^\ast_+$.
In Section \ref{sec:ellipticgenera5d}, we first specialize to an anti-self-dual charge vector $\bm{Q}\in \Lambda^\ast_-$.
For such charges, the missing non-BPS states can be directly identified as excitations of the five-dimensional heterotic string obtained by wrapping an M5-brane on the K3-fiber in M-theory. 
We will first introduce the tools that allow us to count a certain subset of the non-BPS excitations of 
this five-dimensional heterotic string. 
To this end, we will invoke the duality between M-theory on $X_3 \times S^1_M$ and Type IIA string theory on $X_3$. 
We will identify the coefficients of the elliptic genus of the five-dimensional heterotic string at left-moving excitation level $n$ and with charge vector $\bm{Q} \in \Lambda^\ast_{-}$ with 
certain D4-D2-D0 Donaldson--Thomas (DT) invariants in Type IIA string theory on $X_3$. The existence of these states then follows from the correspondence with Noether--Lefschetz numbers \cite{Katz:1999xq,Pandharipande:2014qoa,MR3114953,Bouchard:2016lfg}.
This reasoning is similar in spirit to the way the existence of a tower of excitations of a six-dimensional heterotic string was deduced in \cite{Lee:2018spm,Lee:2019tst}, but the discussion in five dimensions differs at the technical level.
As our main conclusion, there exists a distinguished set of non-BPS states at left-moving excitation level $n_{\text{\tiny L}}$ and with (anti-self-dual) charge vector $\bm{Q}\in \Lambda^\ast_-$ obeying the relation
\bea \label{eq_nQrelation}
n_{\text{\tiny L}} = - \frac{1}{2} Q_i \eta^{ij} Q_j =: - \frac{1}{2} \bm{Q}^2     \,,
\eea
where $\eta^{ij}$ is the inverse of the intersection \eqref{etadef} form on the K3-fiber.

However, there can be charge vectors $\bm{Q}^2 < 0$ that do not belong only to $\Lambda^\ast_-$. As we argue in Section \ref{sec:nonBPS-moregeneral} for these charges there also exist string excitations at left-moving excitation level $n_{\text{\tiny L}}$ satisfying \eqref{eq_nQrelation}.
To show this, we will make use of the duality of M-theory on $X_3$ and a five-dimensional heterotic string, which in the simplest cases is the heterotic string compactified on K3$_\het \times S^1_\het$.

Finally, in Section \ref{sec:tWGCfornonBPS}, we will show that the towers of states with the property $n_{\text{\tiny L}} = - \frac{1}{2} \bm{Q}^2$ indeed satisfy the asymptotic RFC, \eqref{eq:5dRFC}, in the weak coupling limit, and hence provide the states missing at the BPS level.

\subsubsection{Elliptic Genera of Strings in Five Dimensions and Super-Extremal States with \texorpdfstring{$\bm{Q}_+ =0$}{}}
\label{sec:ellipticgenera5d}

In this section, we will identify
 non-BPS particle states with $\bm{Q}\in \Lambda^\ast_-$ obeying the relation \eqref{eq_nQrelation} in the five-dimensional effective theory of M-theory compactified on $X_3$.
These states are excitations of
the BPS string obtained by wrapping an M5-brane on the divisor $\mathbf{S} \subset X_3$ given by the generic K3-fiber. 
This heterotic string is a special example of a Maldacena--Strominger--Witten (MSW) string~\cite{Maldacena:1997de} and has an associated $\mathcal{N}=(0,4)$ worldsheet theory description.

M-theory on $X_3 \times S^1_M$ is dual to Type IIA string theory on $X_3$. Wrapping the five-dimensional heterotic MSW string $r$ times on $S^1_M$ gives rise to a set of BPS states in four dimensions.
More precisely, the states with KK number $n$ along $S^1_M$ and of charge $\bm{Q} \in \Lambda^\ast$ give BPS bound states
of D4-D2-D0 charge vector $\gamma = (r,\bm{Q},n)$.
The BPS counting of such states is, in turn, captured by the modified elliptic genus~\cite{Gaiotto:2006wm}
\begin{equation}
Z_{\mathbf{S}}^{(r)}(\tau,\bar{\tau}, \bm{z}, \mathcal{B}) = \text{Tr}_{\text{\tiny RR}}F_{\text{\tiny R}}^2(-1)^{F_{\text{\tiny R}}} q^{L_0-\frac{c_{\text{\tiny L}}}{24}}\bar{q}^{\bar{L}_0 - \frac{c_{\text{\tiny R}}}{24}} e^{2\pi i z^i Q_i} \,, 
\label{eq:MSWeg}
\end{equation}
where the trace is taken over the Ramond--Ramond (RR) Hilbert space of the heterotic MSW string with wrapping number $r$ on $S^1_M$. 
Here $F_{\text{\tiny R}}$ is the fermion number in the right-moving sector, $c_{\text{\tiny L}}$ and $c_{\text{\tiny R}}$ are the central charges of the MSW string worldsheet theory, $q= e^{2\pi i \tau}$ with $\tau$ being the modular parameter associated with the torus worldsheet, $z^i$ are the fugacity parameters associated with the left-moving currents $\U(1)^i$ in the worldsheet superconformal algebra, and $Q_i\in \mathbb{Z}$.  
The left-moving  $\U(1)$ currents are of course associated precisely with the asymptotically weakly coupled $\U(1)^i$ gauge group factors. 
Moreover, $\mathcal{B}$ is a background B-field that takes values in $\Lambda_{\mathbb{R}}$ and which will play no role for us.

Let us now discuss the modular properties of the elliptic genus~\eqref{eq:MSWeg}. 
As argued in~\cite{Gaiotto:2006wm} and references therein,\footnote{These results have been derived for lattice polarized K3 fibrations whose degenerations are all irreducible. Based on experience with the six-dimensional elliptic genus \cite{Alim:2012ss,Klemm:2012sx,Huang:2015sta,Kim:2016foj,DelZotto:2016pvm,Gu:2017ccq,DelZotto:2017mee,DelZotto:2018tcj}, it is expected that the modular properties change relatively mildly if one allows for reducible special fibers. In particular our main conclusion, the existence of states satisfying \eqref{eq_nQrelation}, must still hold (as happens in the six-dimensional context \cite{Lee:2018spm,Lee:2018urn}) in presence of such special fibers, which only affect the strongly coupled sector in the limit, as explained in Section \ref{sec:WeakCouplK3lim}. With this understanding, we will continue to work in the framework of~\cite{Gaiotto:2006wm} in the sequel. We will comment more on fibrations with fibers degenerating into several components at the end of this section.} \eqref{eq:MSWeg} behaves as a non-holomorphic Jacobi form of weight $(-3/2,1/2)$ in $(\tau,\bar{\tau})$, and in our five-dimensional heterotic MSW string setting, it enjoys an expansion of the form~\cite{Gaiotto:2006wm,Bouchard:2016lfg} 
\begin{equation} \label{eq_ellgenus1}
Z_{\mathbf{S}}^{(r)}(\tau, \bar{\tau}, \bm{z}, \mathcal{B}) = \sum_{\boldsymbol{\mu} \in \Lambda^*/r \Lambda} Z_{\boldsymbol{\mu}}(\tau) \Theta_{\boldsymbol{\mu},r}^*(\tau,\bar{\tau}, \bm{z},\mathcal{B}) \,.
\end{equation}
Here 
 $\Theta_{\boldsymbol{\mu},r}^*(\tau,\bar{\tau},\bm{z},\mathcal{B})$  is the complex conjugate of the Siegel theta series associated to the coset $\boldsymbol{\mu}+\sqrt{r}\Lambda \subset \Lambda^*/\sqrt{r}$ with $\boldsymbol{\mu} \in \Lambda^*/r \Lambda$, 
and $Z_{\boldsymbol{\mu}}(\tau)$ is a vector-valued-modular form that is the generating series for the degeneracy of BPS states $\Omega(\bm{\gamma})$ with charge $\bm{\gamma} = (r,\bm{Q},n)$, 
\begin{equation} \label{eq_Zmutau}
Z_{\boldsymbol{\mu}}(\tau) = \sum_{n=0}^\infty \Omega(\bm{\gamma}) q^{n +\bm{Q}^2/2r -1} \fstop
\end{equation}
Note that the BPS indices $\Omega(\bm{\gamma})$ are conjecturally related to the Donaldson--Thomas invariants associated with the charges $\bm{\gamma}$.

The Siegel theta functions can be expressed as~\cite{Bouchard:2016lfg,MR1625724}
\begin{equation} \label{eq_Siegltheta}
\Theta_{\boldsymbol{\mu},r}^*(\tau,\bar{\tau},\bm{z},\mathcal{B}) = \sum_{\boldsymbol{\lambda} \in\boldsymbol{\mu} + \sqrt{r}\Lambda} e^{-\pi i \tau (\boldsymbol{\lambda} +\mathcal{B})_-^2 -\pi i \bar{\tau} (\boldsymbol{\lambda} +\mathcal{B})^2_+ + 2\pi i \left(\boldsymbol{\lambda} + \frac{\mathcal{B}}{2}\right)\cdot z}\,, 
\end{equation}
where we introduce the notation $\boldsymbol{\lambda}_\pm$ to denote the projection of a vector $\boldsymbol{\lambda} \in \Lambda_{\mathbb{R}}$ into the sublattice $\Lambda_\pm$ inside the orthogonal decomposition \eqref{eq:Lambdapm}. 

Let us now specialize to an anti-self-dual charge vector $\bm{Q} \in \Lambda^\ast_-$ and wrapping number $r=1$. From 
\eqref{eq_ellgenus1}, \eqref{eq_Zmutau} and \eqref{eq_Siegltheta}, we infer the existence of a state at KK number $n = -\frac{1}{2} \bm{Q}^2$ provided the associated BPS numbers 
$\Omega(\boldsymbol{\gamma})$ are non-vanishing.
The multiplicities of these states appear as the expansion coefficients of the term $q^{n-1} e^{2 \pi z^i Q_i} = q^{n -\frac{c_{\text{\tiny L}}}{24}} e^{2 \pi z^i Q_i}$ for $n = -\frac{1}{2} \bm{Q}^2$ in the expansion of the elliptic genus
\eqref{eq_ellgenus1}.\footnote{\label{foonote_exp}The contribution $q^n e^{2 \pi z^i Q_i}$ comes from \eqref{eq_Siegltheta} for $\boldsymbol{\lambda}_- = \bm{Q}$, $\boldsymbol{\lambda}_+=0$, ${\cal B}=0$, and the term $q^{-1}$ comes from \eqref{eq_Zmutau} provided $\Omega(\boldsymbol{\gamma}) \neq 0$ for $n = -\frac{1}{2} \bm{Q}^2$.}

Furthermore, by level matching, we can identify, for such anti-self-dual charges, the KK number $n$ with the left-moving excitation level $n_{\text{\tiny L}}$.
Similar to the six/five-dimensional context of \cite{Lee:2018spm,Lee:2018urn}, this establishes not only the existence of a BPS state in the four-dimensional theory, but also of a string excitation at left-moving excitation $n_{\text{\tiny L}}=n = -\frac{1}{2} \bm{Q}^2$, $\bm{Q} \in \Lambda_-^\ast$ (and suitable right-moving oscillator number such as to level match left- and right-movers). Such states are non-BPS in five dimensions.

It therefore remains to argue for the non-vanishing of the BPS index,
\bea \label{Omeganonzero}
\Omega(\bm{\gamma}) \neq 0 \quad \text{for} \quad  \bm{\gamma} = (1, \bm{Q}, n) \quad \text{at} \quad n= - \frac{\bm{Q}^2}{2}\,,  \,  \bm{Q} \in \Lambda_-^\ast \,.
\eea
For the MSW string associated with the K3-fiber of 
$\rho: X_3 \rightarrow \mathbb{P}^1$,
~\cite{Bouchard:2016lfg} has argued that the BPS indices of D4-D2-D0 bound states are determined by the Fourier coefficients of the same vector-valued-modular form 
that yields all Noether--Lefschetz numbers and Gopakumar--Vafa invariants for $\rho$-vertical curve classes in $X_3$~\cite{MR3114953,MR3508473}. 
More precisely,
the authors~\cite{Bouchard:2016lfg} proved for the $r=1$ heterotic MSW strings the expression\footnote{For $r>1$, there are generalized expressions for $Z_{\bm{\mu}}$ in terms of $\Phi_{\bm{\mu}}$~\cite{Bouchard:2016lfg}, 
but these are not relevant to us.}
\begin{equation}
Z_{\boldsymbol{\mu}}(\tau) = \eta^{-24}(\tau)  \Phi_{\boldsymbol{\mu}}(\tau)\,
\label{eq:NLBPSindex}
\end{equation}
for the generating function \eqref{eq_Zmutau}, where $\eta(\tau)$ is the Dedekind-eta function and
$\Phi_{\boldsymbol{\mu}}$ is a vector-valued-modular form component 
whose expansion coefficients are related to certain Noether--Lefschetz numbers.
Therefore we can deduce \eqref{Omeganonzero} by analyzing the coefficients of $\Phi_{\boldsymbol{\mu}}(\tau)$.
In the following, we will consider the basics of Noether--Lefschetz theory that will us to do this.\footnote{For a brief review of Noether--Lefschetz theory, we refer the reader to Section 1 in~\cite{MR3508473}.}  

Given the moduli space of $\Lambda$-polarized K3 surfaces $\mathcal{M}_\Lambda$, the authors of~\cite{MR3114953} defined the Noether--Lefschetz numbers 
as the classical intersection numbers of Noether--Lefschetz divisors with the embedding of the K3-fibration base curve into $\mathcal{M}_\Lambda$. 
These geometrical invariants $NL_{(h,\bm{Q})} \in \mathbb{Z}$ are labeled by the entries $(h,\bm{Q})$ that are defined in the lattice $\mathbb{Z}\times \Lambda^\ast$. 
Intriguingly, there is a link between modularity and Noether--Lefschetz theory that is described by a vector-valued-modular form
\begin{equation}
\Phi(q) = \sum_{\boldsymbol{\mu}} \Phi_{\boldsymbol{\mu}}(q) e_{\boldsymbol{\mu}} \in  \mathbb{C} \left[\left[q^{\frac{1}{2\Delta(\Lambda)}}\right]\right] \otimes  \mathbb{C}[\Lambda^\ast/\Lambda]\,,
\label{eq:VVMF}
\end{equation}
where $\{e_{\boldsymbol{\mu}}\}$ is a basis of the vector space $\mathbb{C}[\Lambda^\ast/\Lambda]$. 
Moreover, $\Phi$ has weight $11-\text{rk}(\Lambda)/2$ and transforms under the dual representation of the Weil representation $\rho^\ast_\Lambda$ of the metaplectic group 
$\text{Mp}(2,\mathbb{Z})$, which is the double cover of $\text{SL}(2,\mathbb{Z})$. 
The $\Phi_{\boldsymbol{\mu}}(\tau)$
are the objects appearing in \eqref{eq:NLBPSindex}.
Their expansion  coefficients are related to the Noether--Lefschetz numbers $NL_{(h,\bm{Q})}$ 
via
\begin{equation}
NL_{(h,\bm{Q})} = \text{Coeff}\left(\Phi_{\boldsymbol{\mu}}, q^{\Delta_{\text{NL}}}\right)\,, 
\label{eq:NLrel}
\end{equation}
where $\Delta_{\text{NL}} = \Delta_{(h,\bm{Q})} /2 \vert \Delta(\Lambda) \vert$ 
and $\Delta_{(h,\bm{Q})}$ is the discriminant defined via the following matrix
\begin{equation}
\Delta_{(h,\bm{Q})} = (-1)^{\text{rk}\Lambda} \det
\begin{pmatrix} 
\eta_{ij} &  Q_i \\
   Q_j & 2h-2\end{pmatrix}\,.
\end{equation}
Thus, the corresponding expression for the power coefficient in~\eqref{eq:NLrel} reads
\begin{equation}
\Delta_{\text{NL}} = \frac{1}{2}\eta^{ij}Q_i Q_j+1 - h \,.
\end{equation}
Here the relevant point is that two lattice invariants of $\mathbb{Z} \times \Lambda^*$  determine the Noether--Lefschetz numbers, 
which are the discriminant $\Delta(h,\bm{Q})$ and the coset $\boldsymbol{\mu}$ in the abelian group $(\Lambda^\ast/\Lambda)/\pm$. 
Furthermore, we have the Noether--Lefschetz constraints~\cite{MR3114953,MR3508473}:
\begin{itemize}
\item If $\Delta(h,\bm{Q}) < 0$, then $NL_{(h,\bm{Q})}=0$. 
\item If $\Delta(h,\bm{Q})  = 0$, then $NL_{(h,\bm{Q})} =-2$.  
\item If  $\Delta(h,\bm{Q}) > 0$, then $NL_{(h,\bm{Q})} \in \mathbb{Z}$.  
\end{itemize}

Let us come back to the BPS index $\Omega(\bm{\gamma})$ counting for $\bm{\gamma} = (1,\bm{Q},n)$ and the relation
\eqref{eq:NLBPSindex}.
We need to show that 
$\Phi_{\boldsymbol{\mu}}(\tau)$ has a non-vanishing coefficient at order $q^0$, because combined with the expansion
$\eta^{-24}(\tau) = q^{-1} + 24 + {\cal O}(q)$ this gives a non-vanishing coefficient for $Z_{\boldsymbol{\mu}}(\tau)$ at order $q^{-1}$ (see Footnote \ref{foonote_exp}).
Now, the constant contribution at order $q^0$ in $\Phi$ 
is associated to the vanishing of the discriminant $\Delta(h,\bm{Q})=0$, which simply
relates $2h-2 = \bm{Q}^2$ and so $\bm{Q}^2 \in 2\mathbb{Z}$. 
Therefore, $\bm{Q}$ is a representative of the trivial coset $\mathbf{0} \in \Lambda^*/\Lambda$, which is always present in  $\Lambda^*/\Lambda$. 
Indeed, we are interested in states with $-2n = \bm{Q}^2$, which are clearly of this form.
Furthermore the coefficient
at $q^0$ is non-vanishing because it is given by 
$NL_{(h,\bm{Q})} =-2$.

Expanding the component of \eqref{eq:NLBPSindex} associated with the latter class, 
we find that
\begin{equation}
Z_{\mathbf{0}}(\tau)  =  \eta^{-24}(\tau)  \Phi_{\mathbf{0}} (\tau) = \left[ q^{-1} + 24  + \mathcal{O}(q)   \right] \left[-2 + \mathcal{O}(q) \right] = -2 q^{-1} + \mathcal{O}(q^0)\,.
\end{equation}
Hence, the BPS index $\Omega(\bm{\gamma})$ for the set of charges $\bm{\gamma}$ with $n = -\frac{\bm{Q}^2}{2}$ is always non-trivial.

As an illustrative example, consider the well-known $STU$ model in which the generic K3-fiber admits a compatible elliptic fibration over a base $\mathbb P^1_f$. The polarization lattice  $\Lambda$ is a copy of the hyperbolic lattice $U$ and is self-dual, $\Lambda = \Lambda^\ast$.
This model has been discussed from the perspective of the Emergent String Conjecture and the appearance of towers of asymptotically light states in \cite{Lee:2019wij}, to which we refer for more details.
The dual heterotic string is compactified on K3$_\het \times S^1_\het$ with a gauge bundle that breaks $E_8 \times E_8$ completely. 
The polarization lattice can be identified with the lattice of curves $l C_U + k C_T$, where 
\bea \label{CUCTdef}
C_U = \mathbb S + T^2 \coma C_T = T^2
\eea
with $T^2$ denoting the generic elliptic fiber of the K3 and ${\mathbb S}$ the curve associated with the holomorphic section of the elliptic K3. These curves indeed generate the hyperbolic lattice $U$ because they satisfy 
\bea \label{CUint}
C_U \cdot_{\text{\tiny K3}} C_U = 0 = C_T \cdot_{\text{\tiny K3}} C_T \coma  C_U \cdot_{\text{\tiny K3}} C_T =1 \,.
\eea

States with charge vector $\bm{Q} = (l,k)$ in this lattice correspond to states with KK momentum number $n_{\KK} = k$ and winding number $w$ given by $w= l$ along $S^1_\het$.

Of particular interest are the rays generated by charge vectors of negative self-intersection, along which no BPS towers exist.
Since in this section we are focusing on directions along the anti-self-dual part of the lattice, let 
us therefore consider  the direction $k=-l$, corresponding to the charge vector $\bm{Q} = l (1,-1)$. 
Such states indeed lie along $\Lambda^\ast_-$ and hence fall into the class of states analyzed in this section.
BPS states of this type exist only for $l=1$ and correspond to M2-branes wrapped once on the curve $\mathbb P^1_f$ in M-theory.
For these states, the Gopakumar--Vafa index is $N^0_{\mathbb P^1_f} = -2$ (which agrees with the Euler characteristic of the base of the K3-fibration, i.e., of the moduli space of $\mathbb P^1_f$ inside $X_3$), while $N^0_{l \mathbb P^1_f} =0$ for $l > 1$. 
The latter assertions can be verified through Noether--Lefschetz Theory as follows. 
Since $\Lambda^\ast/\Lambda \simeq\{\boldsymbol{0}\}$, we only require to fix a single modular form of weight 10 in \eqref{eq:VVMF}.
Using that the Noether--Lefschetz number $NL_{(h,l,k)}= -2$ for $\Delta_{(h,l,k)}=0$, we obtain~\cite{MR2669707}
\begin{equation}
\Phi_{\boldsymbol{0}} = -2 E_4 E_6\,,
\end{equation}
where $E_k$ is the $k$-th Eisenstein series. The Gromov--Witten/Noether--Lefschetz correspondence theorem states that  
\begin{equation}
N_{l C_U + k C_T}^0 = \sum_{h \geq0} r_h^0 NL_{(h,l,k)}\,,
\end{equation} 
where $r^0_h$ are the coefficients of $\eta^{24}(\tau)$ at level $h-1$~\cite{MR3508473}. 
Indeed, this gives $N^0_{l \mathbb P^1_f} = 0$ for $l > 1$ because 
$NL_{(h,l,k)}=0$ for $\Delta_{(h,l,k)} < 0$.

On the other hand, the heterotic string does admit a whole tower of (massive) non-BPS excitations for arbitrary $k=-l$.
If we focus on states with vanishing excitation number in the right-moving sector, 
the level matching condition for the string with KK number $n_{\KK}$ and winding number $w$ along $S^1_\het$ gives
\bea \label{levelmatching-STU}
n_{\text{\tiny L}} =  - n_{\KK} \, w  - \delta_{\text{I}} = - k l - \delta_{\text{I}}  \,,
\eea
where $n_{\text{\tiny L}}$ is the left-moving excitation level and
$\delta_{\text{I}}$ represents the (in general unknown) contribution from the dimensions compactified along K3$_\het$ together with the sector charged under the $E_8 \times E_8$ lattice.
States with $k=-l$ and carrying no $E_8 \times E_8$ charge therefore satisfy the level-matching condition if their excitation level is $n_{\text{\tiny L}}=l^2 = -\frac{1}{2} \bm{Q}^2$ provided $\delta_{\text{I}}=0$. 
These states are therefore special examples of states of the form \eqref{eq_nQrelation}.

So far, we have considered the elliptic genera of heterotic MSW strings arising from K3 fibrations with degenerations of Type I in the Kulikov classification. 
To our knowledge, there is no well-established version of Noether--Lefschetz theory for the cases when the  K3 fiber in $\rho : X_3 \rightarrow \mathbb{P}^1$ 
splits into several components over the degeneration points. 
In fact, the elliptic genus $Z_{\mathbf{S}}^{(r)}$ of a general MSW string wrapping a reducible divisor $\mathbf{S}\subset X_3$  
behaves as a mock modular form~\cite{Alexandrov:2016tnf}.
For instance, in the case $r=1$ the holomorphic vector-valued modular form $Z_{\boldsymbol{\mu}}(\tau)$ is replaced by
\begin{equation}
\widehat{Z}_{\boldsymbol{\mu}}(\tau, \bar{\tau}) = Z_{\boldsymbol{\mu}}(\tau) - \frac{1}{2} R_{\boldsymbol{\mu}}(\tau,\bar{\tau})\,, 
\label{eq:nonHolEG}
\end{equation}
where $R_{\boldsymbol{\mu}}(\tau,\bar{\tau})$ is a non-holomorphic function of $\tau$ and $\bar{\tau}$ that can be constructed
 in terms of the elliptic genera associated to the irreducible components of $\mathbf{S}$.\footnote{For an explicit description of $R_{\boldsymbol{\mu}}(\tau,\bar{\tau})$, see reference~\cite{Alexandrov:2016tnf}.}
 The upshot is that $Z_{\boldsymbol{\mu}}$ is not modular, while $\widehat{Z}_{\boldsymbol{\mu}}$ transforms as a vector-valued modular form, at the expense of non-holomorphicity introduced by 
 a suitable $R_{\boldsymbol{\mu}}$. 
 Moreover, for $r>1$,  the Siegel theta functions are replaced by mock Siegel theta functions, which we will not discuss here further. 

To illustrate the mock modularity induced by splitting K3 fibers, we will discuss in Section~\ref{sec:examples} the special case when $X_3$ also admits an elliptic fibration. 
There we can extract the holomorphic piece from \eqref{eq:nonHolEG} 
via the genus zero free energy counting BPS states of curves $r \beta + n \mathcal{E}$, where $\beta$ is a base curve in $X_3$ and $\mathcal{E}$ the elliptic fiber of $X_3$.
 Upon $T^2$-duality, such BPS states map into bound states of D4-branes wrapping $r$-times the elliptic surface over $\beta$ with D0-brane charge $n$~\cite{Klemm:2012sx,Alim:2010cf}. 

\subsubsection{Super-Extremal States with \texorpdfstring{$\bm{Q} _+ \neq 0$}{}}
\label{sec:nonBPS-moregeneral}

Up to this point, we have focused on charge vectors $\bm{Q} \in \Lambda_-^*$.\footnote{Recall that $\bm{Q} \in \Lambda_+^*$ is covered by the existence of BPS towers in the K3 fiber.} However,  $\bm{Q}$ can also receive contributions from both $\Lambda_+^*$ and $\Lambda_-^*$. As long as $\bm{Q}^2>0$ the existence of a tower of BPS states along the ray defined by $\bm{Q}$ is ensured by non-vanishing BPS invariants for the curve $\Sigma$ satisfying $\Sigma\cdot_{\text{\tiny K3}} \Sigma>0$. The interesting case is, thus, again $\bm{Q}^2<0$. When $\bm{Q}$ has a contribution in $\Lambda_+^*$, we cannot simply apply the argument of the previous section to argue for a state with left-moving excitation level $n_{\text{\tiny L}} = -\frac12 \bm{Q}^2$. To see this, we note that $n$ appearing, e.g., in \eqref{Omeganonzero} can only be identified as $n_{\text{\tiny L}}$ if $\bm{Q}_+=0$ but in general has to be identified with 
\begin{equation}
    n = n_{\text{\tiny L}} -n_{\text{\tiny R}}\,,
\end{equation}
since by \eqref{eq_ellgenus1} $\frac{1}{2} \bm{Q}_+^2$ corresponds to the right-moving excitation level $n_{\text{\tiny R}}$. Thus, even though Noether--Lefschetz theory still ensures the existence of a state with $n=-\frac{1}{2} \bm{Q}^2$, this does not guarantee the existence of a state with $n_{\text{\tiny L}} =-\frac{1}{2} \bm{Q}^2 $ if $\bm{Q}_+\neq 0$. The proof of the self-repulsiveness condition in the Section \ref{sec:tWGCfornonBPS}, however, requires $n_{\text{\tiny L}} =-\frac{1}{2} \bm{Q}^2 $ rather than merely 
$n_{\text{\tiny L}} - n_{\text{\tiny R}} = -\frac{1}{2} \bm{Q}^2$.

Nonetheless we can argue for the existence of a state with  $n_{\text{\tiny L}} =-\frac{1}{2} \bm{Q}^2 $ in the excitation spectrum of the heterotic string as follows.
In the previous section, we invoked the duality between M-theory on $X_3 \times S^1_M$ and Type IIA string theory on $X_3$ to deduce the existence of suitable excitations of the heterotic string in 5d from the existence of BPS states in 4d.
Instead, consider now the duality between 
M-theory on the K3-fibered Calabi--Yau 3-fold $X_3$ and the heterotic string compactified on K3$_\het \times S^1_\het$.
This way we can identify wrapped M2-brane states in 5d with 6d heterotic string states with winding number along the circle $S^1_\het$.
This works most easily if 
the K3-fibration admits a compatible elliptic fibration. Let us for now assume that this is the case; accordingly, the generic K3-fiber allows for a fibration with generic fiber $T^2$ and holomorphic section $\mathbb S$ over the base $\mathbb{P}^1_f$.
K3-fibrations  with only a genus-one fibration with $N$-section or without a compatible torus fibration at all will be discussed at the end of this section.

There are now two types of charge vectors with $\bm{Q}_+ \neq 0$ to consider.
For the first type of states, $\bm{Q}_+$ receives a contribution from a non-zero winding number $w$ of a heterotic string on $S^1_\het$.
More precisely,
a heterotic string state in the ground state sector with winding number $w \neq 0$, KK number $n_{\KK}$ and charge $\bm{Q}_{\text{\tiny 6d}}$ under the six-dimensional gauge group maps to a state obtained from an M2-brane wrapped along the curve
\bea\label{eq_Qcurve}
\bm{Q} = w C_U + n_{\KK} C_T + \bm{Q}_{\text{\tiny 6d}} = \frac{(w+n_{\KK})}{2} C_+ +\frac{(w-n_{\KK})}{2} C_- + \bm{Q}_{\text{\tiny 6d}} \,.
\eea
Here $C_U = {\mathbb S} + T^2$ and $C_T = T^2$ defined as in \eqref{CUCTdef} generate a copy of the hyperbolic lattice $U$ and we have rewritten $\bm{Q}$ also in terms of the (anti-)self-dual combinations 
\bea
C_{\pm} = (C_U \pm T^2) \in \Lambda^\ast_{\pm} \,,
\eea
with $C_\pm \cdot_{{\text{\tiny K3}} } C_\pm = \pm 2$ and $C_+  \cdot_{{\text{\tiny K3}} } C_-=0$.
From \eqref{CUint} it follows that $\bm{Q}^2 =  2 n_{\KK}\, w + \bm{Q}_{\text{\tiny 6d}}^2$. 
For charges with $\bm{Q}^2 < 0$ the existence of a BPS state along the curve $\bm{Q}$ is a priori not guaranteed. In particular, there may not exist a tower of BPS states along the ray with charge $\bm{Q}$.
The remedy is to consider instead a state with the given KK and winding number at a suitable excitation level $n_{\text{\tiny L}} \neq 0$. More precisely, under the duality of M-theory on $X_3$ and the heterotic string on K3$_\het \times S^1_\het$,
a heterotic string state with winding number $w$, KK number $n_{\KK}$ and charge $Q_{\text{\tiny 6d}}$ at left-moving excitation level $n_{\text{\tiny L}}$ maps to an M2-brane state associated with the curve
\bea \label{Sigmadefa}
\Sigma = \bm{Q} + \frac{n_{\text{\tiny L}}}{w} C_T =  w C_U + (n_{\KK} + \frac{n_{\text{\tiny L}}}{w}) C_T + \bm{Q}_{\text{\tiny 6d}} \,,
\eea
whenever $\frac{n_{\text{\tiny L}}}{w}$ is integral and for $w \neq 0$. The smallest value of $n_{\text{\tiny L}}$ for which such a state is guaranteed to exist is obtained for $\Sigma^2 =0$, i.e.
\bea\label{eq_nL}
n_{\text{\tiny L}} =  -  n_{\KK}\, w - \frac{1}{2} \bm{Q}^2_{\text{\tiny 6d}} = - \frac{1}{2} \bm{Q}^2 \,.
\eea
This uses that curves with $\Sigma^2 =0$ inside the K3 fiber have non-vanishing BPS invariants, which in turn follows from the Noether--Lefschetz argument presented in the previous section.\footnote{The Gromov--Witten/Noether--Lefschetz correspondence theorem states that all curve classes with same intersection form and discriminant class in $\Lambda^*$ have the same associated  Gopakumar--Vafa invariant value~\cite{MR3508473}. This implies that $N_\Sigma^0 =N_{C_T}^0=-\chi(X_3)$, since $C_T^2 =0$, where  $\chi(X_3)$ is generically non-vanishing. In fact, the refined pairs/Noether--Lefschetz correspondence proposed in \cite{MR3524167} predicts the appearance of non-trivial BPS states of the form $(-\chi(X_3)+8)[0,0]+[1/2,0]+[0,1/2]$, where $[j_{\text{\tiny L}},j_{\text{\tiny R}}]$ are representations of  $\SU(2)_{\text{\tiny L}}\times \SU(2)_{\text{\tiny R}}$, which is the little group of the Lorentz group in five dimensions.}
Note that \eqref{eq_nL} is consistent with the level matching-condition \eqref{levelmatching-STU} for the heterotic string on K3$_\het \times S^1_\het$. However, the ansatz \eqref{Sigmadefa} requires $n_{\text{\tiny L}}$ being divisible by $w$. If this is not the case for the solution \eqref{eq_nL}, we can suitably rescale the charge as $\widetilde{\bm{Q}} = w \bm{Q}$ and consider 
$\Sigma = \widetilde{\bm{Q}} + \frac{n_{\text{\tiny L}}}{w^2} C_T$ such that $\Sigma^2 =0$. This gives
\bea \label{nLproperty}
n_{\text{\tiny L}} = - \frac{1}{2} \widetilde{\bm{Q}}^2 \,,
\eea
which is divisible by $w$. For the purpose of showing the existence of a super-extremal tower along the ray $\bm{Q}$, this rescaling from $\bm{Q}$ to $\widetilde{\bm{Q}} = w \bm{Q}$ is inessential.
As we will see in the next section, such states are marginally super-extremal. 

The second type of states have $\bm{Q}_+ \neq 0$, but $w=0$, i.e., the charge with respect to $C_+$ and $C_-$ are equal and opposite.
For vanishing winding number, we must consider instead 
excitations of the six-dimensional heterotic string
with KK number $n_{\KK}$. As is clear from the level-matching condition \eqref{eq_nL}, for such states the left-moving excitation number is unaffected by $n_{\KK}$. We can therefore simply ask for which (minimal) excitation level, a state with charge $\bm{Q}_{\text{\tiny 6d}}$ and vanishing charge along $C_+$ and $C_-$ exists. The answer is the same as for the string directly defined in five dimensions that was deduced in the previous so that once more a state with the characteristic property \eqref{nLproperty} is guaranteed to exist.

If the K3-fibration only admits a compatible genus-one, rather than an elliptic, fibration, the class ${\mathbb S}$
in the definition of $C_U$ entering \eqref{eq_Qcurve} is a $N$-section so that $C_U \cdot_{{\text{\tiny K3}}} C_T = N$ and   $C_U^2 \neq 0$ in general. The heterotic dual involves a circle compactification with Wilson line for the KK $\U(1)$, but this does not affect the conclusions of the discussion in the elliptic case.

It remains to address K3-fibrations whose generic K3-fiber is not itself elliptically or genus-one fibered.
Our analysis carries over whenever the geometry is related through a suitable transition to a K3-fibration admitting a compatible genus-one fibration.
From the heterotic perspective, such a transition means that we start with a 6d heterotic string compactified on an additional $S^1_{\het}$. We then go to a special point in the K\"ahler (i.e., vector multiplet) moduli space that leads to a non-perturbative enhancement involving a combination of the KK and winding $\U(1)$; the transition corresponds to turning on  a non-trivial rank-changing gauge background which only leaves one combination of the KK and winding $\U(1)$s intact \cite{Kachru:1995wm}.
For example, in the model with elliptic fibration of the type described around \eqref{eq_Qcurve} such a special point is the self-dual radius of $S^1_{\het}$ and we break the enhanced $\SU(2)$ gauge group through a choice of non-abelian gauge bundle. In the notation introduced in \eqref{eq_Qcurve}, this corresponds, in the M-theory picture, to shrinking the curve $C_-$ followed by a complex structure transition that higgses the enhanced $\SU(2)$ gauge group. The unbroken $\U(1)$ in this case is associated to $C_+$. 
Any state not charged under $\U(1)_-$ is unaffected by the transition. These are
the states with $n_{\KK} = w$ in the parent theory.
The existence of such states with $n_{\text{\tiny L}} = - \frac{1}{2} {\bm{Q}}^2$ even after the transition is therefore a consequence of the analysis
around \eqref{eq_nL} applied to states with $n_{\KK} = w$. This reasoning generalizes whenever a parent theory is available and allows us to deduce states of the form $n_{\text{\tiny L}} = - \frac{1}{2} {\bm{Q}}^2$ and with $\bm{Q}_+ \neq 0$ even in absence of a compatible genus-one fibration. In view of heterotic/M-theory duality, we find it plausible that such a parent theory indeed always exists, but we leave this as a question for further investigations.

\subsubsection{Tower WGC for K3 Limits}
\label{sec:tWGCfornonBPS}

We now show that the tower of non-BPS states satisfying the relation
\eqref{eq_nQrelation} realize the asymptotic Repulsive Force Conjecture.

Consider the excitations of the heterotic string at \textit{left-moving} excitation level $n_{\text{\tiny L}}$ carrying charges $Q_i$ within the charge lattice $\Lambda^\ast$.
In the weak coupling limit of the heterotic string, the mass is given by 
\begin{equation}\label{eq:masshetK3} 
    M_{n_{\text{\tiny L}}, \bm{Q}}^2 = 8\pi (n_{\text{\tiny L}}-a)T_s + \Delta_{\text{\tiny CB}} = 16\pi^2(4\pi)^{-2/3}\left((n_{\text{\tiny L}}-a)\widehat{\mathcal{V}}_{\bf{S}}+\frac{1}{4}Q_i Q_j \hat{v}^i \hat{v}^j\right) M_\Pl^2 
\end{equation}
with $a=1$ the zero-point energy of the critical heterotic string.
This mass formula has two contributions:
The first contribution manifestly depends on the excitation level along the worldsheet and is proportional to the tension of the heterotic string,
\begin{equation}
    T_s = 2\pi \mathcal{V}_{\bf{S}}M_{\text{\tiny 11d}}^2 = 2\pi (4\pi)^{-2/3}\widehat{\mathcal{V}}_{\bf{S}} M_\Pl^2\fstop
\end{equation}
Here $\widehat{\cal V}_{\bf S}$ denotes the volume, rescaled as in \eqref{rescaledvolumes}, of the generic K3-fiber ${\bf S}$.
The second contribution, $\Delta_{\text{\tiny CB}}$, represents a moduli dependent shift which can be interpreted as a universal, i.e., $n_{\text{\tiny L}}$ independent, mass contribution along the five-dimensional Coulomb 
branch. As such, it must be visible already at the level of the supergravity modes, which correspond to the modes at the massless level $n_{\text{\tiny L}}=a$. 

To gain some intuition for the precise form of $\Delta_{\text{\tiny CB}}$, let us specialize for the moment to a situation in which $X_3$ admits a torus fibration compatible with the K3-fibration $\rho$. In this case, the five-dimensional effective theory can be thought of as the circle compactification of a six-dimensional theory given by F-theory on the base $B_2$ of the torus fibration. The existence of the K3-fibration implies that $B_2$ must be rationally fibered, and a D3-brane along the rational fiber defines the uplift of the five-dimensional heterotic string to six dimensions. 
Its excitations at level $n_{\text{\tiny L}}=a$ give rise to the massless supergravity spectrum in six dimensions. 
After circle compactification, these have a dual description in terms of M2-branes wrapping fibral curves whose volumes can be interpreted as Coulomb branch parameters of the gauge theories. The mass of an M2-brane on such a curve $\mathcal{C}^i$ is given by 
\begin{equation}
    M_{\text{\tiny M2}} = 2\pi Q_i v^i M_{\text{\tiny 11d}}\,,
\end{equation}
such that 
\begin{equation}\label{eq:M2branmass}
    \frac{M_{\text{\tiny M2}}^2}{M_{\Pl}^2} = 4\pi^2 (4\pi)^{-2/3} Q_i Q_j \hat{v}^i \hat{v}^j\,. 
\end{equation}
We can identify this with the expression of the mass of the five-dimensional string excitations at level $n_{\text{\tiny L}}=a$,
\begin{equation}
   \frac{M_{a, \bm{Q}}^2}{M_{\Pl}^2} = 4 \pi^2 (4\pi)^{-2/3} Q_i Q_j \hat{v}^i \hat{v}^j\,.
\end{equation}
Extrapolating to the full excitation tower explains the precise value for the Coulomb branch shift in the final mass formula \eqref{eq:masshetK3}, which is valid for general K3-fibrations.

Let us now analyze in more detail
the special subset of states for which
\begin{equation} \label{eq:qvsm2}
- \frac{1}{2} Q_i \eta^{ij} Q_j = n_{\text{\tiny L}} 
\end{equation}
introduced in \eqref{eq_nQrelation}.
The states with this property have the highest charge-to-mass ratio per excitation level and are therefore candidates to satisfy the asymptotic WGC or RFC.
To verify that these states indeed satisfy  the RFC, we evaluate both sides of \eqref{eq:5dRFC-b}, taking $n_{\text{\tiny L}}\gg a$ so that we can neglect the shift due to the vacuum energy. We can start by defining $\varpi = 16\pi^2(4\pi)^{-2/3}$ and rewriting \eqref{eq:masshetK3} as
\begin{equation} \label{eq:massK3-b}
       \frac{M_{n_{\text{\tiny L}}, \bm{Q}}^2}{M_\Pl^2} = \varpi n_{\text{\tiny L}}\left(\widehat{\mathcal{V}}_{\text{\tiny K3}}+\frac{1}{4n_{\text{\tiny L}}}Q_i Q_j \hat{v}^i \hat{v}^j\right)\fstop
\end{equation}

To evaluate the right-hand side (RHS) of \eqref{eq:5dRFC-b}, we first compute the inverse of the gauge kinetic matrix, $f^{\alpha \beta}$, in the weak coupling limit \eqref{JscalingK3a}. According to \eqref{finv}, we need the 
inverse of $\mathcal{V}_{\alpha\beta}$ evaluated in \eqref{eq:volumes-K3fibr}. In the limit, it takes the form 
\begin{equation}
    \mathcal{V}^{\alpha\beta} \simeq \frac{1}{{v}^0\eta_{kl}{v}^k{v}^l }\left(\begin{array}{cc} -({v}^0)^2 & {v}^i {v}^0 \\ {v}^j {v}^0 & (\eta_{kl}{v}^k{v}^l) \eta^{ij}-{v}^i{v}^j \end{array}\right) = \frac{1}{2}\frac{1}{\mathcal{V}^{1/3}}\left(\begin{array}{cc} -({\hat v}^0)^2 & {\hat v}^i {\hat v}^0 \\ {\hat v}^j {\hat v}^0 & 2\frac{\widehat{\mathcal{V}}}{{\hat v}^0} \eta^{ij}-{\hat v}^i{\hat v}^j \end{array}\right)\coma
\end{equation}
where $\eta^{ij}$ is the inverse of $\eta_{ij}$ defined in \eqref{etadef}. In the weak coupling limit,  $\mathcal{V} \simeq v^0\mathcal{V}_{\Kt} = \frac{1}{2}v^0\eta_{kl}v^kv^l$ so that we can write the gauge couplings matrix as
\begin{equation}\label{eq:invfabK3}
     f^{\alpha\beta} = \mathcal{V}^{1/3}\left(\frac{1}{2}\frac{v^\alpha v^\beta }{\mathcal{V}}-\mathcal{V}^{\alpha\beta}\right) \simeq  \frac{1}{\mathcal{V}^{2/3}}\left(\begin{array}{cc} (v^0)^2 & 0 \\ 0 & v^i v^j-\frac{\mathcal{V}}{v^0} \eta^{ij} \end{array}\right)=  \begin{pmatrix} \left(\hat v^0\right)^2 & 0 \\ 0 & \hat{v}^i \hat {v}^j - \frac{\eta^{ij}}{\hat{v}^0}\end{pmatrix}\coma
\end{equation}
using the rescaled coordinates in \eqref{eq:rescaledcoord}.

With these expressions at hand, and using the mass formula \eqref{eq:massK3-b}, the RHS of \eqref{eq:5dRFC-b} evaluates to
\begin{equation}
\text{RHS} = \frac{2}{3} +\frac{4}{3} - \frac{\varpi n_{\text{\tiny L}}}{\hat{v}^0}\frac{M_\Pl^2}{M_{n_{\text{\tiny L}}, \bm{Q}}^2}-\frac{\varpi^2 n_{\text{\tiny L}}^2}{\hat{v}^0}\frac{M_\Pl^4}{M_{n_{\text{\tiny L}}, \bm{Q}}^4}\frac{1}{4n_{\text{\tiny L}}}Q_i Q_j \hat{v}^i\hat{v}^j\left(1+\frac{1}{2n_{\text{\tiny L}}}\left(Q_k \eta^{kl} Q_l\right)\right)\fstop
    \label{eq:RHScont}
\end{equation}
We can rewrite the second contribution to the RHS of \eqref{eq:RHScont} using $\widehat{\mathcal{V}} \simeq \hat{v}^0\widehat{\mathcal{V}}_{\Kt}$ so that 
\begin{equation}
    \frac{\varpi n_{\text{\tiny L}}}{\hat{v}^0}\frac{M_\Pl^2}{M_{n_{\text{\tiny L}}, \bm{Q}}^2} \simeq \frac{\widehat{\mathcal{V}}_{\Kt}}{\widehat{\mathcal{V}}_{\Kt}+\frac{1}{4n_{\text{\tiny L}}}Q_i Q_j \hat{v}^i\hat{v}^j} = 1 - \frac{\frac{1}{4n_{\text{\tiny L}}}Q_i Q_j \hat{v}^i\hat{v}^j}{\widehat{\mathcal{V}}_{\Kt}+\frac{1}{4n_{\text{\tiny L}}}Q_i Q_j \hat{v}^i\hat{v}^j}\,.
\end{equation}
Eventually, this gives
\begin{equation}
    \text{RHS} \simeq 1+ \frac{\frac{1}{4n_{\text{\tiny L}}}Q_i Q_j \hat{v}^i\hat{v}^j}{\widehat{\mathcal{V}}_{\Kt}+\frac{1}{4n_{\text{\tiny L}}}Q_i Q_j \hat{v}^i\hat{v}^j}-\frac{\varpi^2 n_{\text{\tiny L}}^2}{\hat{v}^0}\frac{M_\Pl^4}{M_{n_{\text{\tiny L}}, \bm{Q}}^4}\frac{1}{4n_{\text{\tiny L}}}Q_i Q_j \hat{v}^i\hat{v}^j\left(1+\frac{1}{2n_{\text{\tiny L}}}\left(Q_k \eta^{kl} Q_l\right)\right)\fstop
    \label{eq:RHScont-final}
\end{equation}
Here and in the sequel, we use the symbol $\simeq$ to indicate that we have used the relation $\mathcal{V} \simeq v^0\mathcal{V}_{\Kt} = \frac{1}{2}v^0\eta_{kl}v^kv^l$, which holds asymptotically in the weak coupling limit.

We are interested in the RFC for the candidate states whose charges $Q_i$ satisfy the relation \eqref{eq:qvsm2}.
For such states,
the terms in brackets cancel
and 
\begin{equation}
   \left. \text{RHS}\right|_{\eqref{eq:qvsm2}} \simeq 1+ \frac{\frac{1}{4n_{\text{\tiny L}}}Q_i Q_j \hat{v}^i\hat{v}^j}{\widehat{\mathcal{V}}_{\Kt}+\frac{1}{4n_{\text{\tiny L}}}Q_i Q_j \hat{v}^i\hat{v}^j} \,.
    \label{eq:RHScont-final-2}
\end{equation}
Let us now compute the left-hand side (LHS) of \eqref{eq:5dRFC-b}, again for the states 
\eqref{eq:qvsm2}: 

\begin{equation}
\begin{split}
\text{LHS}|_{\eqref{eq:qvsm2}}
        &
        \simeq 1 + \frac{\frac{1}{4n_{\text{\tiny L}}}Q_i Q_j \hat{v}^i\hat{v}^j}{\widehat{\mathcal{V}}_{\Kt}+\frac{1}{4n_{\text{\tiny L}}}Q_i Q_j \hat{v}^i\hat{v}^j} \,.
    \end{split}
\end{equation}

It is then clear that the LHS and the RHS of \eqref{eq:5dRFC-b} match as expected, in the asymptotic weak coupling limit and for
$n_{\text{\tiny L}} \gg a$. 
Reinstating the vacuum shift $a$, a lengthy but straightforward computation shows  that the LHS is bounded below by the RHS. More concretely, one finds
\begin{equation} 
\left.\text{RHS}\right|_{\eqref{eq:qvsm2}} \simeq  \text{LHS}|_{\eqref{eq:qvsm2}}-\frac{\widehat{\mathcal{V}}_{\Kt}^2}{ \left(\frac{1}{4 n_{\text{\tiny L}}}Q_i Q_j \hat{v}^i\hat{v}^j+\widehat{\mathcal{V}}_{\Kt}\right)^2}\frac{a}{n_{\text{\tiny L}}} + \mathcal{O}\left(\left(\frac{a}{n_{\text{\tiny L}}}\right)^2\right)\coma
\end{equation}
hence proving that the string excitations with property
\eqref{eq:qvsm2} indeed satisfy the asymptotic tower RFC.

Taking into account also the BPS towers along all rays in the charge lattice $\Lambda^\ast$ with $\bm{Q}^2 \geq 0$, 
the preceding analysis establishes the existence of asymptotically self-repulsive towers along {\it all} directions in the charge lattice $\Lambda^\ast$ associated with asymptotically weakly coupled gauge group factors.
We recall that this lattice is of rank $(1,k_{\rm 0+ \rm I})$ for $k_{\rm 0+ \rm I} \leq 19$  and includes a sublattice of $E_8 \oplus E_8$, the heterotic gauge group present already in ten dimensions. 
The additional generators are associated with suitable unbroken linear combinations of KK and winding number $\U(1)$s of the dual five-dimensional heterotic string theory, and the self-repulsive towers 
identified above include states charged under these gauge groups.

\subsection{(Non-)BPS Towers in Type \texorpdfstring{$T^4$}{} Limits}
\label{sec:tWGClimT4}

We now briefly turn to the asymptotic RFC for limits of Type $T^4$.
We recall from Section \ref{sec:WeakCouplT4lim} that the gauge group factors with a weak coupling limit are encoded in the polarization lattice $\Lambda$ of the fibration. States charged under such $\U(1)$s are hence associated with the directions in the dual lattice $\Lambda^\ast$ of curve classes within the generic abelian surface fiber. 

Consider first charge vectors in the self-dual part of the lattice, or those of vanishing norm. 
The associated rays in the charge lattice
support BPS towers from (multi-)wrapped M2-branes, which automatically satisfy the asymptotic RFC.
As for K3 fibrations, this matches the conclusions of  \cite{Alim:2021vhs} because the associated curves are in the movable cone of $X_3$.
Note that such BPS states exist even though, in many instances, the genus-zero BPS numbers may vanish due to an underlying higher supersymmetry, as discussed in this present context in \cite{Lee:2019wij}; in this case the vanishing of the genus-zero index merely reflects the Bose-Fermi degeneracy resulting from the higher supersymmetry.

Of particular interest 
are therefore the remaining, anti-self-dual charge vectors. As stressed in Section \ref{sec:WeakCouplT4lim}, such directions do not support holomorphic curve classes in the generic abelian surface fiber.
The duality frame set by the M5-brane along the surface fiber defines a five-dimensional Type II compactification on $S^1 \times Z$ with $Z$ an in general non-geometric background \cite{Lee:2019wij}. 
The charges associated with the weakly coupled gauge group factors can be interpreted as Kaluza--Klein or winding numbers of this dual Type II compactification. 
By complete analogy with the discussion around \eqref{levelmatching-STU},
it is therefore clear that string excitations
at excitation level $n= - \frac{1}{2} \bm{Q}^2$ exist. This distinguished set of non-BPS states again furnishes a tower of asymptotically self-repulsive states.

In the spirit of Section \ref{sec:ellipticgenera5d}, explicitly verifying the existence of this tower in the geometry requires computing a possibly refined version of the D4-D2-D0 Donaldson--Thomas invariants.\footnote{For example, \cite{Oberdieck:2017pqm} computes the relative GV invariants for Schoen manifolds.} We leave this interesting direction as a task for the future.

\section{Example}
\label{sec:examples}

We illustrate the possible weak coupling limits and their associated super-extremal towers by means of a Calabi--Yau 3-fold $X_3$ which admits both a K3-fibration
${\rho}: X_3 \rightarrow \mathbb{P}^1$ and
a compatible elliptic fibration $\pi: X_3 \to B_2$.
The elliptic fibration is constructed as a generic Weierstrass model over the base 
$B_2 =\text{Bl}(\mathbb{F}_2)$, the blowup of the Hirzebruch surface $\mathbb{F}_2$ in one point.
Since $B_2$ is rationally fibered, $X_3$ admits also a compatible K3-fibration.

The resulting Calabi--Yau $X_3$ can be described torically via the following data:
\begin{align}
\begin{blockarray}{crrrrrrrrl}
	&&&&&\mathcal{C}^0&\mathcal{C}^1 & \mathcal{C}^2 & \mathcal{C}^3\\
\begin{block}{c(rrrr|rrrr)l}
         D_1& -2 &  -3 &  -1 & -2 & 1 & -1 &0& 1   \\ 
         D_2& -2&  -3 &  -1 & 1  &0  &1 & 0 & -1 \\ 
         D_3& -2&  -3 &  0 & -1 &-2 &1 &0 & 0  \\
	D_4& -2&  -3 &  0&  1 &0 &0& 0 & 1  \\
	D_5& -2 & -3& 1 & 0  & 1&0 & 0 & 0  \\
        D_6&   1&   0& 0  & 0& 0&0 &  2& 0  \\
          D_7& 0&  1 & 0 & 0 &0 &0& 3 & 0 \\
          	D_8& -2 & -3& 0 & 0 & 0&-1 & 1 & -1 \\
\end{block}
\end{blockarray}\,.
\label{eqn:ellipticFibOverBlF2toricdata}
\end{align}
 Assigning projective coordinates $[s:t:u:v: w: x: y: z]$ to the toric divisors $\{D_i\}_{i=1,\ldots,8}$ in the same corresponding ordering, we obtain
  the Stanley--Reissner ideal
  \begin{equation}
  \text{SR} = \{t u, u v, s w, t w, s v, x y z\}\,.
  \end{equation}
The Euler number of $X_3$ is $\chi(X_3)=-420$.
The Mori cone is simplicial and generated by the curves $\cC^i$, $i=0,1,2,3$. The dual K\"ahler cone generators $J_i$ are expressed in terms of the toric divisors, for instance, as
\bea
J_0 = D_1 + D_2 \coma J_1 =D_2 + D_4 \coma J_2 = \frac{1}{2} D_6 \coma J_3 = D_4  \,.
\eea
In particular, $D_1$ and $D_2$ are among the generators of the cone of effective divisors.
Furthermore,
$c_2(X_3) \cdot J_\alpha = (24,48,82,36)$.
This identifies $J_0$ as the divisor associated with the K3-fiber of $\rho$, and $J_2 = S_0 + \pi^*c_1(B_2)$ with $S_0$ being the zero-section of the elliptic fibration. 
Its dual Mori cone generator
$\cC^2$ therefore corresponds to the class generic elliptic fiber.

The generic rational fiber of $B_2$
lies in the class $\cC^1 + \cC^3$. The base of this rational fibration is the base $\mathbb P^1$ of the K3-fibration $\rho$; its class coincides with $\cC^0$. Over a special point on the base $\mathbb P^1$, the rational fiber of $B_2$ splits into two rational curves in class  $\cC^1$ and  $\cC^3$, each of self-intersection $-1$ on $B_2$.
The elliptic fibration over each of these two curves defines a rational elliptic surface, or $dP_9$, of Euler characteristic $12$.
As a result, the K3-fibration undergoes a Kulikov Type II degeneration, in which the generic K3 fiber class splits as
\bea
\mathbf{S}_0 = \mathbf{S}_1 \cup \mathbf{S}_2 \,.
\eea
We identify the class of $\mathbf{S}_1$ and $\mathbf{S}_2$ with the toric divisor classes $D_1$ and $D_2$. 

To the given basis $\{\mathcal{C}^\alpha\}_{\alpha=0,\dots,3}$ of the Mori we can now associate a basis $\{\U(1)^\alpha\}$ of the Abelian gauge factors and hence a basis of charges $\{Q_\alpha\}$ that parametrize the charge lattice. We notice that $\mathcal{C}^2$ is the only Mori cone generator that is also a movable curve. Hence the results of \cite{Alim:2021vhs} imply that there is an infinite tower of BPS states with charge 
\begin{equation}
    \bm{Q}=(Q_0, Q_1 , Q_2, Q_3) = (0,0,n,0)\,. 
\end{equation}
In fact, since $\mathcal{C}^2$ is the elliptic fiber class, the genus-zero BPS invariants along this direction are 
\bea \label{GVTorus-ex}
N^{0}_{(0,0,n,0)} = -\chi(X_3) = 420 \,. 
\eea
On the other hand, the rays in the charge lattice with $Q_2=0$ do not support towers of BPS states and hence invite an application of Claim~\ref{mainclaim}.

To this end we should first consider which linear combinations of $\U(1)^0, \U(1)^1$, and $\U(1)^3$ admit weak coupling limits. Let us begin with the K3-fibration $\rho$ and its associated weak coupling limit of Type K3. The dual of the polarization lattice
is spanned by the generators of the $\rho$-relative Mori cone that lie in the generic K3-fiber. This identifies 
\bea
\Lambda^\ast = \langle \cC^2, \cC^1 + \cC^3 \rangle \simeq U\,,
\eea
where $U$ is the hyperbolic lattice of signature $(1,1)$.
According to the general discussion of Section \ref{sec:WeakCouplK3lim}, the two K\"ahler cone generators $J_1$ and $J_3$ dual to the curves $\cC^1$ and $\cC^3$ in the generic rational fiber must satisfy a homological relation of the form
\eqref{identity}.
Indeed, from the intersection form 
\begin{equation}
  \begin{split}
      \mathcal{I}(X_3) = &\,7 J_2^3+2  J_2^2\cdot J_0+ 4  J_2^2\cdot J_1+3  J_2^2\cdot J_3+2 J_1^2 \cdot J_2+J_3^2 \cdot J_2+J_0\cdot  J_1\cdot  J_2\\
                                &+J_0 \cdot J_2 \cdot J_3+2 J_1 \cdot  J_2 \cdot J_3\,
  \end{split}
\end{equation}
it follows that 
 \begin{equation}
     J_3\cdot J_0\cdot J_\alpha = J_1  \cdot J_0 \cdot J_\alpha\coma \forall \alpha =0,\ldots,3 \,.
     \label{eq:relationdivexamp}
 \end{equation}
%
%
%
%

An infinite distance limit of Type K3 is parametrized as
\begin{equation}
    J = \lambda \tilde{v}^0 J_0 + \frac{1}{\sqrt{\lambda}}\tilde{v}^i J_i\coma \lambda \to \infty \,.
\end{equation}
In terms of the rescaled Mori cone volumes $\hat{\tilde{v}}^\alpha=\frac{\tilde{v}^\alpha}{\mathcal{V}^{1/3}}$, the gauge kinetic matrix $f_{\alpha\beta}$ at leading order in $\lambda$ takes the form
\begin{equation}
    f_{\alpha\beta} = \lambda\frac{\hat{\tilde{v}}_0^2 \hat{\tilde{v}}_2^2 }{\left(\hat{\tilde{v}}_0 \hat{\tilde{v}}_2 \left(\hat{\tilde{v}}_1+\hat{\tilde{v}}_2+\hat{\tilde{v}}_3\right)\right)^{4/3}}\left(
\begin{array}{cccc}
 0 & 0 & 0 & 0 \\
 0 & 1 & 1 & 1 \\
 0 & 1 & \frac{\hat{\tilde{v}}_1^2+2 \left(\hat{\tilde{v}}_2+\hat{\tilde{v}}_3\right) \hat{\tilde{v}}_1+2 \hat{\tilde{v}}_2^2+\hat{\tilde{v}}_3^2+2 \hat{\tilde{v}}_2 \hat{\tilde{v}}_3}{\hat{\tilde{v}}_2^2} & 1 \\
 0 & 1 & 1 & 1 \\
\end{array}
\right)+\mathcal{O}\left(1/\sqrt{\lambda}\right)\fstop
\end{equation}
We notice that the second and the fourth rows (or columns), associated to the divisors $J_1$ and $J_3$ satisfying \eqref{eq:relationdivexamp}, are identical. At leading order, the rank of the matrix $f_{ij}$ is therefore reduced, as expected from the discussion in Section \ref{sec:WeakCouplK3lim}. In particular,
the space of asymptotically weakly coupled abelian gauge symmetries is spanned by the combination
\begin{equation}
    \U(1)_+ = \U(1)^1 + \U(1)^3\coma
\end{equation}
together with $\U(1)^2$, while any $\U(1)$ involving the orthogonal combination $ \U(1)^1 - \U(1)^3$ as well as $\U(1)^0$ cannot become asymptotically weakly coupled in the limit of Type K3. In particular, $\U(1)^1$ and $\U(1)^3$ individually do not admit weak coupling limits as in \eqref{qfabq}. Hence, we do not expect to find any super-extremal non-BPS string excitations with charge $\bm{Q}=(0,n,0,0)$ or $\bm{Q}=(0,0,0,n)$ for $n>1$. Instead a tower of super-extremal excitations charged under $\U(1)^1$ or $\U(1)^3$ must have $Q_1=Q_3$. And indeed,  $\U(1)_+$ and $\U(1)^2$ are precisely the abelian gauge symmetries under which the curve classes in the dual polarization lattice $\Lambda^\ast$ are charged. From the heterotic perspective, these are the $\U(1)$s associated to winding and momentum along the heterotic $S^1$. For these $\U(1)$s the existence of states satisfying \eqref{eq_nQrelation} can be established from the elliptic genus as in Section \ref{sec:ellipticgenera5d}. 
 
To this end, we discuss the elliptic genera of MSW strings realized by the divisor $J_0=D_1+D_2$ and its splitting components.
 First, we compute Gopakumar--Vafa invariants $N^0_C$ for curve classes $C= \beta + n \mathcal{E}$, 
 which give rise to the elliptic genera of strings in six dimensions realized by D3-branes wrapping $\pi_\ast\beta$ as in~\cite{Alim:2012ss,Klemm:2012sx,Huang:2015sta,Kim:2016foj,DelZotto:2016pvm,Gu:2017ccq,DelZotto:2017mee,DelZotto:2018tcj,Lee:2018spm,Lee:2018urn}.
 Considering the curve $\beta =\mathcal{C}^1$,  we obtain the six-dimensional elliptic genus of the E-string, given by~\cite{Klemm:1996hh}
  \begin{equation}
  Z_E (\tau) = \frac{E_4(\tau)}{\eta^{12}(\tau)}=q^{-\frac{1}{2}}(1 + 252q +5130q^2 +\mathcal{O}(q^3) )= q^{-\frac{1}{2}} \sum_d N_{\mathcal{C}^1+ d \mathcal{C}^2 }^0 q^d\,,
  \end{equation}
 where $E_{k}$ is the $k$-th Eisenstein series. 
The same meromorphic modular form $Z_E(\tau)$ is obtained for the curve class $\beta = \mathcal{C}^3$ instead. 
 Moreover, the six-dimensional elliptic genus of the heterotic string is
  \begin{equation}
  Z_{\text{het}}(\tau) = -\frac{23}{12} \frac{E_4 E_6 }{\eta^{24}} -\frac{1}{12} \frac{E_4^2 E_2}{\eta^{24}} = -2q^{-1} - \chi(X_3) +\mathcal{O}(q)= q^{-1} \sum_d N_{\mathcal{C}^1+\mathcal{C}^3 +d \mathcal{C}^2 }^0 q^d\,.
 \end{equation}
 As discussed at the end of Section \ref{sec:ellipticgenera5d}, 
 from the latter expression we derive the holomorphic piece for the heterotic MSW string from \eqref{eq:nonHolEG}.
 Since $\Lambda^* = U$, the discriminant group $\Lambda^*/\Lambda$ only contains the trivial class.
 Hence, the non-holomorphic completion constrains the five-dimensional heterotic elliptic genus $Z^{(1)}_{J_0}(\tau,\bar{\tau},\bm{z},\mathcal{B}) = \widehat{Z}_{\mathbf{0}}(\tau,\bar{\tau}) \Theta_{\mathbf{0},1}^*(\tau,\bar{\tau},\bm{z},\mathcal{B})$ to take the form 
 \begin{equation}
 \widehat{Z}_{\mathbf{0}} (\tau,\bar{\tau})=  -\frac{23}{12} \frac{E_4 E_6 }{\eta^{24}} -\frac{1}{12} \left(\frac{E_4}{\eta^{12}}\right)^2 \widehat{E}_2\,,
 \label{eq:nonHolHet}
 \end{equation}
 where $\widehat{E}_2 = E_2 - 3/\pi \text{Im}(\tau)$ is the non-holomorphic second Eisenstein series, which is also a mock modular form. 
 Notice that the quadratic factors $E_4/ \eta^{12}$ in \eqref{eq:nonHolHet} are meromorphic modular forms corresponding to the MSW strings deriving from the $dP_9$ surfaces~\cite{Klemm:2012sx} given by $D_1$ and $D_2$; 
their quadratic product is expected to be present in the non-holomorphic contribution since $J_0 = D_1 +D_2$~\cite{Alexandrov:2016tnf}. 
Using~\eqref{eq:nonHolHet}, similar arguments as discussed in Section \ref{sec:ellipticgenera5d} can be repeated to argue for the existence of a non-trivial tower of states satisfying \eqref{eq_nQrelation}.

Since $X_3$ admits in addition an elliptic fibration, we can also consider a limit of Type $T^2$, for example by imposing the scaling
\begin{equation} \label{toruslimit-ex}
    J = \sqrt{\lambda} \sum_{i \neq 2} \tilde{v}^i J_i + \frac{1}{\lambda} \tilde{v}^2 J_2 \coma \lambda \to \infty \fstop
\end{equation}
The only asymptotically weakly coupled gauge group in such a limit is $\U(1)_{\cal E} = \U(1)^2$, under which the generic elliptic fiber class is charged. Following our general discussion in Section~\ref{sec_TowerWGC},  the super-extremal states come from the BPS states counted by \eqref{GVTorus-ex}. These states are the KK modes for the asymptotic decompactification to six dimensions in the limit \eqref{toruslimit-ex}.

Since these two classes of limits exhaust the possible weak coupling limits, the $\U(1)$s which cannot undergo any weak coupling limit are
all linear combinations outside the span of $\U(1)^2$ and $\U(1)_+$.
Curves charged under such $\U(1)$s do not lie in the movable cone. Furthermore, the associated directions in the charge lattice do not support a BPS tower, nor can we identify a stringy non-BPS tower. On the other hand,  $\U(1)_+$ and $\U(1)^2$ do admit a weak coupling limit, and we can identify an associated stringy tower of non-BPS and, respectively, BPS super-extremal states. This demonstrates Claim~\ref{mainclaim} in the present explicit example.

 \newpage

\section{Conclusions and Discussion}
\label{sec:discussion}

We have investigated the asymptotic form of the WGC in M-theory compactifications on Calabi--Yau 3-folds. Our main motivation was to understand when non-BPS states furnish towers of super-extremal states along directions in the charge lattice where no tower of BPS states exists according to the analysis of \cite{Alim:2021vhs}.

The tower WGC is best motivated in asymptotic weak coupling limits.
A natural starting point for our analysis are therefore asymptotically weak coupling limits for the gauge theories. More precisely, our goal was to prove the tower WGC for those linear combinations of $\U(1)$ factors that can undergo a limit in which the ratio between the WGC scale, $\Lambda_\text{\tiny{WGC}}$, and the quantum gravity cut-off vanishes. As our main result we have shown Claim \ref{mainclaim} stating that whenever a direction in the charge lattice is not populated by a tower of BPS states, there either does not exist a weak coupling limit for the dual gauge group factors, or one can identify a tower of asymptotically super-extremal non-BPS states arising as excitations of a critical string.

An important ingredient in arriving at this conclusion is our classification of all weak coupling limits for $\U(1)$ gauge factors in five-dimensional M-theory, building on the classification of infinite distance limits in \cite{Lee:2019wij}. According to this reference, any infinite distance limit in the vector multiplet moduli space of M-theory on Calabi--Yau 3-folds is associated to a limit where either a $T^2$-, a K3-, or a $T^4$-fiber shrinks. Our analysis characterizes the lattice of states charged under a potentially weakly coupled $\U(1)$
very compactly as follows:
It is the lattice spanned by all curve classes on the 3-fold which lie in the generic fiber or which arise at a degeneration of the fiber at finite distance in its moduli space. For Type $T^2$ limits, this means that only $\U(1)$s which play the role of a KK $\U(1)$ in some asymptotic duality frame can become weakly coupled. For Type K3 limits, the resulting structure is richer: In physics terms, asymptotically weakly coupled $\U(1)$s embed into a dual perturbative heterotic gauge group, including possible KK or winding $\U(1)$s.
 Geometrically, we have to distinguish between irreducible special K3 fibers and degenerations in which the K3 fiber splits into several components. 
In the context of semi-stable degenerations, the first type of degenerations are Kulikov Type I degenerations at finite distance, which contribute to the charge lattice coupling to potentially weakly coupled $\U(1)$s; the second type of degeneration is of Kulikov Type II/III at infinite distance and does not yield $\U(1)$s that can become weakly coupled.  Although we argued that our conclusions hold more generally than in semi-stable settings, for instance in those exemplified in \cite{Braun:2016sks}, the importance of K3-fibrations for the taxonomy of weak coupling limits certainly motivates a more in-depth analysis of such fiber degenerations beyond Kulikov type.

Armed with the classification of the $\U(1)$s admitting a weak coupling limit, we have shown that there always exists either a BPS tower of super-extremal states (the KK/decompactification case) or a tower of super-extremal non-BPS states arising as excitations of a perturbative, critical string. In the latter case, we made use of the modular properties of the elliptic genus of the five-dimensional string \cite{Bouchard:2016lfg,Gaiotto:2006wm} and the 
connection between Donaldson--Thomas invariants and Noether--Lefschetz theory \cite{Katz:1999xq,Pandharipande:2014qoa}.
This establishes the existence of a tower of states for which the square of the quantized charge is equal to twice the (negative of the) left-moving excitation level. 
Strictly speaking, the results of \cite{Gaiotto:2006wm} apply to geometries in which all K3-fibers are irreducible, which is not the case in the presence of Type II/III degenerations.  These degenerations signal the presence of non-perturbative strings, such as E- or M-strings. By experience from six dimensions \cite{Lee:2018spm,Lee:2018urn}, however, the presence of the non-perturbative strings should not affect the existence of the states required for the tWGC. 
For the example of a K3-fibration with a compatible elliptic fibration, we have illustrated that the departure from modularity in the presence of Type II Kulikov degenerations is rather mild, as in six dimensions. This example is a special case of the more drastic mock modularity that could be expected beyond elliptic K3 surfaces with Type II degenerations, and it would be interesting to extend the analysis of the five-dimensional elliptic genus to include such types of degenerations.

Most of our analysis focused on the Type K3 limits, and we have treated the Type $T^4$ limits largely by analogy. We have argued that such abelian surface limits behave similarly (given that they are also more constrained due to enhanced supersymmetry), but a systematic study of the underlying geometries could nevertheless be interesting and would provide a more complete and satisfactory discussion. 

Our findings can be viewed as a five-dimensional analogue of the results of  \cite{Cota:2022yjw} for four-dimensional $\mathcal{N}=1$ theories. In both settings, the only non-BPS tower satisfying the asymptotic tWGC are given by the excitations of critical strings, since otherwise the relevant gauge groups cannot undergo an effective weak coupling limit. Based on these results for the asymptotic tWGC it is natural to ask about the fate of the tWGC away from weak coupling limits. In principle, there are two ways to address this question: First, one can follow the non-BPS states corresponding to super-extremal string excitations
along our motion in moduli space away from asymptotic weak coupling and trace their charge-to-mass ratio. As we re-enter the bulk of the moduli space, string loop corrections to the physical charge and the mass become important and can, in principle, change the charge-to-mass ratio. In \cite{Klaewer:2020lfg} the analogous effect in four-dimensional ${\cal N}=1$ theories was investigated. Compared to this analysis, the five-dimensional case is considerably simpler, since here one does not have to worry about perturbative $\alpha'$-corrections. This makes it possibly more feasible to verify the tWGC including string loop corrections, at least as long as a weak coupling limit exists in principle. Notice that, as long as we stay sufficiently close to the asymptotic region in moduli space, the string excitations still have masses below the species scale and can hence still be treated as particle-like excitations in an effective theory of supergravity. Therefore also the WGC should still be equivalent to the RFC. Using this, one could, as was done in the analogous four-dimensional ${\cal N}=1$ setting of \cite{Klaewer:2020lfg}, infer the string loop corrections to the black hole extremality bound from the corrections to the RFC, and it would be interesting to do so.

What remains more mysterious is the tower WGC for the $\U(1)$s which do not admit a weak coupling limit. 
We conclude this paper with some, partially more speculative, remarks related to this open question.
First, recall that in many instances an infinite tower of (super-) extremal states can indeed be firmly established away from weak coupling.
This is the case whenever the charge lattice supports BPS towers. Combining the reasoning of \cite{Alim:2021vhs} and our findings, these directions in the charge lattice include all curves in the movable cone of a Calabi--Yau $X_3$ \cite{Alim:2021vhs} that are not contained in a non-degenerate torus or surface fiber or a degenerate such fiber undergoing only a finite distance degeneration in its moduli space.\footnote{In addition there can be infinite towers of BPS states along curves that shrink at a boundary where also a divisor shrinks to zero size or strictly super-extremal towers in a cone slightly larger than  the movable cone \cite{Alim:2021vhs}.}
Furthermore, there can, of course, exist a {\it finite} number of BPS states along other directions in the charge lattice even though they do not support a tower.
For example, the base $\mathbb P^1$ of a surface fibration falls into this category and admits a single super-extremal BPS state, but not an infinite tower.

Similarly, based on our analysis, we certainly cannot \textit{exclude} the existence of super-extremal {\it non-BPS} states charged under $\U(1)$s without a weak coupling limit. 
To name but one possible source, such states might, in principle, come from M2-branes wrapped on non-holomorphic curves.
In particular, it is very conceivable that there exists a {\it finite} number of charged states with very low mass, which would then be highly super-extremal, also along directions without BPS representatives. Since the mass of these states is well below the quantum gravity cut-off, they can be included in the low-energy EFT in a sensible way. 

The crucial question is rather whether one expects towers of non-BPS super-extremal states along directions without weak coupling regimes, and how to make sense of them. The reason one might be skeptical is that one would expect an infinite {\it tower} of charged states to appear, if at all, at the mass scale set by $\Lambda_\text{\tiny{WGC}}$, 
and hence above the cutoff of the EFT, as we now explain.


First, we concede that 
from the point of view of the WGC alone, it would certainly be consistent if the actual mass of the super-extremal states was below $\Lambda_\text{\tiny WGC}$. In that case, even if the ratio \eqref{WGCtosp} is always at least of $\mathcal{O}(1)$ for a given $\U(1)$, there could exist a tower of super-extremal states below the quantum gravity cut-off in this direction of the charge lattice. These states would, however, contribute to the number of light species and hence lower the species scale. In particular, a {\it tower} of such states would considerably lower the species scale. In the vicinity of, e.g., a decompactification limit, this would spoil the relation between species scale and higher-dimensional Planck scale. Extrapolating from such situations it seems unnatural to have such a highly super-extremal tower also in absence of a decompactification limit.\footnote{To prove that also in the interior of the moduli space there cannot be such a tower of super-extremal states below the species scale it would be useful to have a closed expression for the moduli dependence of the species scale and hence of the number of light species (BPS and non-BPS) at any point in the moduli space of the five-dimensional EFT similar to what \cite{vandeHeisteeg:2022btw} proposed for ${\cal N}=2$ theories in four dimensions.} A second argument is that the existence of a highly super-extremal infinite tower of states parametrically below $\Lambda_\text{\tiny{WGC}}$ would run counter to the expectation that for higher and higher masses these states should become black holes  - which would be in tension with the fact that they would still be super-extremal.

These two arguments appear to favor the scenario that the WGC tower sits at 
$\Lambda_\text{\tiny{WGC}}$, at least for high masses. In particular, this \textit{is} the case for the tower of super-extremal states for $\U(1)$s with a weak coupling limit, as our results show.

Now, for a $\U(1)$ for which the ratio \eqref{WGCtosp} never vanishes, a tower at  $\Lambda_\text{\tiny{WGC}}$ has a mass at or above the quantum gravity cut-off. While BPS towers of this type are protected by the BPS condition, for a non-BPS tower (not inherited from a weak coupling limit) there would be almost no chance of identifying the tower states; after all, there is no way to include them as particles in the EFT. This goes against the original logic of the WGC to constrain the EFT. Therefore, for a $\U(1)$ for which the ratio \eqref{WGCtosp} is always at least of order one, the tWGC does not give a constraint on the spectrum of the low-energy effective theory of gravity. 
Of course, this does not mean that the tWGC may not hold for generically strongly coupled gauge theories, but if it does, it would not have any immediate implications for the particle spectrum of the EFT, and verifying it would in any case require control of a strongly coupled theory above its quantum gravity cutoff.
From this point of view, it is tempting to speculate that away from weak coupling (or at least when there is no chance to reach weak coupling asymptotically) the WGC may not have to be interpreted in its tower version.

On the other hand, the {\it asymptotic} tower WGC is on firm grounds, and 
we believe that the results in this work on its realization in the non-BPS sector in M-theory serve as an important further piece of evidence in its favor.

\subsection*{Acknowledgments}

We thank Murad Alim, Rafael \'Alvarez-Garc\'ia, Ben Heidenreich, Damian van de Heisteeg, Amir Kashani-Poor, Albrecht Klemm, Seung-Joo Lee, Wolfgang Lerche, Fernando Marchesano, Luca Martucci, Boris Pioline, Matt Reece, Tom Rudelius and Cumrun Vafa for helpful discussions. 
C. F. C., A. M. and T. W. are supported in part by Deutsche Forschungsgemeinschaft under Germany's Excellence Strategy EXC 2121  Quantum Universe 390833306 and by Deutsche Forschungsgemeinschaft through a German-Israeli Project Cooperation (DIP) grant ``Holography and the Swampland”. M. W. is supported in part by a grant from the Simons Foundation (602883, CV) and also by the NSF grant PHY-2013858.

\printbibliography

@article{Agmon:2022thq,
    author = "Agmon, Nathan Benjamin and Bedroya, Alek and Kang, Monica Jinwoo and Vafa, Cumrun",
    title = "{Lectures on the string landscape and the Swampland}",
    eprint = "2212.06187",
    archivePrefix = "arXiv",
    primaryClass = "hep-th",
    month = "12",
    year = "2022"
}

@article{Marchesano:2022axe,
    author = "Marchesano, Fernando and Melotti, Luca",
    title = "{EFT strings and emergence}",
    eprint = "2211.01409",
    archivePrefix = "arXiv",
    primaryClass = "hep-th",
    month = "11",
    year = "2022"
}

@article{Grimm:2018ohb,
    author = "Grimm, Thomas W. and Palti, Eran and Valenzuela, Irene",
    title = "{Infinite Distances in Field Space and Massless Towers of States}",
    eprint = "1802.08264",
    archivePrefix = "arXiv",
    primaryClass = "hep-th",
    doi = "10.1007/JHEP08(2018)143",
    journal = "JHEP",
    volume = "08",
    pages = "143",
    year = "2018"
}

@article{Maldacena:1997de,
    author = "Maldacena, Juan Martin and Strominger, Andrew and Witten, Edward",
    title = "{Black hole entropy in M theory}",
    primaryClass = "hep-th", eprint = "hep-th/9711053",
    archivePrefix = "arXiv",
    doi = "10.1088/1126-6708/1997/12/002",
    journal = "JHEP",
    volume = "12",
    pages = "002",
    year = "1997"
}

@article{Bouchard:2016lfg,
    author = "Bouchard, Vincent and Creutzig, Thomas and Diaconescu, Duiliu-Emanuel and Doran, Charles and Quigley, Callum and Sheshmani, Artan",
    title = "{Vertical D4\textendash{}D2\textendash{}D0 Bound States on K3 Fibrations and Modularity}",
    eprint = "1601.04030",
    archivePrefix = "arXiv",
    primaryClass = "hep-th",
    doi = "10.1007/s00220-016-2772-y",
    journal = "Commun. Math. Phys.",
    volume = "350",
    number = "3",
    pages = "1069--1121",
    year = "2017"
}

@book{HulekKlaus2011MSoA,
series = {De Gruyter Expositions in Mathematics ; 12},
publisher = {De Gruyter},
isbn = {9783110891928},
year = {2011},
title = {Moduli Spaces of Abelian Surfaces : Compactification, Degenerations and Theta Functions},
edition = {Reprint 2011},
language = {eng},
address = {Berlin ; Boston},
author = {Hulek, Klaus},
}

@article{Palti:2020mwc,
    author = "Palti, Eran",
    title = "{A Brief Introduction to the Weak Gravity Conjecture}",
    doi = "10.31526/lhep.2020.176",
    journal = "LHEP",
    volume = "2020",
    pages = "176",
    year = "2020"
}

@article{Kaya:2022edp,
    author = "Kaya, Sami and Rudelius, Tom",
    title = "{Higher-group symmetries and weak gravity conjecture mixing}",
    eprint = "2202.04655",
    archivePrefix = "arXiv",
    primaryClass = "hep-th",
    doi = "10.1007/JHEP07(2022)040",
    journal = "JHEP",
    volume = "07",
    pages = "040",
    year = "2022"
}

@article{Heidenreich:2020ptx,
    author = "Heidenreich, Ben and Rudelius, Tom",
    title = "{Infinite distance and zero gauge coupling in 5D supergravity}",
    eprint = "2007.07892",
    archivePrefix = "arXiv",
    primaryClass = "hep-th",
    reportNumber = "ACFI-T20-08",
    doi = "10.1103/PhysRevD.104.106016",
    journal = "Phys. Rev. D",
    volume = "104",
    number = "10",
    pages = "106016",
    year = "2021"
}

@article{Palti:2021ubp,
    author = "Palti, Eran",
    title = "{Stability of BPS states and weak coupling limits}",
    eprint = "2107.01539",
    archivePrefix = "arXiv",
    primaryClass = "hep-th",
    doi = "10.1007/JHEP08(2021)091",
    journal = "JHEP",
    volume = "08",
    pages = "091",
    year = "2021"
}

@article{Wiesner:2022qys,
    author = "Wiesner, Max",
    title = "{Light Strings and Strong Coupling in F-theory}",
    eprint = "2210.14238",
    archivePrefix = "arXiv",
    primaryClass = "hep-th",
    month = "10",
    year = "2022"
}

@article{Kachru:1995wm,
    author = "Kachru, Shamit and Vafa, Cumrun",
    title = "{Exact results for N=2 compactifications of heterotic strings}",
    eprint = "hep-th/9505105",
    archivePrefix = "arXiv",
    reportNumber = "HUTP-95-A016",
    doi = "10.1016/0550-3213(95)00307-E",
    journal = "Nucl. Phys. B",
    volume = "450",
    pages = "69--89",
    year = "1995"
}

@article{Grimm:2022sbl,
    author = "Grimm, Thomas W. and Lanza, Stefano and Li, Chongchuo",
    title = "{Tameness, Strings, and the Distance Conjecture}",
    eprint = "2206.00697",
    archivePrefix = "arXiv",
    primaryClass = "hep-th",
    doi = "10.1007/JHEP09(2022)149",
    journal = "JHEP",
    volume = "09",
    pages = "149",
    year = "2022"
}

@article{vandeHeisteeg:2022btw,
    author = "van de Heisteeg, Damian and Vafa, Cumrun and Wiesner, Max and Wu, David H.",
    title = "{Moduli-dependent Species Scale}",
    eprint = "2212.06841",
    archivePrefix = "arXiv",
    primaryClass = "hep-th",
    month = "12",
    year = "2022"
}

@article{Castellano:2022bvr,
    author = "Castellano, Alberto and Herr\'aez, Alvaro and Ib\'a\~nez, Luis E.",
    title = "{The Emergence Proposal in Quantum Gravity and the Species Scale}",
    eprint = "2212.03908",
    archivePrefix = "arXiv",
    primaryClass = "hep-th",
    month = "12",
    year = "2022"
}

@article{Pandharipande:2014qoa,
    author = "Pandharipande, R. and Thomas, R. P.",
    title = "{The Katz-Klemm-Vafa conjecture for $K3 $ surfaces}",
    eprint = "1404.6698",
    archivePrefix = "arXiv",
    primaryClass = "math.AG",
    doi = "10.1017/fmp.2016.2",
    journal = "Forum Math. Pi",
    volume = "4",
    pages = "e4",
    year = "2016"
}

@article{Montero:2022prj,
    author = "Montero, Miguel and Vafa, Cumrun and Valenzuela, Irene",
    title = "{The Dark Dimension and the Swampland}",
    eprint = "2205.12293",
    archivePrefix = "arXiv",
    primaryClass = "hep-th",
    month = "5",
    year = "2022"
}

@article{Cota:2022yjw,
    author = "Cota, Cesar Fierro and Mininno, Alessandro and Weigand, Timo and Wiesner, Max",
    title = "{The Asymptotic Weak Gravity Conjecture for Open Strings}",
    eprint = "2208.00009",
    archivePrefix = "arXiv",
    primaryClass = "hep-th",
    year = "2022",
    doi = "10.1007/JHEP11(2022)058",
    journal = "JHEP",
    volume = "11",
    pages = "058",
}

@article{Alim:2010cf,
    author = "Alim, Murad and Haghighat, Babak and Hecht, Michael and Klemm, Albrecht and Rauch, Marco and Wotschke, Thomas",
    title = "{Wall-crossing holomorphic anomaly and mock modularity of multiple M5-branes}",
    eprint = "1012.1608",
    archivePrefix = "arXiv",
    primaryClass = "hep-th",
    reportNumber = "BONN-TH-2010-13, LMU-ASC-102-10",
    doi = "10.1007/s00220-015-2436-3",
    journal = "Commun. Math. Phys.",
    volume = "339",
    number = "3",
    pages = "773--814",
    year = "2015"
}

@article{Alexandrov:2016tnf,
    author = "Alexandrov, Sergei and Banerjee, Sibasish and Manschot, Jan and Pioline, Boris",
    title = "{Multiple D3-instantons and mock modular forms I}",
    eprint = "1605.05945",
    archivePrefix = "arXiv",
    primaryClass = "hep-th",
    reportNumber = "L2C:16-056, IPHT-T16-037, CERN-TH-2016-121, TCDMATH-16-08",
    doi = "10.1007/s00220-016-2799-0",
    journal = "Commun. Math. Phys.",
    volume = "353",
    number = "1",
    pages = "379--411",
    year = "2017"
}

@article{Lee:2021qkx,
    author = "Lee, Seung-Joo and Weigand, Timo",
    title = "{Elliptic K3 surfaces at infinite complex structure and their refined Kulikov models}",
    eprint = "2112.07682",
    archivePrefix = "arXiv",
    primaryClass = "hep-th",
    doi = "10.1007/JHEP09(2022)143",
    journal = "JHEP",
    volume = "09",
    pages = "143",
    year = "2022"
}

@incollection{MR3524167,
	author = {Katz, S. and Klemm, A. and Pandharipande, R.},
	booktitle = {K3 surfaces and their moduli},
	doi = {10.1007/978-3-319-29959-4_6},
	mrclass = {14J28 (14N35)},
	mrnumber = {3524167},
	mrreviewer = {Dragos Nicolae Oprea},
	note = {With an appendix by R. P. Thomas},
	pages = {111--146},
	publisher = {Birkh\"{a}user/Springer, [Cham]},
	series = {Progr. Math.},
	title = {On the motivic stable pairs invariants of {$K3$} surfaces},
	url = {https://doi.org/10.1007/978-3-319-29959-4_6},
	volume = {315},
	year = {2016},
	eprint = "1407.3181",
    archivePrefix = "arXiv",
    primaryClass = "math.AG",
	}

@article{MR1625724,
	author = {Borcherds, Richard E.},
	doi = {10.1007/s002220050232},
	fjournal = {Inventiones Mathematicae},
	issn = {0020-9910},
	journal = {Invent. Math.},
	mrclass = {11F37 (11F22 14J28 17B67 57R57)},
	mrnumber = {1625724},
	mrreviewer = {I. Dolgachev},
	number = {3},
	pages = {491--562},
	title = {Automorphic forms with singularities on {G}rassmannians},
	url = {https://mathscinet.ams.org/mathscinet-getitem?mr=1625724},
	volume = {132},
	year = {1998}}

@article{Oehlmann:2019ohh,
    author = "Oehlmann, Paul-Konstantin and Schimannek, Thorsten",
    title = "{GV-Spectroscopy for F-theory on genus-one fibrations}",
    eprint = "1912.09493",
    archivePrefix = "arXiv",
    primaryClass = "hep-th",
    doi = "10.1007/JHEP09(2020)066",
    journal = "JHEP",
    volume = "09",
    pages = "066",
    year = "2020"
}

@article{Kashani-Poor:2019jyo,
    author = "Kashani-Poor, Amir-Kian",
    title = "{Determining F-theory Matter Via Gromov-Witten Invariants}",
    eprint = "1912.10009",
    archivePrefix = "arXiv",
    primaryClass = "hep-th",
    doi = "10.1007/s00220-021-04145-4",
    journal = "Commun. Math. Phys.",
    volume = "386",
    number = "2",
    pages = "1155--1207",
    year = "2021"
}

@article{Gaiotto:2006wm,
    author = "Gaiotto, Davide and Strominger, Andrew and Yin, Xi",
    title = "{The M5-Brane Elliptic Genus: Modularity and BPS States}",
    primaryClass = "hep-th", eprint = "hep-th/0607010",
    archivePrefix = "arXiv",
    doi = "10.1088/1126-6708/2007/08/070",
    journal = "JHEP",
    volume = "08",
    pages = "070",
    year = "2007"
}

@article{MR2669707,
	author = {Klemm, A. and Maulik, D. and Pandharipande, R. and Scheidegger, E.},
	doi = {10.1090/S0894-0347-2010-00672-8},
	fjournal = {Journal of the American Mathematical Society},
	issn = {0894-0347},
	journal = {J. Amer. Math. Soc.},
	mrclass = {14N35 (11F23 14D20 14N10)},
	mrnumber = {2669707},
	mrreviewer = {Dragos Nicolae Oprea},
	number = {4},
	pages = {1013--1040},
	title = {Noether-{L}efschetz theory and the {Y}au-{Z}aslow conjecture},
	url = {https://doi.org/10.1090/S0894-0347-2010-00672-8},
	volume = {23},
	year = {2010}}

@article{Heidenreich:2017sim,
    author = "Heidenreich, Ben and Reece, Matthew and Rudelius, Tom",
    title = "{The Weak Gravity Conjecture and Emergence from an Ultraviolet Cutoff}",
    eprint = "1712.01868",
    archivePrefix = "arXiv",
    primaryClass = "hep-th",
    doi = "10.1140/epjc/s10052-018-5811-3",
    journal = "Eur. Phys. J. C",
    volume = "78",
    number = "4",
    pages = "337",
    year = "2018"
}

@article{MR3508473,
	Author = {Pandharipande, R. and Thomas, R. P.},
	Doi = {10.1017/fmp.2016.2},
	Fjournal = {Forum of Mathematics. Pi},
	Journal = {Forum Math. Pi},
	Mrclass = {14N35 (14J28)},
	Mrnumber = {3508473},
	Mrreviewer = {Dragos Nicolae Oprea},
	Pages = {e4, 111},
	Title = {The {K}atz-{K}lemm-{V}afa conjecture for {$K3$} surfaces},
	Url = {https://mathscinet.ams.org/mathscinet-getitem?mr=3508473},
	Volume = {4},
	Year = {2016}}

@incollection{MR3114953,
	Author = {Maulik, Davesh and Pandharipande, Rahul},
	Booktitle = {A celebration of algebraic geometry},
	Mrclass = {14N35 (14J10 14J28)},
	Mrnumber = {3114953},
	Mrreviewer = {Satoshi Minabe},
	Pages = {469--507},
	Publisher = {Amer. Math. Soc., Providence, RI},
	Series = {Clay Math. Proc.},
	Title = {Gromov-{W}itten theory and {N}oether-{L}efschetz theory},
	Url = {https://mathscinet.ams.org/mathscinet-getitem?mr=3114953},
	Volume = {18},
	Year = {2013}}

@article{Heidenreich:2019zkl,
    author = "Heidenreich, Ben and Reece, Matthew and Rudelius, Tom",
    title = "{Repulsive Forces and the Weak Gravity Conjecture}",
    eprint = "1906.02206",
    archivePrefix = "arXiv",
    primaryClass = "hep-th",
    doi = "10.1007/JHEP10(2019)055",
    journal = "JHEP",
    volume = "10",
    pages = "055",
    year = "2019"
}

@article{Montero:2016tif,
    author = "Montero, Miguel and Shiu, Gary and Soler, Pablo",
    title = "{The Weak Gravity Conjecture in three dimensions}",
    eprint = "1606.08438",
    archivePrefix = "arXiv",
    primaryClass = "hep-th",
    reportNumber = "IFT-UAM-CSIC-16-060, FTUAM-16-25, MAD-TH-16-03",
    doi = "10.1007/JHEP10(2016)159",
    journal = "JHEP",
    volume = "10",
    pages = "159",
    year = "2016"
}

@article{Heidenreich:2016aqi,
    author = "Heidenreich, Ben and Reece, Matthew and Rudelius, Tom",
    title = "{Evidence for a sublattice weak gravity conjecture}",
    eprint = "1606.08437",
    archivePrefix = "arXiv",
    primaryClass = "hep-th",
    doi = "10.1007/JHEP08(2017)025",
    journal = "JHEP",
    volume = "08",
    pages = "025",
    year = "2017"
}

@article{Andriolo:2018lvp,
    author = "Andriolo, Stefano and Junghans, Daniel and Noumi, Toshifumi and Shiu, Gary",
    title = "{A Tower Weak Gravity Conjecture from Infrared Consistency}",
    eprint = "1802.04287",
    archivePrefix = "arXiv",
    primaryClass = "hep-th",
    reportNumber = "KOBE-COSMO-18-01, MAD-TH-17-07",
    doi = "10.1002/prop.201800020",
    journal = "Fortsch. Phys.",
    volume = "66",
    number = "5",
    pages = "1800020",
    year = "2018"
}

@article{vanBeest:2021lhn,
    author = "van Beest, Marieke and Calder\'on-Infante, Jos\'e and Mirfendereski, Delaram and Valenzuela, Irene",
    title = "{Lectures on the Swampland Program in String Compactifications}",
    eprint = "2102.01111",
    archivePrefix = "arXiv",
    primaryClass = "hep-th",
    doi = "10.1016/j.physrep.2022.09.002",
    journal = "Phys. Rept.",
    volume = "989",
    pages = "1--50",
    year = "2022"
}

@article{Brennan:2017rbf,
    author = "Brennan, T. Daniel and Carta, Federico and Vafa, Cumrun",
    title = "{The String Landscape, the Swampland, and the Missing Corner}",
    eprint = "1711.00864",
    archivePrefix = "arXiv",
    primaryClass = "hep-th",
    reportNumber = "IFT-UAM-CSIC-17-105",
    doi = "10.22323/1.305.0015",
    journal = "PoS",
    volume = "TASI2017",
    pages = "015",
    year = "2017"
}

@article{Palti:2019pca,
    author = "Palti, Eran",
    title = "{The Swampland: Introduction and Review}",
    eprint = "1903.06239",
    archivePrefix = "arXiv",
    primaryClass = "hep-th",
    reportNumber = "MPP-2019-53",
    doi = "10.1002/prop.201900037",
    journal = "Fortsch. Phys.",
    volume = "67",
    number = "6",
    pages = "1900037",
    year = "2019"
}

@article{Harvey:1995fq,
    author = "Harvey, Jeffrey A. and Moore, Gregory W.",
    title = "{Algebras, BPS states, and strings}",
    primaryClass = "hep-th", eprint = "hep-th/9510182",
    archivePrefix = "arXiv",
    reportNumber = "EFI-95-64, YCTP-P16-95",
    doi = "10.1016/0550-3213(95)00605-2",
    journal = "Nucl. Phys. B",
    volume = "463",
    pages = "315--368",
    year = "1996"
}

@article{Heidenreich:2021yda,
    author = "Heidenreich, Ben and Reece, Matthew and Rudelius, Tom",
    title = "{The Weak Gravity Conjecture and axion strings}",
    eprint = "2108.11383",
    archivePrefix = "arXiv",
    primaryClass = "hep-th",
    reportNumber = "ACFI-T21-10",
    doi = "10.1007/JHEP11(2021)004",
    journal = "JHEP",
    volume = "11",
    pages = "004",
    year = "2021"
}

@article{Kulikov1,
doi = {10.1070/IM1977v011n05ABEH001753},
url = {https://dx.doi.org/10.1070/IM1977v011n05ABEH001753},
year = {1977},
month = {10},
publisher = {},
volume = {11},
number = {5},
pages = {957},
author = {Viktor S Kulikov},
title = {Degenerations of K3 Surfaces and Enriques Surfaces},
journal = {Mathematics of the USSR-Izvestiya}
}

@article{Gendler:2022ztv,
    author = "Gendler, Naomi and Heidenreich, Ben and McAllister, Liam and Moritz, Jakob and Rudelius, Tom",
    title = "{Moduli Space Reconstruction and Weak Gravity}",
    eprint = "2212.10573",
    archivePrefix = "arXiv",
    primaryClass = "hep-th",
    reportNumber = "ACFI-T22-10",
    month = "12",
    year = "2022"
}

@article{Kulikov2,
doi = {10.1070/IM1981v017n02ABEH001361},
url = {https://dx.doi.org/10.1070/IM1981v017n02ABEH001361},
year = {1981},
month = {4},
publisher = {},
volume = {17},
number = {2},
pages = {339},
author = {Vik S Kulikov},
title = {On Modifications of Degenerations of Surfaces with $\kappa=0$},
journal = {Mathematics of the USSR-Izvestiya},
}

@article{Braun:2016sks,
    author = "Braun, Andreas P. and Watari, Taizan",
    title = "{Heterotic-Type IIA Duality and Degenerations of K3 Surfaces}",
    eprint = "1604.06437",
    archivePrefix = "arXiv",
    primaryClass = "hep-th",
    reportNumber = "IPMU16-0055",
    doi = "10.1007/JHEP08(2016)034",
    journal = "JHEP",
    volume = "08",
    pages = "034",
    year = "2016"
}

@article{PerssonPink,
    author = "Persson, U. and Pinkham, H.",
    title = "{Degenerations of surfaces with trivial canonical bundle}",
    journal = "Ann of Math. (2) 113",
    volume = "01",
    pages = "45--66",
    year = "1981",
    doi = "10.2307/1971133"
}

@article{Dvali:2010vm,
    author = "Dvali, Gia and Gomez, Cesar",
    title = "{Species and Strings}",
    eprint = "1004.3744",
    archivePrefix = "arXiv",
    primaryClass = "hep-th",
    reportNumber = "CERN-PH-TH-2010-069, IFT-UAM-CSIC-10-25",
    month = "4",
    year = "2010"
}

@article{Marchesano:2022avb,
    author = "Marchesano, Fernando and Wiesner, Max",
    title = "{4d strings at strong coupling}",
    eprint = "2202.10466",
    archivePrefix = "arXiv",
    primaryClass = "hep-th",
    reportNumber = "IFT-UAM/CSIC-22-13",
    doi = "10.1007/JHEP08(2022)004",
    journal = "JHEP",
    volume = "08",
    pages = "004",
    year = "2022"
}

@article{Lanza:2020qmt,
    author = "Lanza, Stefano and Marchesano, Fernando and Martucci, Luca and Valenzuela, Irene",
    title = "{Swampland Conjectures for Strings and Membranes}",
    eprint = "2006.15154",
    archivePrefix = "arXiv",
    primaryClass = "hep-th",
    doi = "10.1007/JHEP02(2021)006",
    journal = "JHEP",
    volume = "02",
    pages = "006",
    year = "2021"
}

@article{Lanza:2021udy,
    author = "Lanza, Stefano and Marchesano, Fernando and Martucci, Luca and Valenzuela, Irene",
    title = "{The EFT stringy viewpoint on large distances}",
    eprint = "2104.05726",
    archivePrefix = "arXiv",
    primaryClass = "hep-th",
    doi = "10.1007/JHEP09(2021)197",
    journal = "JHEP",
    volume = "09",
    pages = "197",
    year = "2021"
}

@article{Klaewer:2020lfg,
    author = "Klaewer, Daniel and Lee, Seung-Joo and Weigand, Timo and Wiesner, Max",
    title = "{Quantum corrections in 4d $N$ = 1 infinite distance limits and the weak gravity conjecture}",
    eprint = "2011.00024",
    archivePrefix = "arXiv",
    primaryClass = "hep-th",
    reportNumber = "CTPU-PTC-20-24, IFT-UAM/CSIC-20-148, MITP/20-064, ZMP-HH/20-21",
    doi = "10.1007/JHEP03(2021)252",
    journal = "JHEP",
    volume = "03",
    pages = "252",
    year = "2021"
}

@article{Lee:2019tst,
    author = "Lee, Seung-Joo and Lerche, Wolfgang and Weigand, Timo",
    title = "{Modular Fluxes, Elliptic Genera, and Weak Gravity Conjectures in Four Dimensions}",
    eprint = "1901.08065",
    archivePrefix = "arXiv",
    primaryClass = "hep-th",
    doi = "10.1007/JHEP08(2019)104",
    journal = "JHEP",
    volume = "08",
    pages = "104",
    year = "2019"
}

@article{Arkani-Hamed:2006emk,
    author = "Arkani-Hamed, Nima and Motl, Lubos and Nicolis, Alberto and Vafa, Cumrun",
    title = "{The String landscape, black holes and gravity as the weakest force}",
    primaryClass = "hep-th", eprint="hep-th/0601001",
    archivePrefix = "arXiv",
    reportNumber = "HUTP-05-A0057",
    doi = "10.1088/1126-6708/2007/06/060",
    journal = "JHEP",
    volume = "06",
    pages = "060",
    year = "2007"
}

@book{Polchinski:1998rr,
    author = "Polchinski, J.",
    title = "{String theory. Vol. 2: Superstring theory and beyond}",
    doi = "10.1017/CBO9780511618123",
    isbn = "978-0-511-25228-0, 978-0-521-63304-8, 978-0-521-67228-3",
    publisher = "Cambridge University Press",
    series = "Cambridge Monographs on Mathematical Physics",
    month = "12",
    year = "2007"
}

@article{Klemm:1996hh,
    author = "Klemm, Albrecht and Mayr, Peter and Vafa, Cumrun",
    editor = "Froehlich, J. and Rittenberg, V. and Schwimmer, A.",
    title = "{BPS states of exceptional noncritical strings}",
    primaryClass = "hep-th", eprint="hep-th/9607139",
    archivePrefix = "arXiv",
    reportNumber = "CERN-TH-96-184, HUTP-96-A031",
    doi = "10.1016/S0920-5632(97)00422-2",
    journal = "Nucl. Phys. B Proc. Suppl.",
    volume = "58",
    pages = "177",
    year = "1997"}

@article{Klemm:2012sx,
    author = "Klemm, Albrecht and Manschot, Jan and Wotschke, Thomas",
    title = "{Quantum geometry of elliptic Calabi-Yau manifolds}",
    eprint = "1205.1795",
    archivePrefix = "arXiv",
    primaryClass = "hep-th",
    month = "5",
    year = "2012"
}

@article{DelZotto:2018tcj,
    author = "Del Zotto, Michele and Lockhart, Guglielmo",
    title = "{Universal Features of BPS Strings in Six-dimensional SCFTs}",
    eprint = "1804.09694",
    archivePrefix = "arXiv",
    primaryClass = "hep-th",
    doi = "10.1007/JHEP08(2018)173",
    journal = "JHEP",
    volume = "08",
    pages = "173",
    year = "2018"
}

@article{Huang:2015sta,
    author = "Huang, Min-xin and Katz, Sheldon and Klemm, Albrecht",
    title = "{Topological String on elliptic CY 3-folds and the ring of Jacobi forms}",
    eprint = "1501.04891",
    archivePrefix = "arXiv",
    primaryClass = "hep-th",
    reportNumber = "USTC-ICTS-15-02, BONN-TH-2015-01",
    doi = "10.1007/JHEP10(2015)125",
    journal = "JHEP",
    volume = "10",
    pages = "125",
    year = "2015"
}

@article{Kim:2016foj,
    author = "Kim, Hee-Cheol and Kim, Seok and Park, Jaemo",
    title = "{6d strings from new chiral gauge theories}",
    eprint = "1608.03919",
    archivePrefix = "arXiv",
    primaryClass = "hep-th",
    reportNumber = "SNUTP16-002",
    month = "8",
    year = "2016"
}

@article{DelZotto:2016pvm,
    author = "Del Zotto, Michele and Lockhart, Guglielmo",
    title = "{On Exceptional Instanton Strings}",
    eprint = "1609.00310",
    archivePrefix = "arXiv",
    primaryClass = "hep-th",
    doi = "10.1007/JHEP09(2017)081",
    journal = "JHEP",
    volume = "09",
    pages = "081",
    year = "2017"
}

@article{Katz:1999xq,
    author = "Katz, Sheldon H. and Klemm, Albrecht and Vafa, Cumrun",
    title = "{M theory, topological strings and spinning black holes}",
    eprint = "hep-th/9910181",
    archivePrefix = "arXiv",
    reportNumber = "HUTP-99-A056, IASSNS-HEP-98-107, OSU-M-99-9",
    doi = "10.4310/ATMP.1999.v3.n5.a6",
    journal = "Adv. Theor. Math. Phys.",
    volume = "3",
    pages = "1445--1537",
    year = "1999"
}

@article{DelZotto:2017mee,
    author = "Del Zotto, Michele and Gu, Jie and Huang, Min-Xin and Kashani-Poor, Amir-Kian and Klemm, Albrecht and Lockhart, Guglielmo",
    title = "{Topological Strings on Singular Elliptic Calabi-Yau 3-folds and Minimal 6d SCFTs}",
    eprint = "1712.07017",
    archivePrefix = "arXiv",
    primaryClass = "hep-th",
    doi = "10.1007/JHEP03(2018)156",
    journal = "JHEP",
    volume = "03",
    pages = "156",
    year = "2018"
}

@article{Alim:2012ss,
    author = "Alim, Murad and Scheidegger, Emanuel",
    title = "{Topological Strings on Elliptic Fibrations}",
    eprint = "1205.1784",
    archivePrefix = "arXiv",
    primaryClass = "hep-th",
    doi = "10.4310/CNTP.2014.v8.n4.a4",
    journal = "Commun. Num. Theor. Phys.",
    volume = "08",
    pages = "729--800",
    year = "2014"
}

@article{Gu:2017ccq,
    author = "Gu, Jie and Huang, Min-xin and Kashani-Poor, Amir-Kian and Klemm, Albrecht",
    title = "{Refined BPS invariants of 6d SCFTs from anomalies and modularity}",
    eprint = "1701.00764",
    archivePrefix = "arXiv",
    primaryClass = "hep-th",
    doi = "10.1007/JHEP05(2017)130",
    journal = "JHEP",
    volume = "05",
    pages = "130",
    year = "2017"
}

@article{Oberdieck:2017pqm,
      author         = "Oberdieck, Georg and Pixton, Aaron",
      title          = "{Gromov-Witten theory of elliptic fibrations: Jacobi
                        forms and holomorphic anomaly equations}",
      journal        = "Geom. Topol.",
      volume         = "23",
      year           = "2019",
      pages          = "1415-1489",
      doi            = "10.2140/gt.2019.23.1415",
      eprint         = "1709.01481",
      archivePrefix  = "arXiv",
      primaryClass   = "math.AG",
      SLACcitation   = "%%CITATION = ARXIV:1709.01481;%%"
}

@article{Schoen,
    author = "Schoen, C.",
    title = "{On fiber products of rational elliptic surfaces with section}",
    journal = "Math. Z.",
    volume = "197",
    pages = "177-199",
    year = "1988"
}

@article{Lee:2018urn,
    author = "Lee, Seung-Joo and Lerche, Wolfgang and Weigand, Timo",
    title = "{Tensionless Strings and the Weak Gravity Conjecture}",
    eprint = "1808.05958",
    archivePrefix = "arXiv",
    primaryClass = "hep-th",
    reportNumber = "CERN-TH-2018-190",
    doi = "10.1007/JHEP10(2018)164",
    journal = "JHEP",
    volume = "10",
    pages = "164",
    year = "2018"
}

@article{Lee:2018spm,
    author = "Lee, Seung-Joo and Lerche, Wolfgang and Weigand, Timo",
    title = "{A Stringy Test of the Scalar Weak Gravity Conjecture}",
    eprint = "1810.05169",
    archivePrefix = "arXiv",
    primaryClass = "hep-th",
    reportNumber = "CERN-TH-2018-220",
    doi = "10.1016/j.nuclphysb.2018.11.001",
    journal = "Nucl. Phys. B",
    volume = "938",
    pages = "321--350",
    year = "2019"
}

@article{Heidenreich:2015nta,
    author = "Heidenreich, Ben and Reece, Matthew and Rudelius, Tom",
    title = "{Sharpening the Weak Gravity Conjecture with Dimensional Reduction}",
    eprint = "1509.06374",
    archivePrefix = "arXiv",
    primaryClass = "hep-th",
    doi = "10.1007/JHEP02(2016)140",
    journal = "JHEP",
    volume = "02",
    pages = "140",
    year = "2016"
}

@article{Alim:2021vhs,
    author = "Alim, Murad and Heidenreich, Ben and Rudelius, Tom",
    title = "{The Weak Gravity Conjecture and BPS Particles}",
    eprint = "2108.08309",
    archivePrefix = "arXiv",
    primaryClass = "hep-th",
    reportNumber = "ACFI-T21-09",
    doi = "10.1002/prop.202100125",
    journal = "Fortsch. Phys.",
    volume = "69",
    number = "11-12",
    pages = "2100125",
    year = "2021"
}

@article{Palti:2017elp,
    author = "Palti, Eran",
    title = "{The Weak Gravity Conjecture and Scalar Fields}",
    eprint = "1705.04328",
    archivePrefix = "arXiv",
    primaryClass = "hep-th",
    doi = "10.1007/JHEP08(2017)034",
    journal = "JHEP",
    volume = "08",
    pages = "034",
    year = "2017"
}

@article{Vafa:2005ui,
    author = "Vafa, Cumrun",
    title = "{The String landscape and the swampland}",
    primaryClass = "hep-th", eprint="hep-th/0509212",
    archivePrefix = "arXiv",
    reportNumber = "HUTP-05-A043",
    month = "9",
    year = "2005"
}

@article{Harlow:2022gzl,
    author = "Harlow, Daniel and Heidenreich, Ben and Reece, Matthew and Rudelius, Tom",
    title = "{The Weak Gravity Conjecture: A Review}",
    eprint = "2201.08380",
    archivePrefix = "arXiv",
    primaryClass = "hep-th",
    reportNumber = "ACFI-T22-01",
    month = "1",
    year = "2022"
}

@article{Cheung:2014vva,
    author = "Cheung, Clifford and Remmen, Grant N.",
    title = "{Naturalness and the Weak Gravity Conjecture}",
    eprint = "1402.2287",
    archivePrefix = "arXiv",
    primaryClass = "hep-ph",
    reportNumber = "CALT-68-2879",
    doi = "10.1103/PhysRevLett.113.051601",
    journal = "Phys. Rev. Lett.",
    volume = "113",
    pages = "051601",
    year = "2014"
}

@article{Ooguri:2006in,
    author = "Ooguri, Hirosi and Vafa, Cumrun",
    title = "{On the Geometry of the String Landscape and the Swampland}",
    primaryClass = "hep-th", eprint="hep-th/0605264",
    archivePrefix = "arXiv",
    reportNumber = "CALT-68-2600, HUTP-06-A017",
    doi = "10.1016/j.nuclphysb.2006.10.033",
    journal = "Nucl. Phys. B",
    volume = "766",
    pages = "21--33",
    year = "2007"
}

@article{Dvali:2009ks,
    author = "Dvali, Gia and Lust, Dieter",
    title = "{Evaporation of Microscopic Black Holes in String Theory and the Bound on Species}",
    eprint = "0912.3167",
    archivePrefix = "arXiv",
    primaryClass = "hep-th",
    reportNumber = "CERN-PH-TH-2009-243, MPP-2009-205, LMU-ASC-56-09",
    doi = "10.1002/prop.201000008",
    journal = "Fortsch. Phys.",
    volume = "58",
    pages = "505--527",
    year = "2010"
}

@article{Dvali:2007hz,
    author = "Dvali, Gia",
    title = "{Black Holes and Large N Species Solution to the Hierarchy Problem}",
    eprint = "0706.2050",
    archivePrefix = "arXiv",
    primaryClass = "hep-th",
    doi = "10.1002/prop.201000009",
    journal = "Fortsch. Phys.",
    volume = "58",
    pages = "528--536",
    year = "2010"
}

@article{Lee:2019wij,
    author = "Lee, Seung-Joo and Lerche, Wolfgang and Weigand, Timo",
    title = "{Emergent strings from infinite distance limits}",
    eprint = "1910.01135",
    archivePrefix = "arXiv",
    primaryClass = "hep-th",
    reportNumber = "CERN-TH-2019-159",
    doi = "10.1007/JHEP02(2022)190",
    journal = "JHEP",
    volume = "02",
    pages = "190",
    year = "2022"
}

@article{Grana:2021zvf,
    author = "Gra\~na, Mariana and Herr\'aez, Alvaro",
    title = "{The Swampland Conjectures: A Bridge from Quantum Gravity to Particle Physics}",
    eprint = "2107.00087",
    archivePrefix = "arXiv",
    primaryClass = "hep-th",
    doi = "10.3390/universe7080273",
    journal = "Universe",
    volume = "7",
    number = "8",
    pages = "273",
    year = "2021"
}

@article{Bastian:2020egp,
    author = "Bastian, Brice and Grimm, Thomas W. and van de Heisteeg, Damian",
    title = "{Weak gravity bounds in asymptotic string compactifications}",
    eprint = "2011.08854",
    archivePrefix = "arXiv",
    primaryClass = "hep-th",
    doi = "10.1007/JHEP06(2021)162",
    journal = "JHEP",
    volume = "06",
    pages = "162",
    year = "2021"
}

@article{Gendler:2020dfp,
    author = "Gendler, Naomi and Valenzuela, Irene",
    title = "{Merging the weak gravity and distance conjectures using BPS extremal black holes}",
    eprint = "2004.10768",
    archivePrefix = "arXiv",
    primaryClass = "hep-th",
    doi = "10.1007/JHEP01(2021)176",
    journal = "JHEP",
    volume = "01",
    pages = "176",
    year = "2021"
}

\end{document}